%% file: main.tex
\documentclass[journal=jacsat,manuscript=article]{achemso}

\usepackage[version=3]{mhchem} 
\usepackage{nicefrac}
\usepackage{makecell}
\usepackage{ltablex}
\usepackage{siunitx}
\usepackage{graphicx}
\usepackage{newfloat}
\usepackage{array}      
\usepackage{longtable} 
\usepackage{booktabs}
\usepackage{setspace}
\usepackage{float}
\usepackage{pdflscape}
\usepackage{xltabular}        
\newcolumntype{C}[1]{>{\centering\arraybackslash}m{#1}}
\usepackage{textcomp}
\usepackage{xurl}
\usepackage[hidelinks]{hyperref}
\newcommand{\doi}[1]{\href{https://doi.org/#1}{\nolinkurl{#1}}}

\newcommand*{\SItablefont}{\footnotesize}
\usepackage{subcaption}
\usepackage{rotating}

\newlength{\panelh}
\newcommand{\panelfigdir}{}
\newcounter{panelctr}
\newcommand{\panel}[2]{
  \refstepcounter{panelctr}
  \IfFileExists{\panelfigdir#1}
    {\includegraphics[height=\panelh,width=0.85\linewidth,keepaspectratio]{#1}}
    {\fbox{\parbox[c][\panelh][c]{0.75\linewidth}{\centering\small\textit{Missing:}
      \detokenize{#1}}}}\\[2pt]
  {\small(\alph{panelctr})}\label{#2}}

\newcolumntype{C}[1]{>{\centering\arraybackslash}m{#1}}

\author{Elodie A. Sandraz}
\author{Christopher Wolverton}
\email{c-wolverton@northwestern.edu}
\affiliation[Northwestern]{Department of Materials Science and Engineering, Northwestern University, Evanston, IL,
60208, USA}

\title{Heterostructure morphology stability maps in multiphase nanoparticles}

\abbreviations{}

\begin{document}

\begin{abstract}
  In the rapidly growing field of nanocombinatorics, the advent of megalibraries of multiphase nanoparticles (NPs) has created a need for predictive computational approaches. A notable characteristic of multiphase NPs is the internal interface architecture that arises from the joining of constituent phases at boundaries. We propose a strategy to predict the ground-state interface architecture from the DFT-calculated surface and interfacial energies of all constituent phases as input and use this strategy to create stability maps to predict multiphase NP morphologies. A Potts model, a $q$-state generalization of the Ising model, in which $q-1$ spins represent discretized regions of a constituent phase, provides a flexible, discretized representation of a $(q-1)$-phase nanoparticle with arbitrary interfacial join morphology. Interactions between spins represent interactions between phases, or interfacial energies. To model the surface of a nanoparticle, an additional spin represents the surrounding environment and interaction energies with this spin are surface energies. We apply a Monte Carlo algorithm to search for the lowest-energy spin configuration. The trade-off between surface and interfacial energy minimization drives the formation of an interface morphology. We calculate a biphase nanoparticle interface morphology stability map to study the relation between the three biphase morphologies we found. Next, we use the stability map to predict NP morphologies of sample biphase systems and compare these to experimentally available data.
  We repeat the grid search approach for triphase nanoparticle interface morphologies and discover 11 triphase morphologies, nine more than previously enumerated. We then predict sample triphase systems and finally revisit experimentally reported triphase morphologies.
\end{abstract}

Nanostructures exhibit valuable optical, chemical, electronic, and magnetic properties applicable in catalysis \cite{mirkincatalystdesign}, plasmonics, \cite{Agplasmonic, plasmonics, metalNPplasmwaveguide, polyelemnp_plasmon} medicine and therapeutics, \cite{nanomedicine, magreview} and information storage.\cite{magreview, bomag}
Introducing multiple metals improves the catalytic, \cite{mirkincatalystdesign, quatCO2catalysts, Huang2020High-Index-FacetElectrocatalysts, PNASTHH, PdPtlattice, colloidoptoelect, pinchedplasmonic, multicompplasmonic, PdPtnanodendrites}  optical,\cite{nanointerf, plasmoninterf} electronic, \cite{nanointerf, plasmoninterf} plasmonic, \cite{polyelemnp_plasmon, catalystinterf, MohanPNP} and magnetic \cite{magreview, bomag} properties of monometallic nanostructures, resulting in the formation of multicomponent nanoparticles (NPs). Multicomponent NPs can be single-phase or phase-separate to form multiphase NPs.
The internal interfaces that arise between different phases within these multicomponent NPs hold promise for enhancing electronic, \cite{Mangeri2015InfluenceNanoparticles, nanointerf} plasmonic, \cite{polyelemnp_plasmon, plasmoninterf} and catalytic \cite{PdPtlattice, catalystinterf} properties through techniques such as charge transfer tuning and strain engineering \cite{Mangeri2015InfluenceNanoparticles, plasmoninterf, catalystinterf, MohanPNP, Chen2016PolyelementalLibraries, Gamler2019BuildingDurability} as well as defect engineering.\cite{Gamler2020EffectNanocrystals}
Internal domain morphology and distribution become important in catalytic,\cite{voidcontrol, mutlimetallicplates} energy conversion,\cite{nanointerf} and biomedical applications,\cite{nanointerf, Mangeri2015InfluenceNanoparticles} thus warranting further investigation. Researchers are developing nanoparticle megalibraries to expedite the discovery of new multiphase nanoparticles. \cite{Smith2023Megalibraries:Discovery, Kluender2019CatalystNanomaterials, Smith2023MolecularCatalysts, schaaknanorod, schaakreview, MohanPNP, Chen2016PolyelementalLibraries, schaak2025}
While previous theoretical works have studied sample nanoparticle systems,\cite{MohanPNP, jacs, polymersurfacepassivation} a comprehensive study of how domain interfacial and surface energies interplay to shape internal NP morphology remains unexplored.

While many studies attribute lattice mismatch as the driving force of multicomponent nanoparticle architecture, \cite{PdPtlattice, nanointerf, plasmoninterf, jacs} quantitative examination of the role of interfacial and surface energies is still under study. \cite{MohanPNP, Chen2016PolyelementalLibraries, jacs}
To determine the lowest-energy internal interface morphology within multiphase NPs, Chen \textit{et al.} proposed an idealized geometry-based method.\cite{MohanPNP} Chen \textit{et al.} use eq. \ref{eqn:np_toten}, reproduced here, to calculate NP total energy by summing the surface and interfacial energies of a particle’s domains; or in the case of multiphase NPs, of the constituent phases. In eq. \ref{eqn:np_toten}, $i$ and $j$ represent a constituent phase of an $n$-phase NP, thereby taking on values between $1$ and $n$. The total energy $E$ of the NP is the sum of the surface energies and the interfacial energies of the constituent phases multiplied by their corresponding surface and interfacial areas.
The surface energy of a domain $i$ is the surface energy per unit area of phase $i$, $\gamma_i$, multiplied by the corresponding surface area, $A_i$. The interfacial energy of domains $i$ and $j$, is the interfacial energy per unit area, $\sigma_{ij}$, of phases $i$ and $j$ multiplied by the shared interfacial area, $A_{ij}$.
Evaluating eq. \ref{eqn:np_toten} requires (i) the identity of the phases, $i$ and $j$, (ii) their corresponding surface energies $\gamma_i$, (iii) interfacial energies $\sigma_{ij}$, and (iv) the corresponding surface ($A_i$, $A_j$) and interfacial areas $A_{ij}$. 

\begin{equation}
  E = \sum_{i=1}^{n}{\left(\gamma_i A_i + \left( \sum_{j > i}^{n}{\sigma_{ij}A_{ij}} \right)\right)}
  \label{eqn:np_toten}
\end{equation}

\begin{figure}
    \centering
    \includegraphics[width=0.45\linewidth]{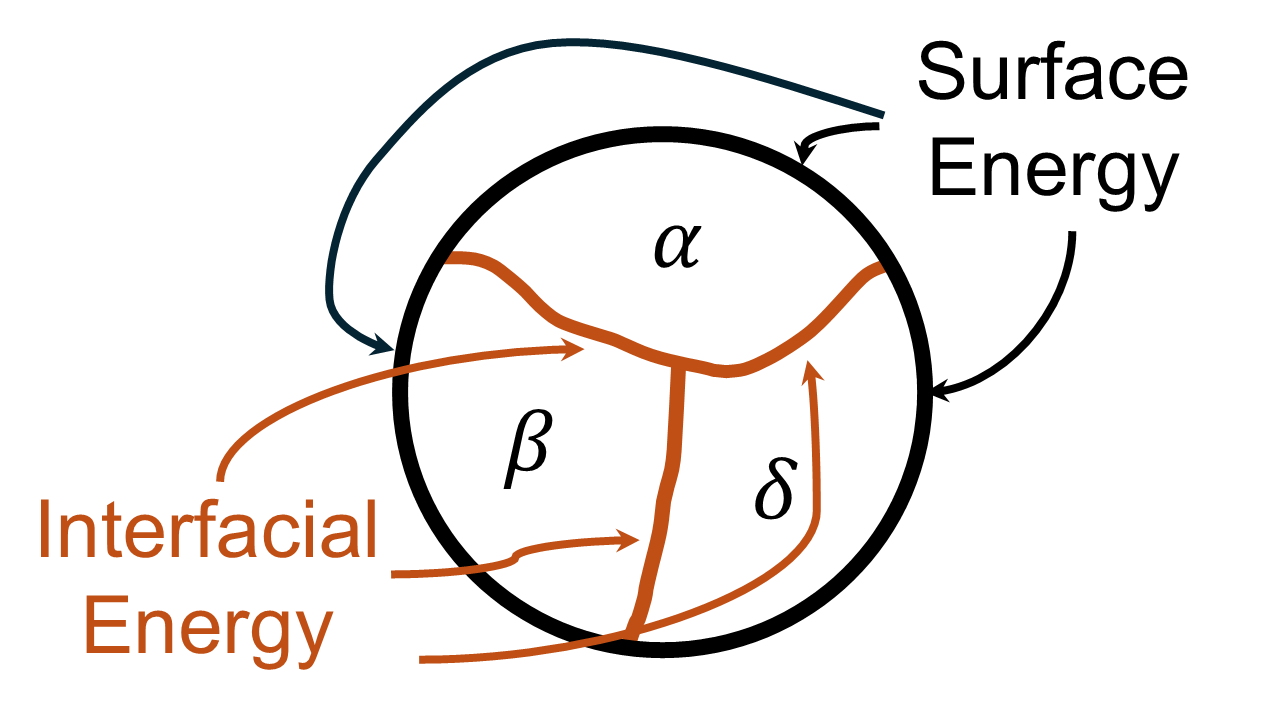}
    \caption{Sketch accompanying eq.~\ref{eqn:np_toten} to calculate the total energy of a multiphase NP. The total energy is the sum of the surface and interfacial energies of the phases $\alpha,\beta,\delta$.}
    \label{fig:NP_tot_en}
\end{figure}

To determine the values of $\gamma_i$ and $\sigma_{ij}$, and hence $E$, we first need the identities of the phases $i$ and $j$. The grand canonical linear programming (GCLP)\cite{gclp} approach implemented within the Open Quantum Materials Database (OQMD)\cite{oqmd1,oqmd2} calculates the $T=0\ \si{\kelvin}$ thermodynamically stable ground-state, bulk phases from the elemental composition, which we use to determine the total number of domains $n$ as well as their identity. We note the OQMD does not include solid solutions, so the results of this approach do not capture their formation within a NP. The results from Ref. \citenum{MohanPNP} suggest that this limitation does not have a significant impact on the predicted internal NP morphology.
We use Density Functional Theory (DFT) to calculate the surface ($\gamma$) and interfacial energies ($\sigma$) of each phase. Although DFT evaluates these energies under $T=0\ \si{\kelvin}$ conditions, using these has previously worked well in modeling experimental NP morphologies. \cite{MohanPNP, JiahongNW}

The total energy of the morphology is then determined by the values of $A_i$ and $A_{ij}$, or the surface and interfacial areas of the constituent phase domains. Prior work obtained $A_i$ and $A_{ij}$ analytically for highly idealized geometries, and the simplifying assumption of equivolume amounts of each phase. The distribution of phases within the NP affects the values of $A_i$ and $A_{ij}$, which impacts the calculated NP total energy. To find the lowest-energy morphology configuration, one can enumerate the possible domain distributions and use simple geometric relations to calculate $A_i$ and $A_{ij}$. When applying eq.~\ref{eqn:np_toten} to each enumerated morphology, Chen \textit{et al.} showed in select triphase cases (\ce{Au-Co-Pd3Sn}, \ce{Ag-Cu-Co}) that the lowest-energy morphology calculated coincides with the experimentally observed morphology,\cite{MohanPNP,JiahongNW} implying that the use of an idealized geometry model in conjunction with DFT to solve eq.~\ref{eqn:np_toten} is a remarkably good simple approximation of the total energy of multicomponent nanoparticles. 

To find the stable NP morphology with minimum total energy, we must minimize $E$ with respect to $A_i$ and $A_{ij}$.
Since $i$ and $j$ are decided by the NP composition, so are $\gamma_i$ and $\sigma_{ij}$. The remaining optimizable parameter is the area that each phase occupies, or the surface and interfacial area terms $A_i$ and $A_{ij}$.
As morphologies become more complicated, such as the tetra-phase nanoparticles illustrated in Figure S1 in Ref. \citenum{MohanPNP}, the analytical calculation of $A_i$ and $A_{ij}$ becomes difficult. Furthermore, this method limits the search for the lowest-energy configuration among the configurations that are manually enumerated; however, it is plausible that the lowest-energy morphology may be missed during manual enumeration.
We need an improved method to explore all non-idealized geometries of interface configurations while minimizing the NP total energy.

We propose the Potts model, a generalized $q$-state Ising model,\cite{potts,Wu1982TheModel,Wu1988PottsTheory} as a discretized, coarse-grained representation of multiphase nanoparticles, and minimize it with a spin-conserved Metropolis Monte Carlo (MC) algorithm\cite{Metropolis1949TheMethod, Metropolis1953EquationMachines}, or equivalently Kawasaki exchange dynamics, to search for the ground-state interface morphology. In our proposed Potts model, each $q$ spin state represents membership to a domain,\cite{Holm2001, Janssens2003RandomGrowth} and the model has previously been applied to a range of boundary-driven phenomena, including grain growth,\cite{graingrowthpotts} Ostwald ripening,\cite{ostwaldpotts, Tikare1998ComparisonProcesses} foam evolution,\cite{Janssens2003RandomGrowth, foam} bubble evolution, crystallization, and nucleation.\cite{Holm2001, Homer2013HybridEvolution} These applications share a common physical basis: the Potts model evolves toward the configuration that minimizes total boundary area,\cite{Holm2001} which is precisely the quantity that governs NP interface morphology in eq. \ref{eqn:np_toten}. MC search is well suited to this minimization because it can efficiently explore high-dimensional configuration spaces\cite{glotzersimrev} without requiring prior knowledge of the energy landscape.

To map a multicomponent NP onto the Potts model, we define a spin state, $q$, as representing the identity of a phase. The $n$-phase NP is mapped onto a Potts lattice surrounded by some environment. Because we represent the environment as an additional spin, we use an $(n + 1)$-spin Potts model, and each lattice point represents a discretized volume of space. The energy of the system becomes a discretized version of eq.~\ref{eqn:np_toten}, or a sum of nearest-neighbor (NN) and next-nearest-neighbor (NNN) interactions described by the Hamiltonian in eq.~\ref{eqn:ham}. The additional shell of NNN interactions mitigates lattice anisotropy effects.\cite{Holm2001} 
The surface and interfacial energies of the constituent phases set the interaction energy between lattice sites, held in the zero-diagonal symmetric matrix, $J_{q_{m}q_{n}}$, which represents interfacial energies when $q_{m},q_{n}$ (spins on lattice sites $m$ and $n$, respectively) are phases and surface energies when either represents the environment.

\begin{equation}
  H = \frac{1}{2}\sum_{m,\ n}^{\mathrm{NN,\ NNN}}{J_{q_m q_n}(1-\delta_{q_m q_n})}
  \label{eqn:ham}
\end{equation}

\begin{figure}[h!]
\centering
\includegraphics[height=10cm]{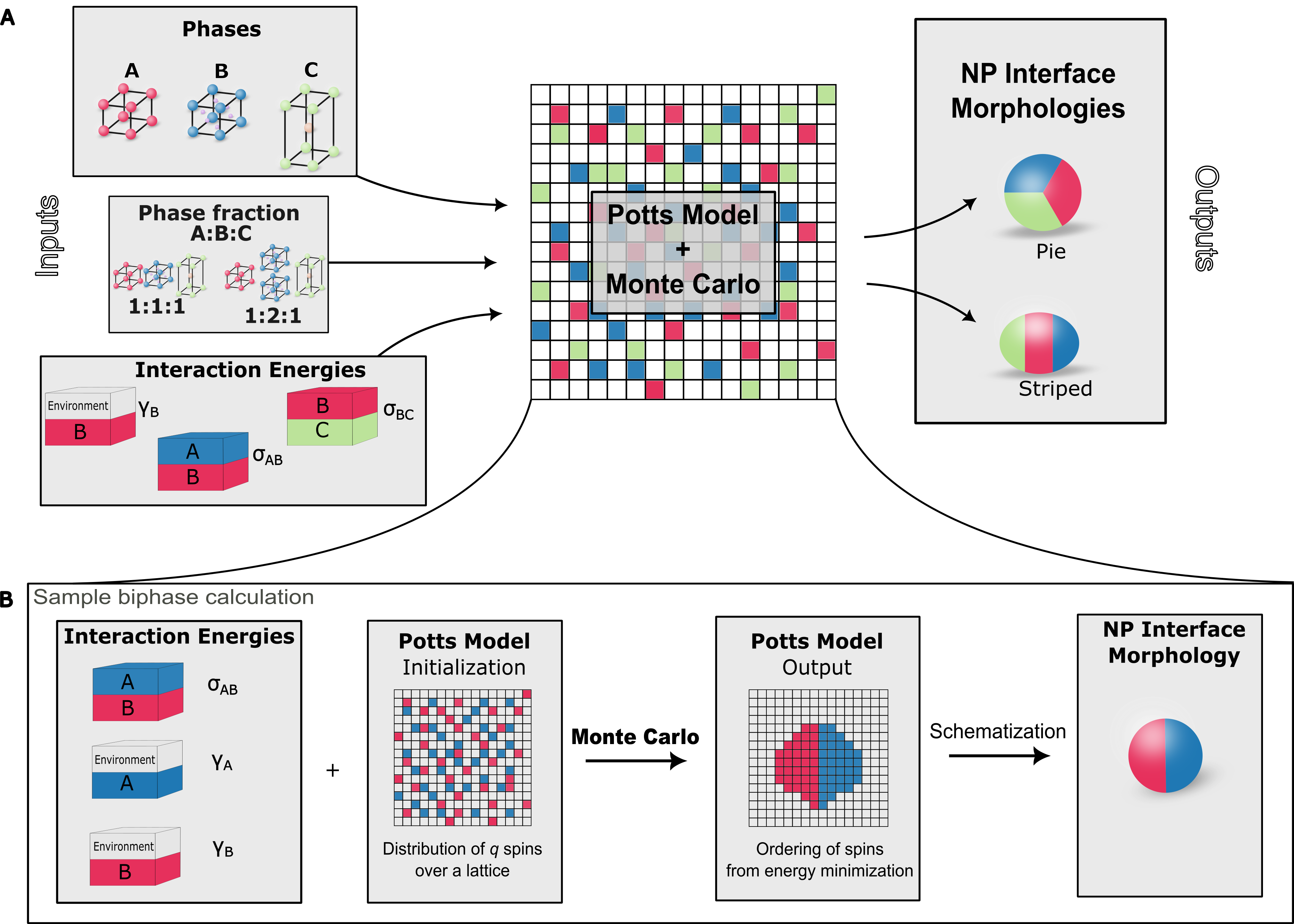}
  \caption{Schematic of the proposed Monte Carlo Potts model applied to multiphase nanoparticle interface morphology determination with illustrated inputs and outputs. (A) The identity of the phases, their volume fractions, and their pairwise interaction energies (i.e. surface and interfacial energies) are input to a Monte Carlo Potts model, which calculates the ground-state interface morphology from the inputs. The Potts model has lattice points representing an integrated region of space of each phase. (B) Schematic of a sample biphase calculation. The Potts model is initialized by distributing spins, each representing membership to a phase A, B, C, etc., across a lattice. The pairwise interaction energies of the spins are surface and interfacial energies. A spin-conserved Metropolis Monte Carlo algorithm is applied to the Potts model to minimize eq. \ref{eqn:np_toten} via its discretized form, the Hamiltonian in eq. \ref{eqn:ham}. The resulting ordered lattice is schematized to determine the NP interface morphology of the constituent phases of the NP.}
  \label{fig:fig1}
\end{figure}

The identity of the phases, represented by values of $q$, is determined via methods such as GCLP, as described previously.\cite{gclp, oqmd1, oqmd2} To capture the phase-separation effects observed in multiphase NPs, like interactions in the Potts model must be energetically favorable. Equation \ref{eqn:np_toten} accounts for this via its zero-energy internal energy of each phase. In the Potts model, this is replicated by using zero-energy like spin interactions, or internal energy of each phase, that do not contribute to the NP total energy. This approach is also used in computational grain evolution studies to reduce computational cost, particularly when the interface is the region of interest.\cite{Janssens2003RandomGrowth} We use the Kronecker delta, $\delta$, to discount the internal energy: $\delta$ takes spin states $q_{m}$ and $q_{n}$ as inputs and is valued at 1 when they are the same and valued at 0 otherwise. Multiplying the interaction energy matrix $J_{q_{m}q_{n}}$ by a $(1-\delta_{q_{m}q_{n}})$ term ensures that only unlike spin interactions contribute to the total energy of the particle, and we multiply the overall sum by a factor of one-half to correct for double counting of pairwise lattice interactions. The result is a discretized total energy equation, eq. \ref{eqn:ham}, in the familiar form of a Monte Carlo Hamiltonian, which we minimize using a Metropolis Monte Carlo algorithm.\cite{Metropolis1949TheMethod,Metropolis1953EquationMachines}

We solve for the ground-state spin configuration -- equivalently, the ground-state NP interface morphology -- using the methodology illustrated in Figure \ref{fig:fig1}. After determining the number and identity of the constituent phases (denoted here as A and B), we initialize a lattice with randomly distributed spins representing finite amounts of A, B, and Environment. We input sample surface and interfacial energies into $J_{q_{m}q_{n}}$, then evolve the lattice under the spin-conserved Metropolis algorithm with eq. \ref{eqn:ham} as the Hamiltonian, using a simulated annealing approach to avoid entrapment in local minima. We categorize the final spin configuration both visually and quantitatively from the spin-state distribution, into a ground-state interface morphology. We repeat this procedure at least five times per input parameter configuration, using five different random initial lattices, to sample the energy landscape at each point. As we show later, this approach recovers complex morphologies that would be difficult to enumerate manually or express with a simple idealized geometric relation for $A_i$ and $A_{ij}$. We first apply this approach to biphase morphologies and find that the inclusion of transitional morphologies causes significant shifts in the stability map boundaries. We then apply a Monte Carlo Potts (MCP) model to triphase particles and find a substantially richer morphology landscape than previously assumed.

\section{Results}
We use the MCP model to calculate NP morphology stability maps, then use them to predict NP morphologies, and subsequently compare the predictions with their experimentally observed counterparts. We first present biphase NPs followed by triphase NPs.

\subsection{Biphase nanoparticle interface morphologies and stability map}
\label{sec:biphase}
We sampled interface morphologies by varying surface and interfacial energies, from which we calculated a stability map of biphase nanoparticle interface morphologies. Creating a biphase stability map requires the surface energies of the two phases and their interfacial energy, so we normalize the surface and interfacial energies by the largest surface energy to reduce the number of parameters to two.
For example, take sample phases A and B, with phase A having the higher surface energy.
The energy  of the system is described using parameters $\nicefrac{\gamma_B}{\gamma_A}$  and $\nicefrac{\sigma_{AB}}{\gamma_A}$, so that the interaction energy matrix $J_{q_{m}q_{n}} = \left(\begin{smallmatrix}
  0 & 1 & \nicefrac{\gamma_B}{\gamma_A} \\
  1 & 0 & \nicefrac{\sigma_{AB}}{\gamma_A} \\
  \nicefrac{\gamma_B}{\gamma_A} & \nicefrac{\sigma_{AB}}{\gamma_A} & 0 \end{smallmatrix} \right)$.
Normalization using this convention imposes the constraint that $\nicefrac{\gamma_B}{\gamma_A}$ must be less than or equal to $1$. 
Indeed, we find the inverted solution for values $\gamma_B ⁄ \gamma_A > 1$ --- the same spin arrangement with the identities of A and B swapped.
In the remainder of this paper, we will label phases alphabetically by decreasing surface energies such that $\gamma_A \ge \gamma_B$. 
MCP assumes phase-segregating phases, which adds the constraint that the interfacial energy parameter, $\sigma_{AB}$, must be greater than the internal bulk energy, $0$. If the interfacial energy were negative, the two phases would mix in our Potts model, which would result in a new phase requiring a new spin representation.
Furthermore, though we expect interfacial energies to be lower than surface energies\cite{interfacedatabase} -- otherwise a surface would always preferably form over an interface, and no interface would form --  which would result in an upper bound of $\nicefrac{\sigma_{AB}}{\gamma_A} < 1$, we later show that exploring values above 1 is necessary to recover all morphology phase boundaries.
The biphase morphology stability map is thus constructed as a function of $\nicefrac{\gamma_B}{\gamma_A}$  and $\nicefrac{\sigma_{AB}}{\gamma_A}$, within the limits of $\nicefrac{\gamma_B}{\gamma_A} \le 1$ and $\nicefrac{\sigma_{AB}}{\gamma_A} > 0$.

We input sample combinations of $\nicefrac{\gamma_B}{\gamma_A}$  and $\nicefrac{\sigma_{AB}}{\gamma_A}$ into MCP in a grid search fashion to construct the biphase NP stability map. We categorize MCP-calculated biphase particles into three different interface morphologies, shown with their sample MCP outputs and idealized schematics in Table \ref{tab:biphase_morphs}. The Core@Shell (CS) consists of a core surrounded by a shell of the other phase, sharing one interface; only the shell domain has a surface.
The Janus (J) morphology is a single particle split into two side-by-side domains that share an interface, with each domain retaining its own surface.
The 2 separate particles (2p) morphology consists of two fully separate single-phase particles, each with its own surface and no shared interface between them. 
The Janus morphology additionally exhibits two transitional architectures, both observed experimentally:\cite{pinchedplasmonic, Najafishirtari2016TheHeterodimers, George2013CONanocrystals, Monji2017ThermodynamicEffect, mirkincatalystdesign, catalystinterf} 
the Lens-type morphology, in which the core domain contacts the environment but exposes less surface than in the idealized Janus case -- placing it on the boundary between Janus and Core@Shell -- 
and the Dumbbell morphology, in which the particle is pinched along the interface, increasing surface area and decreasing interfacial area relative to idealized Janus. Both are topologically equivalent to Janus but capture variations in interface character that the discretized Potts lattice resolves and that the idealized geometry approach cannot. 

\begin{table*}[h!]
\small
  \caption{\ Biphase morphology descriptions. This table contains the names, visualizations, and characteristics of three topologically distinct biphase particle morphologies and two transition morphologies. The names of the morphologies are followed by categorical type: Core@Shell (CS), Janus (J, including the Lens-type and Dumbbell transition morphologies), and $n$ separate particles ($n$P). Then, a sample MCP model output and a schematic of the idealized morphology, and finally the number of surfaces and interfaces each morphology exhibits, which allows for morphology categorization while illustrating the range of architectures biphase morphologies exhibit.}
  \label{tab:biphase_morphs}
  
 \begin{tabular*}{\textwidth}{
  >{\centering\arraybackslash}m{0.1\textwidth}
  >{\centering\arraybackslash}m{0.05\textwidth}
  >{\centering\arraybackslash}m{0.18\textwidth}
  >{\centering\arraybackslash}m{0.18\textwidth}
  >{\centering\arraybackslash}m{0.1\textwidth}
  >{\centering\arraybackslash}m{0.1\textwidth}}
    \hline
    \thead{Morphology} & \thead{Type} & \thead{Sample MCP\\output} & \thead{Idealized\\Morphology} &    \thead{\# of\\surfaces} & \thead{\# of\\interfaces} \\
    \hline
    Core@Shell &  CS &  \includegraphics[width=\linewidth]{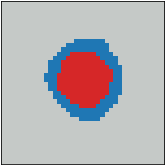} &  \includegraphics[width=\linewidth]{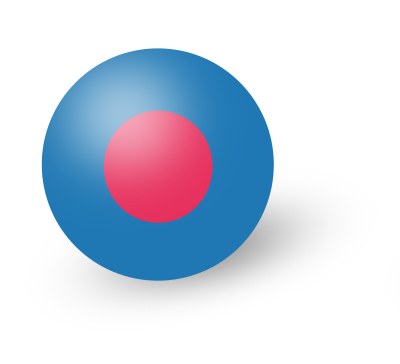}
    &  1 &  1
    \\
    Lens-type \newline (transitional) & J & \includegraphics[width=\linewidth]{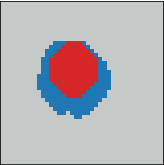} & \includegraphics[width=\linewidth]{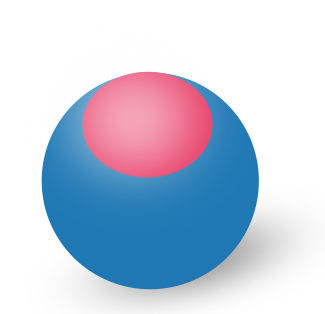} & 2 & 1 \\
    Janus & J & \includegraphics[width=\linewidth]{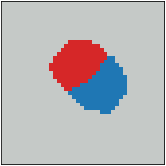} & \includegraphics[width=\linewidth]{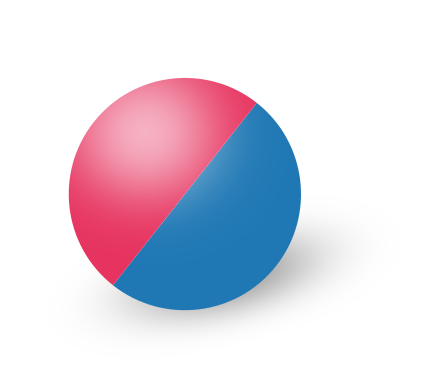}
    & 2 & 1 \\
    \makecell{Dumbbell \\ (transitional)} & J & \includegraphics[width=\linewidth]{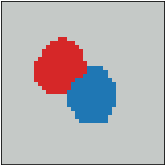} & \includegraphics[width=\linewidth]{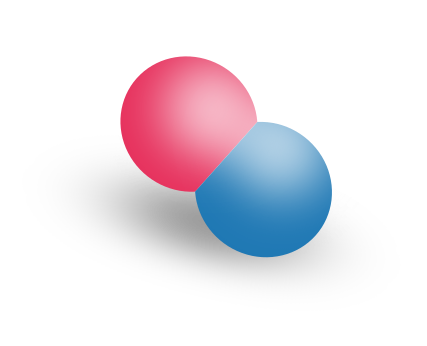}
    & 2 & 1 \\
    2p & \textit{n}P & \includegraphics[width=\linewidth]{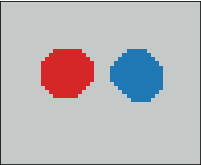} &  \includegraphics[width=\linewidth]{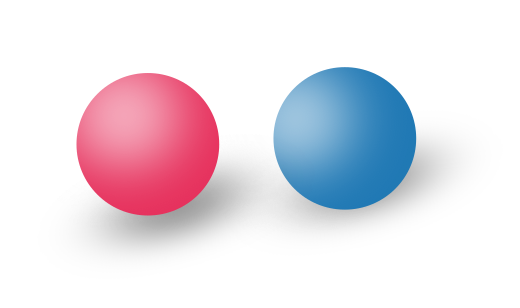}
    & 2 & 0 \\
    \hline
  \end{tabular*}
\end{table*}

\begin{figure*}[h!]
\centering
  \includegraphics[width=14cm]{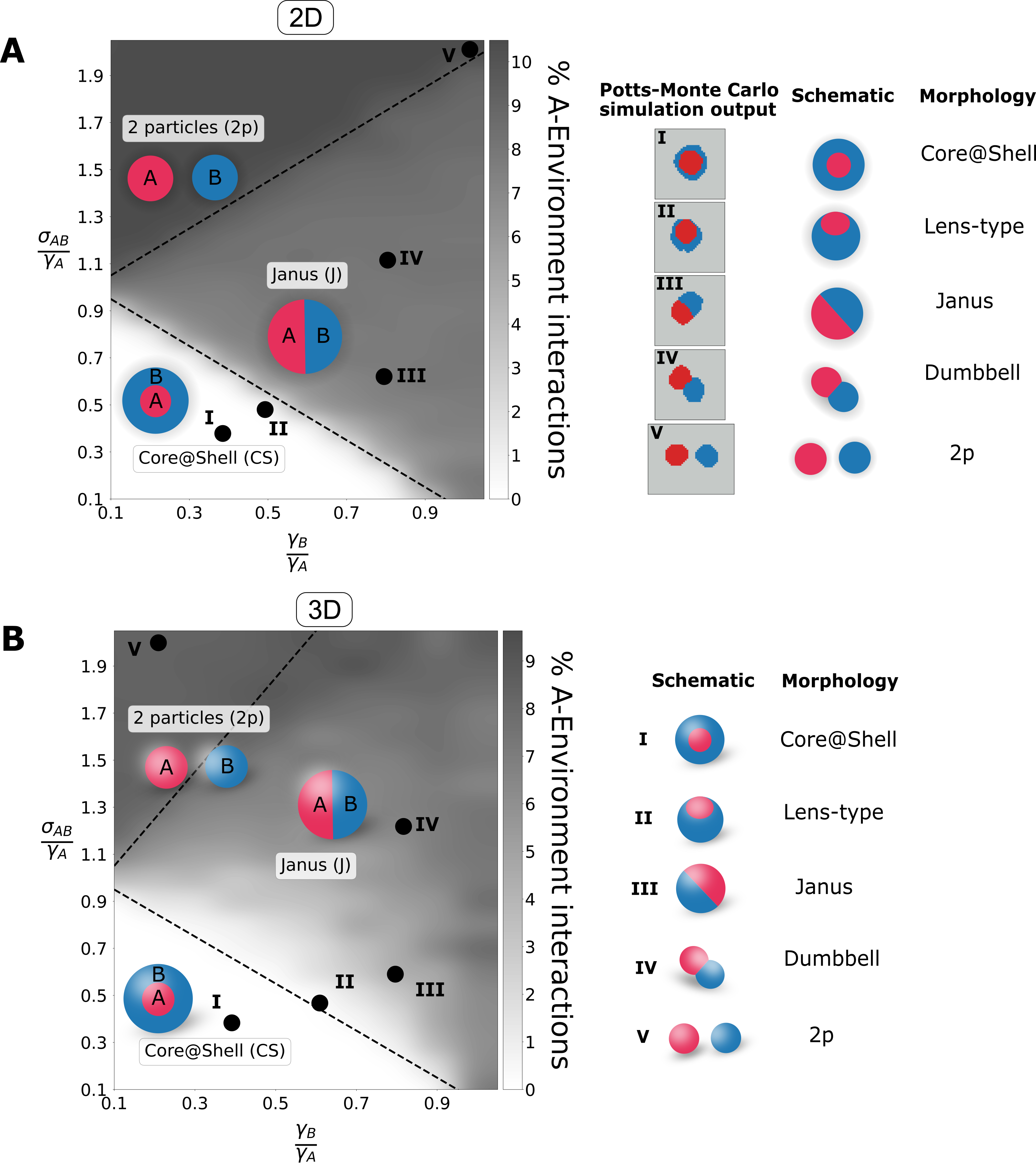}
  \caption{Biphase NP morphology stability maps calculated with MCP on a (A) 2D square lattice and (B) 3D cubic lattice, as a function of $\nicefrac{\gamma_B}{\gamma_A}$ and $\nicefrac{\sigma_{AB}}{\gamma_A}$. The output lattices are described in terms of the number of \ce{A-B}, \ce{A-Environment}, and \ce{B-Environment} NN interactions, normalized by the total number of interactions in the lattice points occupied by the NP; the shading shows the percentage of \ce{A-Environment} interactions. We define NP morphology boundaries (dashed lines) along the discontinuous changes in the amount of \ce{A-Environment} and \ce{A-B} interactions, which correspond to changes in NP morphology. Points I--V mark the sample inputs and the corresponding MCP output, schematic, and morphology (Core@Shell, Lens-type, Janus, Dumbbell, and 2p) shown at right.}
  \label{fig:biphase_interactions}
\end{figure*}

Beyond the classification in Table \ref{tab:biphase_morphs}, the discretized lattice lets us classify each MCP output quantitatively from the number of \ce{A-B}, \ce{A-Environment}, and \ce{B-Environment} nearest-neighbor spin interactions in the final configuration. A morphology is Core@Shell when one of the \ce{A-Environment} or \ce{B-Environment} counts is zero; the phase with zero environment contacts is the core, confirmed by the other phase's environment-contact count being much larger than the \ce{A-B} count.
A morphology falls into the Janus family when all three contact types are nonzero, with their relative magnitudes distinguishing the three variants: Lens-type occurs when one environment-contact count is close to zero and the \ce{A-B} count approaches the Core@Shell value; idealized Janus occurs when the two environment-contact counts are approximately equal; and Dumbbell occurs when both environment-contact counts are again approximately equal but larger than in the idealized Janus case, with a correspondingly smaller \ce{A-B} count. Finally, 2p is identified when the \ce{A-B} count is zero and both environment-contact counts are at their maximum. Figure \ref{fig:biphase_interactions} illustrates this quantitative classification, where we observe discontinuous changes of \ce{A-Environment} interactions along which we can draw a boundary representing the transition from one NP morphology to another.

These morphologies are mapped onto a stability map which captures the effect of surface and interfacial energies on morphology, with phase boundaries representing a change in ground-state morphology. Although these plots are interface morphology phase diagrams, with the boundaries between them representing boundaries between morphologies, we will refer to the diagrams as stability maps and to the morphology boundary curves as phase boundaries for simplicity. We calculated a biphase morphology stability map by inputting combinations of $\nicefrac{\gamma_B}{\gamma_A}$ and $\nicefrac{\sigma_{AB}}{\gamma_A}$, each ranging from 0.005 to 1 as parameters into the MCP model to calculate the location of the phase boundaries between the morphologies discovered.
To investigate the dependence of the formation of two separate particles on interfacial energy, we later input values of $\nicefrac{\sigma_{AB}}{\gamma_A}$ beyond 1 into the MCP model.
MCP morphology outputs reveal internal interface morphology changes with input surface and interfacial energies of the constituent phases, which form the phase boundaries in Figure \ref{fig:biphase_interactions}. We calculate the biphase stability maps using a 2-dimensional square lattice (Fig. \ref{fig:biphase_interactions}A) and 3-dimensional cubic lattice (Fig. \ref{fig:biphase_interactions}B) in the Potts model, which yield slightly shifted phase boundaries owing to the dimensionality of the lattice. The 3-dimensional cubic lattice increases the surface and interfacial areas relative to its 2-dimensional square counterpart, thus increasing the values of $A_i$ and $A_{ij}$ in eq. \ref{eqn:np_toten}, which impact the values of surface ($\gamma_i$) and interfacial ($\sigma_{ij}$) energies at which one morphology is preferred over another.
Beyond this quantitative comparison, the 3D lattice also offers a view of internal interface architecture, complementing the largely two-dimensional snapshots typical of experimental characterization techniques.\cite{Holm2001}

Figure \ref{fig:biphase_interactions} illustrates the effects of surface and interfacial energies on ground-state morphology formation. We will walk through Figure \ref{fig:biphase_interactions}A, beginning in the lower left of the biphase stability map. Point I is in the CS morphology region, which occurs when the interfacial energy is sufficiently smaller (as determined by the phase boundary) than the surface energy of the core. Moving to Point II, the surface energy of the shell ($\gamma_B$) increases relative to the core ($\gamma_A$), leading to the exposure of both surfaces as neither one is wholly preferred.  The initial reduction of the B surface produces the transitional Lens-type morphology. As the B surface shrinks, the A surface increases and a phase boundary forms. Exposure of both surfaces leads to the formation of the Janus morphology. Continuing to Point III, we find the Janus morphology with evenly distributed A and B domains about the center of the particle. From this point, increasing the interfacial energy between A and B brings us to Point IV, where the A and B domains pinch to minimize their interfacial area. The interfacial area decreases, resulting in increased surface areas. A pinched Janus morphology forms, also known as the Dumbbell morphology. Continuing to increase the interfacial energy brings us to Point V, where it becomes more favorable to completely eliminate the interfacial area at the expense of increased surface area. The NP separates into two single-phase particles (2p) with no interface between A and B.
The 2p morphology is only observed when $\sigma_{AB}⁄\gamma_A > \nicefrac{\gamma_B}{\gamma_A} + 1$, or in other words when $\sigma_{AB} > \gamma_A + \gamma_B$.
This MCP-calculated Janus-2p boundary in Fig. \ref{fig:biphase_interactions}A recovers the Dupr\'{e} equation,\cite{workofadhesion} which describes the energy required to separate two surfaces.
We observe the same trends across Points I through V in the 3-dimensional stability map in Figure \ref{fig:biphase_interactions}B.
Although the results presented assume a ratio of $1:1$ amount of spins $A:B$, we repeated the study using ratios varying from $1:2$ to $1:10$ and obtained the same stability map. All morphologies in the biphase stability map have the number of interfaces predicted in Ref. \citenum{MohanPNP}: at most one interface. This study finds the number of surfaces differentiates between biphase morphologies: either $1$ or $2$.

\begin{figure*}[h!]
\centering
  \includegraphics[width=\textwidth]{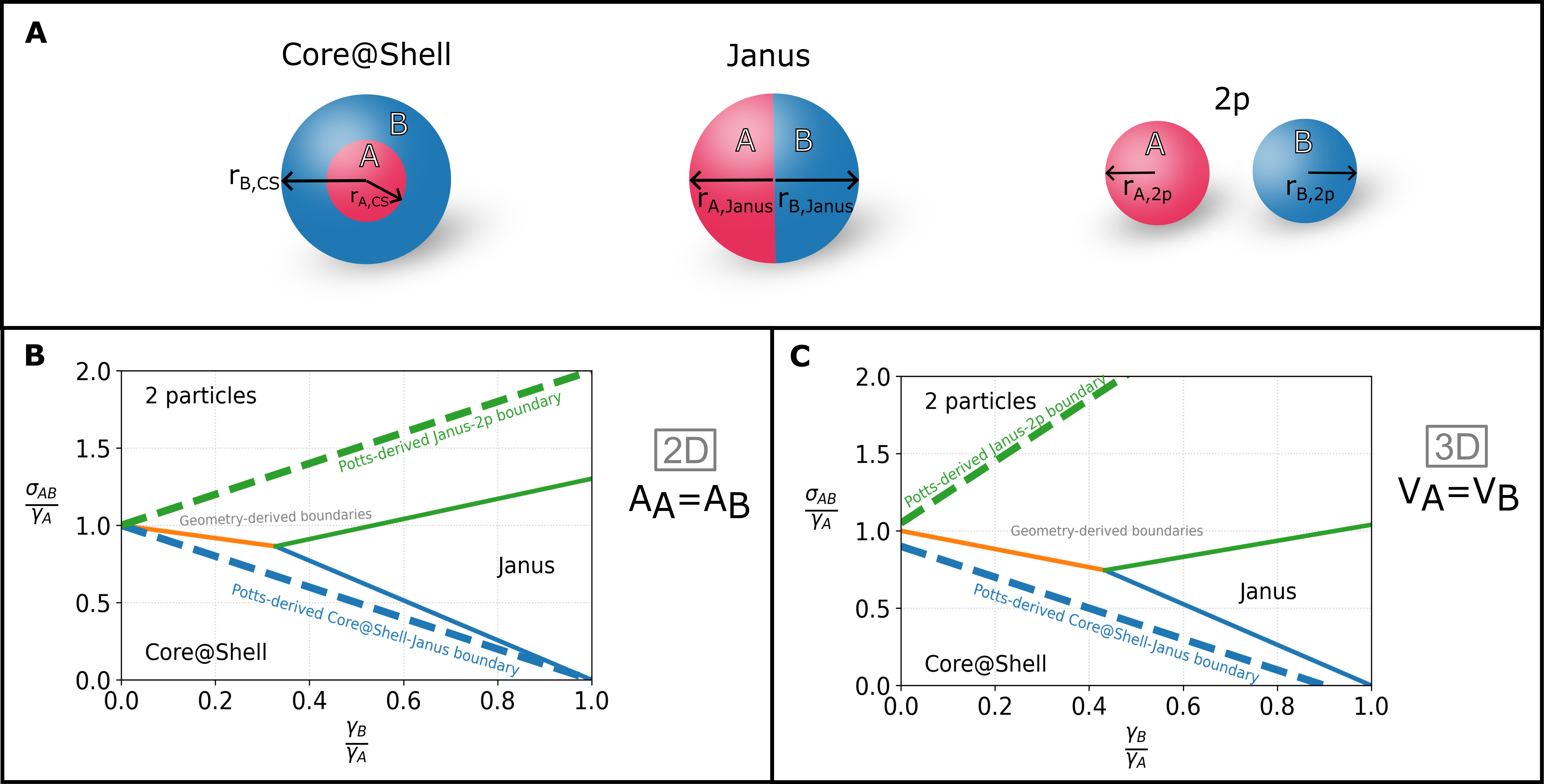}
  \caption{(A) The illustrated variables necessary for calculating eq. \ref{eqn:np_toten} for each morphology using the idealized geometry approach. Phase boundaries occur where at least two morphologies are degenerate in energy. The geometry-derived phase boundaries (solid lines) are compared to the MCP-calculated phase boundaries (dashed lines) under (B) 2D constraints, meaning both MCP and the idealized geometry approach utilize an equal-area constraint, and a square lattice. (C) The 3D case uses an equal-volume constraint and a cubic lattice to reveal further expansion of the Janus region when moving from the idealized geometry to the MCP approach.}
  \label{fig:biphase_diagrams}
\end{figure*}

Because these observed biphase morphologies are geometrically simple, we can solve for the areas $A_i$ and $A_{ij}$ in eq. \ref{eqn:np_toten} using simple geometric relations and the new set of three biphase morphologies for consideration.
We call this the idealized geometry approach, which is described in Ref. \citenum{MohanPNP}.
In Figure \ref{fig:biphase_diagrams}A, we label the necessary geometric dimensions of each morphology.
We use eq. \ref{eqn:np_toten} to calculate the energy of each morphology and select the lowest energy one. Phase boundaries occur where two morphologies are degenerate in energy.
In the 2D case (Fig. \ref{fig:biphase_diagrams}B), each domain maintains equal areas such that $A_A$ = $A_B$, leading to the constraint $A_{A,\ \mathrm{Janus}}=A_{B,\ \mathrm{Janus}}=A_{A,\ \mathrm{Core@Shell}}=A_{B,\ \mathrm{Core@Shell}}=A_{A,\ \mathrm{2p}}=A_{B,\ \mathrm{2p}}$. 
We compare the phase boundaries derived from the idealized geometry approach with the MCP-derived phase boundaries in Figure \ref{fig:biphase_diagrams}B in the 2D case.
We observe a triple point in both stability map calculation approaches, implying there exists some combination of surface and interfacial energies such that all three morphologies are degenerate.
We repeated this approach for the 3D case in Figure \ref{fig:biphase_diagrams}C, with the only change being that each domain maintains the same volume ($V_A$ = $V_B$) across morphologies. The phase boundaries of the 3D consideration are shifted compared to those of the 2D consideration due to the factor differences between the geometric relations of circles ($\pi r^2$) and spheres ($(\nicefrac{4}{3}) \pi r^3$). Additionally, we observe a triple point in the idealized geometry approach but not in the MCP-calculated approach.
Two features of the idealized geometry approach explain the differences in Figs. \ref{fig:biphase_diagrams}B\&C: (1) morphologies do not necessarily exhibit a spherical shape and (2) the approach omits transitional interface morphologies.
For example, the MCP-predicted transition Dumbbell morphology is a Janus morphology, but it is not a perfect sphere. The idealized geometry approach underestimates the surface areas and overestimates the interfacial area. Similarly, the Lens-type morphology has an increased interfacial area compared to the idealized Janus morphology.

The MCP approach requires no a priori knowledge of candidate morphologies and can instead explore all internal interface configurations directly. We find two transitional Janus morphologies (Lens-type and Dumbbell) which expand the Janus morphology space and significantly shift the stability map boundaries. Therefore, it becomes important to consider transitional morphologies when deriving stability maps.

\subsection{Prediction of biphase morphologies}
In this section, we predict biphase NP morphologies using the MCP-calculated biphase stability map from combinations of DFT-calculated surface and interfacial energies. We select phase-separating combinations of isostructural elemental phases, map them onto phases A and B, and use corresponding surface and interfacial energies along the close-packed direction from existing DFT databases.\cite{Tran2016DataCrystals, interfacedatabase}. Other crystallographic directions are compared in the Supplementary Information. We model the NP in vacuum, enabling the use of DFT-calculated surface energies.
We use the above-mentioned DFT interfacial and surface energy databases to calculate the ratios $\nicefrac{\gamma_{B}}{\gamma_A}$ and $\nicefrac{\sigma_{AB}}{\gamma_A}$ to predict the biphase morphology using the stability map.
The resulting morphology predictions are shown in Figure \ref{fig:biphase_pred}, separated by structure (FCC, BCC, and HCP).

Figure \ref{fig:biphase_pred}A lists NP morphology predictions of phase-separating elemental FCC phases. The values used to plot the predictions in Figure \ref{fig:biphase_pred} are listed in Table~\ref{tab:biphase_pred} in the Supplementary Information. MCP predicts all FCC systems as Core@Shell.
The same analysis is conducted for BCC structures using the close-packed direction of [110] in Figure \ref{fig:biphase_pred}B, in which Core@Shell is the most frequently observed morphology, although MCP predicts one Janus NP (\ce{W-Cr}) and one NP on the phase boundary of Core@Shell and Janus (\ce{V-Ta}). 
In Figure \ref{fig:biphase_pred}C, we find HCP systems along the [0001] direction, which also reveal Core@Shell as the most commonly observed NP morphology, followed by two Janus NPs (\ce{Zn-Cd}, \ce{Tl-Cd}) and one on the Core@Shell-Janus phase boundary (\ce{Tc-Co}). Overall, Core@Shell is the most frequently observed morphology when using DFT-calculated surface and interfacial energies, which supports the hypothesis in Ref. \citenum{polymersurfacepassivation} that DFT calculations of surfaces in vacuum will tend to predict Core@Shell NP morphologies.

\begin{figure}[hp!]
    \centering
    \begin{subfigure}[c]{\textwidth}
    \centering
    \includegraphics[width=0.7\textwidth]{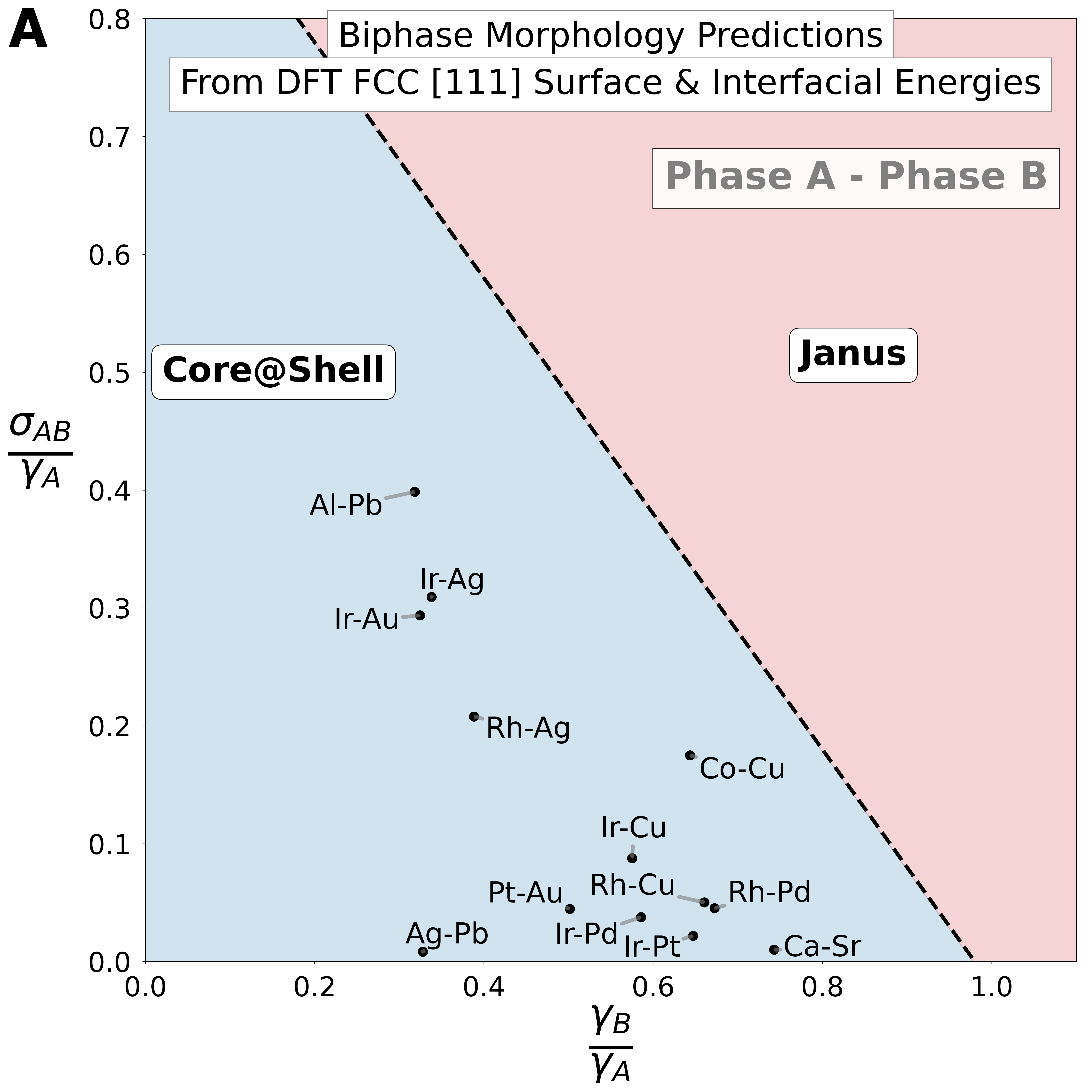}
    \label{fig:first}
\end{subfigure}
\begin{subfigure}[c]{\textwidth}
    \centering
    \includegraphics[width=0.7\textwidth]{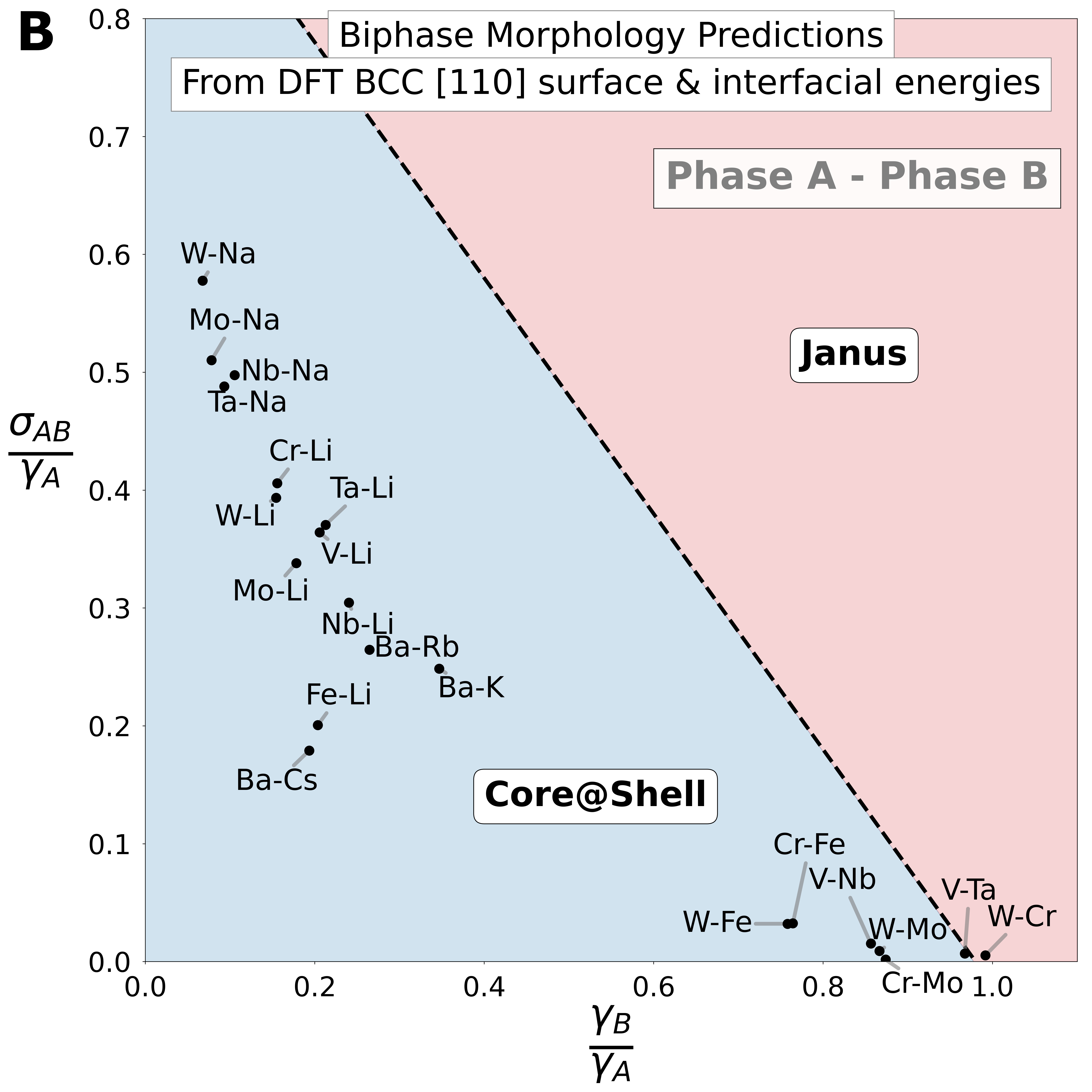}
\end{subfigure}
\end{figure}
\begin{figure}[h!] \ContinuedFloat
\begin{subfigure}[c]{\textwidth}
    \centering
    \includegraphics[width=0.7\textwidth]{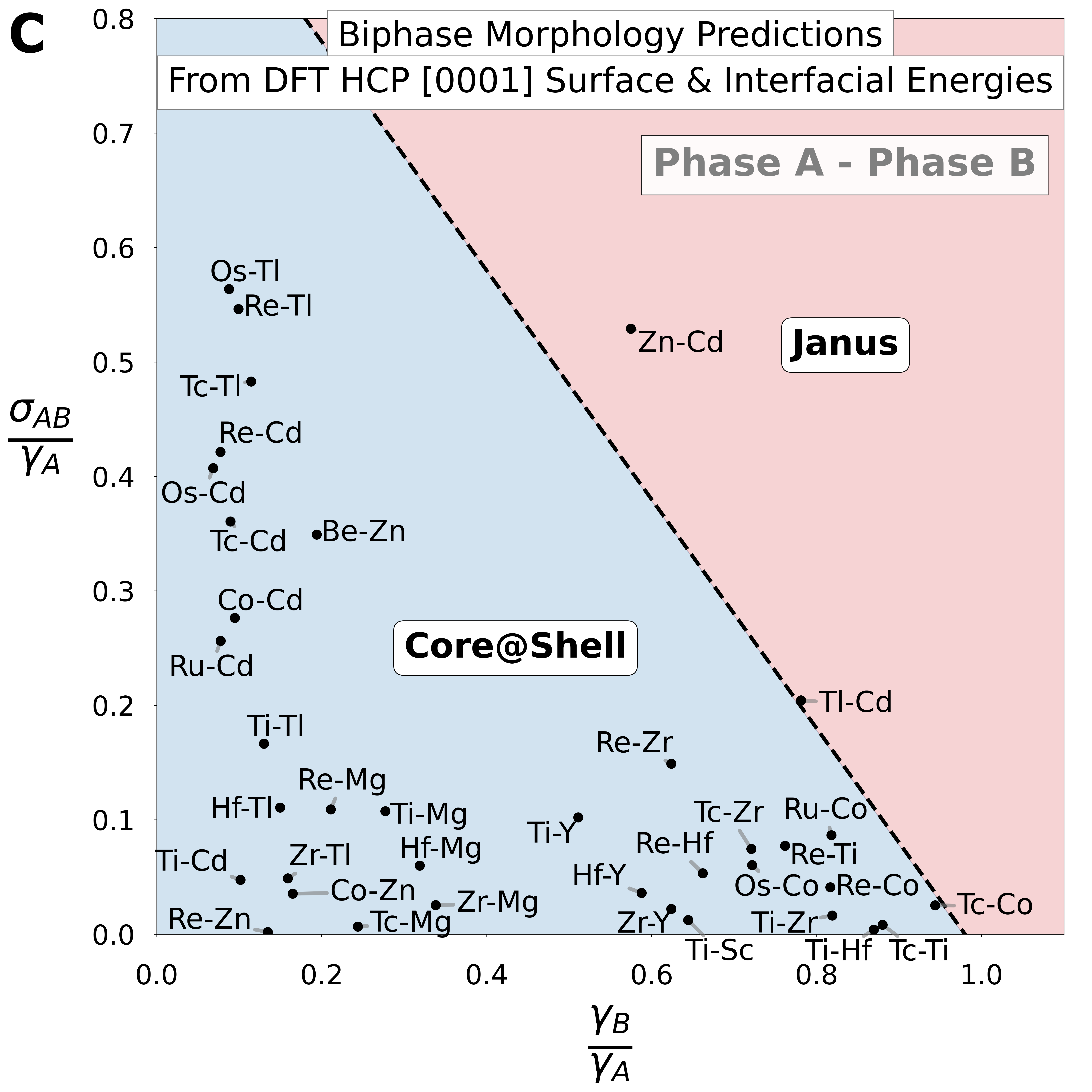}
    \label{fig:third}
\end{subfigure}
\caption{All pairwise combinations of phase-separating isostructural elemental phases drawn from DFT-calculated interfacial energy data in Ref. \citenum{interfacedatabase} are matched to the corresponding surface energies in Ref. \citenum{Tran2016DataCrystals}. The DFT-calculated surface\cite{Tran2016DataCrystals} and interfacial\cite{interfacedatabase} energies of each A-B system are used to calculate the ratios $\nicefrac{\sigma_{AB}}{\gamma_A}$  and  $\nicefrac{\gamma_{B}}{\gamma_A}$, with phase A representing the phase with the higher surface energy of the two. The $(\nicefrac{\gamma_B}{\gamma_A}, \nicefrac{\sigma_{AB}}{\gamma_A})$ point is then plotted on the biphase stability map to predict biphase NP morphologies. The crystallographic direction selected for the surface and interfacial energies is along the phase structure's close-packed direction. The selected elemental phase-separating (A) FCC-FCC systems are predicted using energies along the close-packed [111] direction, (B) BCC-BCC systems along the [110] direction, and finally (C) HCP-HCP systems along the [0001] direction.}
\label{fig:biphase_pred}
\end{figure}

\subsection{Comparison of MCP biphase morphologies with experimental data}
We use the biphase stability map to compare MCP-predicted morphologies with experimentally reported ones and find that six of eight predictions match experiment. We compare the impact of surface and interfacial energy choices on morphology prediction. There are two surface energy choices: DFT-calculated ones\cite{Tran2016DataCrystals} and experimental ones.\cite{expsurfen} The interfacial energies of our systems are generally not available from experiment, so we use DFT-calculated ones from two sources which use different calculation approaches: 1) the database by Salgado-Casanova \textit{et al.}, \cite{interfacedatabase} which separates the coherency strain energy contribution from the interfacial energy, and 2) Refs. \citenum{MohanPNP} and  \citenum{jacs}, which do not explicitly perform this separation. We note that an approximation of eq. \ref{eqn:np_toten} is the neglect of volume-dependence to the NP total energy (e.g. coherency strain), which is discussed in the Supplementary Materials of Ref. \citenum{MohanPNP}. Calculating interfacial energies is complex and requires careful consideration of input structures; therefore we constrain our study to cases for which the interfacial energies are calculated.

\begin{figure}[h!]
\centering
  \includegraphics[width=0.8\textwidth]{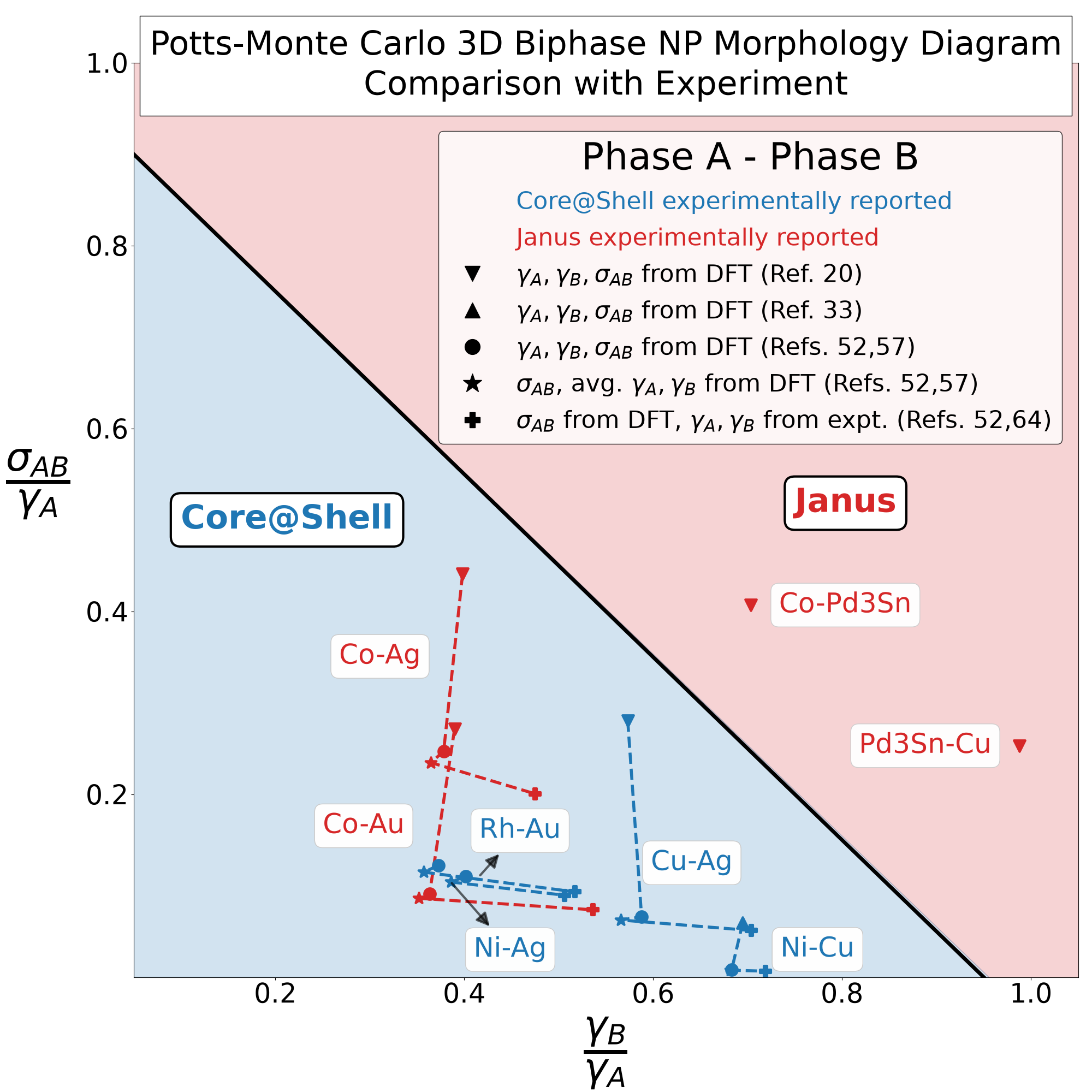}
  \caption{Predicted biphase morphologies corresponding to experimentally reported multiphase NPs. Each \ce{A-B} point represents the morphology formed by phase A and phase B, with phase A being the phase with the higher surface energy. Core@Shell morphologies take the form of A@B. The point color signifies the experimentally reported morphology, which is compared with the morphology phase region color it is located in. Like colors suggest good agreement between MCP and unlike colors imply poor agreement. Predicted morphologies agree with experimentally reported morphologies in six of eight NPs. DFT-calculated interfacial energies for some systems differ so a dashed line representing this range is provided for reference. Each \ce{A-B} system is plotted from surface and interfacial energies, using combinations of DFT-calculated and experimental ones. The FCC structure is utilized for the single-element phases and the L1\textsubscript{2} structure is utilized to represent the \ce{Pd3Sn} phase. The values plotted are reproduced in Table~\ref{tab:biphase_pred}.}
  \label{fig:biphase_exp}
\end{figure}

We applied the MCP model to eight experimentally reported multimetallic nanoparticles: \ce{Co-Ag},\cite{jacs} \ce{Co-Au},\cite{jacs,GhoshChaudhuri2012Core/shellApplications} \ce{Rh-Au},\cite{Ovari2010FormationLEIS} \ce{Ni-Ag},\cite{Pinkas2019SynthesisNanoparticles} \ce{Cu-Ag},\cite{jacs, Pinkas2019SynthesisNanoparticles} \ce{Ni-Cu},\cite{Pinkas2019SynthesisNanoparticles} \ce{Co-Pd3Sn},\cite{MohanPNP} and \ce{Cu-Pd3Sn}.\cite{MohanPNP} These eight systems are plotted on the biphase stability map in Figure \ref{fig:biphase_exp}. Table~\ref{tab:biphase_pred} contains the details of each system in Figure \ref{fig:biphase_exp}, including their surface and interfacial energies. Surface and interfacial energies along the close-packed direction are used in the DFT-calculated cases, whereas the experimental surface energies are an average across all crystallographic directions since these generally cannot be reliably measured along a single crystallographic orientation.\cite{expsurfen} The MCP model results agree with the experimentally reported biphase morphologies in six of eight NP systems (\ce{Rh-Au}, \ce{Ni-Ag}, \ce{Cu-Ag}, \ce{Ni-Cu}, \ce{Co-Pd3Sn}, \ce{Cu-Pd3Sn}). As the predictions are largely in agreement with experiment, we will next illustrate one case for which multiple morphologies are experimentally reported (\ce{Cu-Ag}) and then discuss the two conflicting cases. We note that there are a large number of additional experimentally reported multiphase NP morphologies (reproduced in Tables~\ref{tab:multiphase_1}--\ref{tab:multiphase_4}), but the DFT calculations of these surface and interfacial energies are not available to enable our MCP model prediction.

To understand how interfacial and surface energies compete to influence morphology, we use the \ce{Cu-Ag} NP as an example.
Experimentally, biphase \ce{Cu-Ag} NPs form either a Core@Shell or a random alloy depending on the precursor used\cite{Pinkas2019SynthesisNanoparticles} and are also reported as both 2p\cite{jacs} and Janus\cite{MohanPNP}. Another work reports the \ce{Cu-Ag} NP can present as Core@Shell or Janus depending on the size of \ce{Cu} core used.\cite{Langlois2008GrowthNanoparticles} The wide range of experimental reports suggests the energetic stability is only one factor controlling morphology. We can use the biphase stability map to study how the \ce{Cu-Ag} point could move within the stability map.
\ce{Cu-Ag} is firmly Core@Shell when using DFT-calculated interfacial energy values from Ref. \citenum{interfacedatabase} which separate out the strain energy component. When using experimental surface energy values extrapolated to $T=0\ \si{\kelvin}$ from Ref. \citenum{expt_surface_energy}, $\nicefrac{\gamma_{B}}{\gamma_A}$ increases.
The DFT interfacial energies in Ref. \citenum{MohanPNP} increase $\nicefrac{\sigma_{AB}}{\gamma_A}$ because they do not separate out the coherency strain energy component from the interfacial energy, moving the \ce{Cu-Ag} point towards the phase boundary.
Each calculation scheme reflects how energetic contributions can shift the predicted NP morphology.
We can extend this thinking to study the cases where there is a discrepancy between the MCP-predicted and experimentally-reported NP morphology.

Here we discuss possible reasons for discrepancies between experimentally-reported NP morphologies and the MCP-predicted ones. We divide these explanations into two categories: (I) the experimentally reported morphology is not the energetically stable ground state, and (II) the approximations underlying the input to MCP may be inaccurate. There are three main possible inaccuracies: (A) the input phases may be incorrect, or the DFT-calculated (B) surface energies and (C) interfacial energies may be inaccurate. (A) may occur when using OQMD to determine NP phases due to underlying assumptions of $T=0\ \si{\kelvin}$, bulk materials, and perfectly ordered phases. For example, OQMD does not consider solid solution formation nor phases formed at higher temperatures. In (B), the surface energy databases provide energies at $T=0\ \si{\kelvin}$ and in vacuum. Chen \textit{et al.} report surface passivation effects not considered by DFT calculations of surfaces in vacuum result in a decrease in surface energy difference between two metals \cite{polymersurfacepassivation}. Additionally, although DFT surface energies tend to have generally good agreement with experiment \cite{Tran2016DataCrystals}, in some cases they may be widely inaccurate, such as in \ce{Au}.\cite{np_Au_surf_en} In (C), the interfacial energies are only calculated for coherent interfaces and in some cases do not account for coherency strain. We will focus on the impact of these three inaccuracies resulting from calculation schemes as we discuss the two exceptions in Figure \ref{fig:biphase_exp}.
As we discuss later, comparison of triphase NP predictions with experiment introduces one additional category of explanation, arising from the larger space of possible morphologies in three-phase systems.

In the following sections, we analyze how changes in surface and interfacial energies may impact the morphology predicted by the stability map to study the impact of each inaccuracy on MCP-predicted NP morphology. 
Since MCP predictions agree with experimental reports in six of the eight NPs, only the two discrepancies (\ce{Co-Ag} and \ce{Co-Au}) are discussed in further detail in the sub-sections below.

\subsubsection{\ce{Co-Ag} NP.} 
The \ce{Co-Ag} NP is predicted here as a Core@Shell morphology but is experimentally reported as a Janus morphology \cite{jacs}. This discrepancy could be ascribed to two factors belonging in categories (I) and (II). Most polyelemental NPs are annealed at around 600~\textdegree{}C \cite{Chen2016PolyelementalLibraries,  MohanPNP}; however the \ce{Co-Ag} system was annealed at a lower temperature of 185~\textdegree{}C \cite{jacs}. The lower annealing temperature may imply the prediction error falls into (I) because the \ce{Co-Ag} NP was not able to reach its energetic ground state. As for (II), we note both Refs. \citenum{MohanPNP} and \citenum{interfacedatabase} used an FCC structure for \ce{Co}; however \ce{Co} only undergoes a phase transition from HCP to FCC at 427~\textdegree{}C. The lower annealing temperature of 185~\textdegree{}C could mean that HCP Co more accurately represents the phase present in the NP.
We can instead estimate the new location of the \ce{HCP\ Co-FCC\ Ag} point on the stability map. The HCP surface energy of Co is slightly higher than that of its FCC structure,\cite{Tran2016DataCrystals} thereby increasing $\gamma_A$. The incoherency of the \ce{HCP\ Co} domain with the \ce{FCC\ Ag} domain is also expected to increase the interfacial energy of the \ce{Co-Ag} interface, $\sigma_{AB}$.
For a Janus morphology to form, the increase in interfacial energy must be fractionally larger than the increase in surface energy.
Furthermore, the surface passivation effects not considered by existing DFT surface energy calculations would decrease the surface energy difference\cite{polymersurfacepassivation} between \ce{Co} and \ce{Ag}, which thereby increases the ratio $\nicefrac{\gamma_{B}}{\gamma_{A}}$ towards the Janus region. Finally, other computational studies of \ce{Co-Ag} biphase NP determined the NP lies on the boundary between Core@Shell and Janus, with the transition attributable to the choice of substrate and NP seed. \cite{Hizi2023CombinedSystems} Thus, the morphology formed may depend on the experimental conditions considered.

\subsubsection{\ce{Co-Au} NP.}
The \ce{Co-Au} NP is predicted as a Core@Shell morphology whereas it is experimentally reported as a Janus morphology. The \ce{Co-Au} NP is reported as Janus in two cases using different annealing temperatures: 230~\textdegree{}C and 500~\textdegree{}C,\cite{jacs} which are below and above the \ce{HCP-FCC} transition, respectively. As stated earlier in the \ce{Co-Ag} discussion, the \ce{HCP} surface energy of \ce{Co} is slightly larger than its \ce{FCC} counterpart,\cite{Tran2016DataCrystals} thus we can estimate a slight increase in $\gamma_A$ for the fabrication case at 230~\textdegree{}C.  DFT-calculated surface energies for \ce{Au} are almost half the experimental surface energy of \ce{Au},\cite{expt_surface_energy, np_Au_surf_en} leading to a doubling of $\gamma_B$.
As for the \ce{Co-Au} interfacial energy, only the coherent \ce{FCC-FCC} structure is known. Thus, these three inaccuracies fall into category (II). Incorporating the estimated surface energy adjustments would shift the point further right, closer to the phase boundary between Core@Shell and Janus.

\subsubsection{Summary of biphase morphology comparison to experiment.}
As the above subsections highlight, the morphology stability map is a tool for studying the impact of selected surface and interfacial energies on NP morphologies. 
The stability map may additionally be used to investigate the significance of coherency strain energies as opposed to interfacial energies when designing multiphase NPs, as illustrated by the vertical shift between points using the calculation scheme in Refs. \citenum{MohanPNP},~\citenum{jacs} and the points derived from Salgado-Casanova \textit{et al.}'s DFT-calculation scheme.
Additionally, NPs are often fabricated with a stabilizing surface layer to prevent NP aggregation,\cite{MohanPNP, Chen2016PolyelementalLibraries, Pinkas2019SynthesisNanoparticles} so the surface energies of the phases may be calculated to more accurately reflect the NP environment.
The stability map remains the same in each case, but the points will shift under the different surface and interfacial energy choices. 

\subsection{Triphase nanoparticle interface morphologies}
We next extend the grid search approach used previously in the biphase NP morphology to search for triphase NP interface morphologies.
The triphase NPs consist of three domains (A, B, C) surrounded by some environment, and are thus modeled using six unlike interactions: \ce{A-B} ($\sigma_{AB}$), \ce{A-C} ($\sigma_{AC}$), \ce{B-C} ($\sigma_{BC}$), \ce{A-Environment} ($\gamma_A$), \ce{B-Environment} ($\gamma_B$), and \ce{C-Environment} ($\gamma_C$). As described earlier in the biphase case, we reduce the number of parameters by normalizing the six interactions by the largest surface energy, leaving us with five ratios: $\nicefrac{\gamma_B}{\gamma_A}$, $\nicefrac{\gamma_C}{\gamma_A}$, $\nicefrac{\sigma_{AB}}{\gamma_A}$, $\nicefrac{\sigma_{AC}}{\gamma_A}$, $\nicefrac{\sigma_{BC}}{\gamma_A}$.
We order the phases by decreasing surface energy such that $\gamma_A \ge \gamma_B \ge \gamma_C$ to reduce the number of grid points required to calculate the triphase stability map without loss of generality.
The parameter bounds derived in the biphase NP also apply here, with $\nicefrac{\gamma_B}{\gamma_A}, \nicefrac{\gamma_C}{\gamma_A} \leq 1 $ and $\nicefrac{\sigma_{AB}}{\gamma_A}, \nicefrac{\sigma_{AC}}{\gamma_A}, \nicefrac{\sigma_{BC}}{\gamma_A} \leq 1 $ defining the bounds of the grid search.

We perform MCP calculations for triphase NPs using a grid of 30,250 points and find 11 topologically distinct morphologies, which we list in Table \ref{tab:triphase}. The Pie morphology consists of three radially distributed domains about the center of the particle, such that each domain has a surface and an interface with every other domain. The Striped morphology is divided into three parallel domains such that the outer domains do not share an interface but share an interface with the central domain. Each domain has a surface. The Venn morphology is a transitional morphology similar to Striped, with the central domain reducing to a point such that it does not have a surface. This morphology is transitional because it exists between the Striped morphology and the next morphology, Pokéball, by reducing a surface to a point and thus has a distinct number of surfaces and interfaces. The Pokéball morphology is an extension of the biphase Janus morphology with a third domain lying along the central interface. All domains share an interface with each other.
The Janus with Core@Shell morphology is a Janus morphology with one of the domains containing the third domain. The inner domain only shares an interface with the domain that encapsulates it. 
The Onion morphology is a layered morphology consisting of layered shells surrounding a core.
The Shell with Janus core morphology is a Janus morphology encapsulated by a third domain which forms the shell.
The Shell with 2 cores morphology consists of one domain encapsulating two separated domains such that these two do not share an interface.
The remaining three morphologies are included for completeness, but they represent a summation of single-phase and biphase morphologies. For this reason, we call these morphologies non-contiguous.
The Janus + 1p morphology consists of a Janus biphase particle and a separate single-phase particle.
The Core@Shell + 1p morphology contains a biphase Core@Shell particle and a separate single-phase particle.
The 3p morphology consists of three separate single-phase particles.

\begingroup
\small
\begin{longtable}{C{0.13\textwidth} C{0.07\textwidth} C{0.15\textwidth} C{0.15\textwidth} C{0.07\textwidth}C{0.07\textwidth}
}
\caption{Triphase morphology descriptions. This table contains the names, visualizations, and characteristics of the eleven topologically distinct triphase NP interface morphologies. First, we list the names of each one followed by its categorical type. Then, we show the MCP output and the idealized schematic representation. Finally, we show the respective number of surfaces and interfaces.}
\label{tab:triphase}
\\
\toprule
\thead{Morphology} &
\thead{Type} &
\thead{Sample MCP \\ output} &
\thead{Idealized \\ Morphology} &
\thead{\# of\\surfaces} &
\thead{\# of\\interfaces} \\
\midrule
\endfirsthead

\toprule
\thead{Morphology} &
\thead{Type} &
\thead{Sample MCP \\ output} &
\thead{Idealized \\ Morphology} &
\thead{\# of\\surfaces} &
\thead{\# of\\interfaces} \\
\midrule
\endhead

Pie &
Pi &
\includegraphics[width=0.7\linewidth]{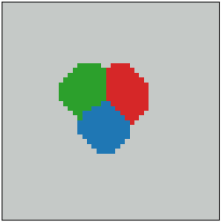} &
\includegraphics[width=0.7\linewidth]{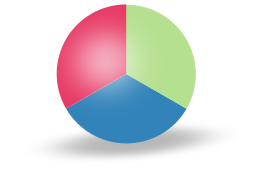} &
3 & 3 \\
Striped &
Str &
\includegraphics[width=0.7\linewidth]
{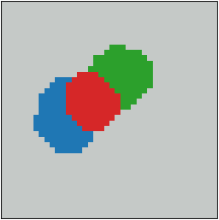} &
\includegraphics[width=0.7\linewidth]{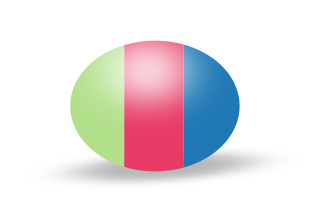} &
3 & 2 \\
\makecell{Venn\\(Transitional)} &
Str-Po &
\includegraphics[width=0.7\linewidth]{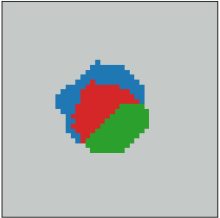} &
\includegraphics[width=0.7\linewidth]{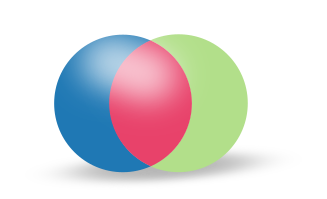} &
2 & 2 \\
Pok\'eball &
Po &
\includegraphics[width=0.7\linewidth]{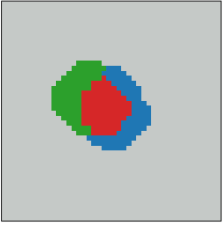} &
\includegraphics[width=0.7\linewidth]{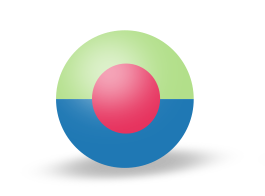} &
2 & 3 \\
Janus with Core@Shell &
J-CS &
\includegraphics[width=0.7\linewidth]{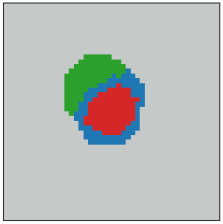} &
\includegraphics[width=0.7\linewidth]{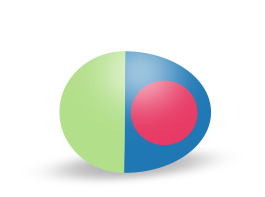} &
2 & 2 \\
Onion &
CS &
\includegraphics[width=0.7\linewidth]{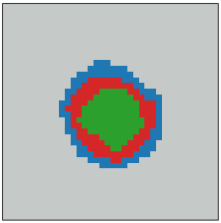} &
\includegraphics[width=0.7\linewidth]{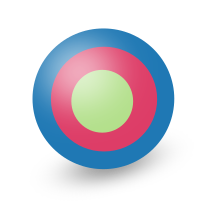} &
1 & 2 \\
Shell with Janus core &
SJ &
\includegraphics[width=0.7\linewidth]{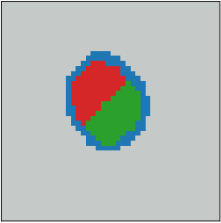} &
\includegraphics[width=0.7\linewidth]{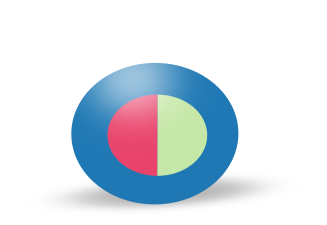} &
1 & 3 \\
Shell with 2 cores &
S2 &
\includegraphics[width=0.71\linewidth]{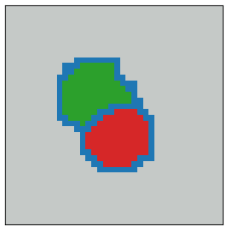} &
\includegraphics[width=0.7\linewidth]{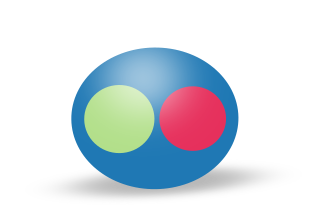} &
1 & 2 \\
Janus + 1p &
J+1p &
\includegraphics[width=0.7\linewidth]{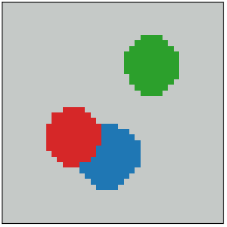} &
\includegraphics[width=0.7\linewidth]{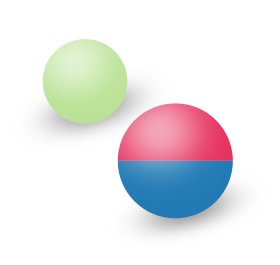} &
3 & 1 \\
Core@Shell + 1p &
CS+1p &
\includegraphics[width=0.71\linewidth]{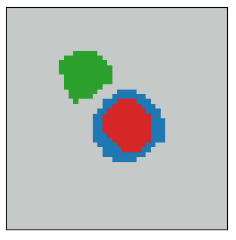} &
\includegraphics[width=0.7\linewidth]{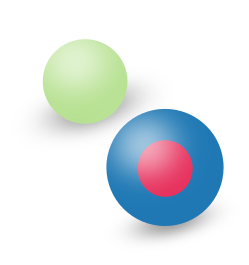} &
3 & 1 \\
3p &
\textit{n}P &
\includegraphics[width=0.7\linewidth]{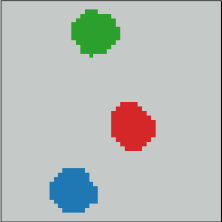} &
\includegraphics[width=0.71\linewidth]{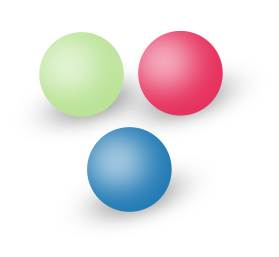} &
3 & 0 \\
\end{longtable}
\endgroup

\begin{figure*}[h!]
 \centering
 \includegraphics[width=0.8\textwidth]{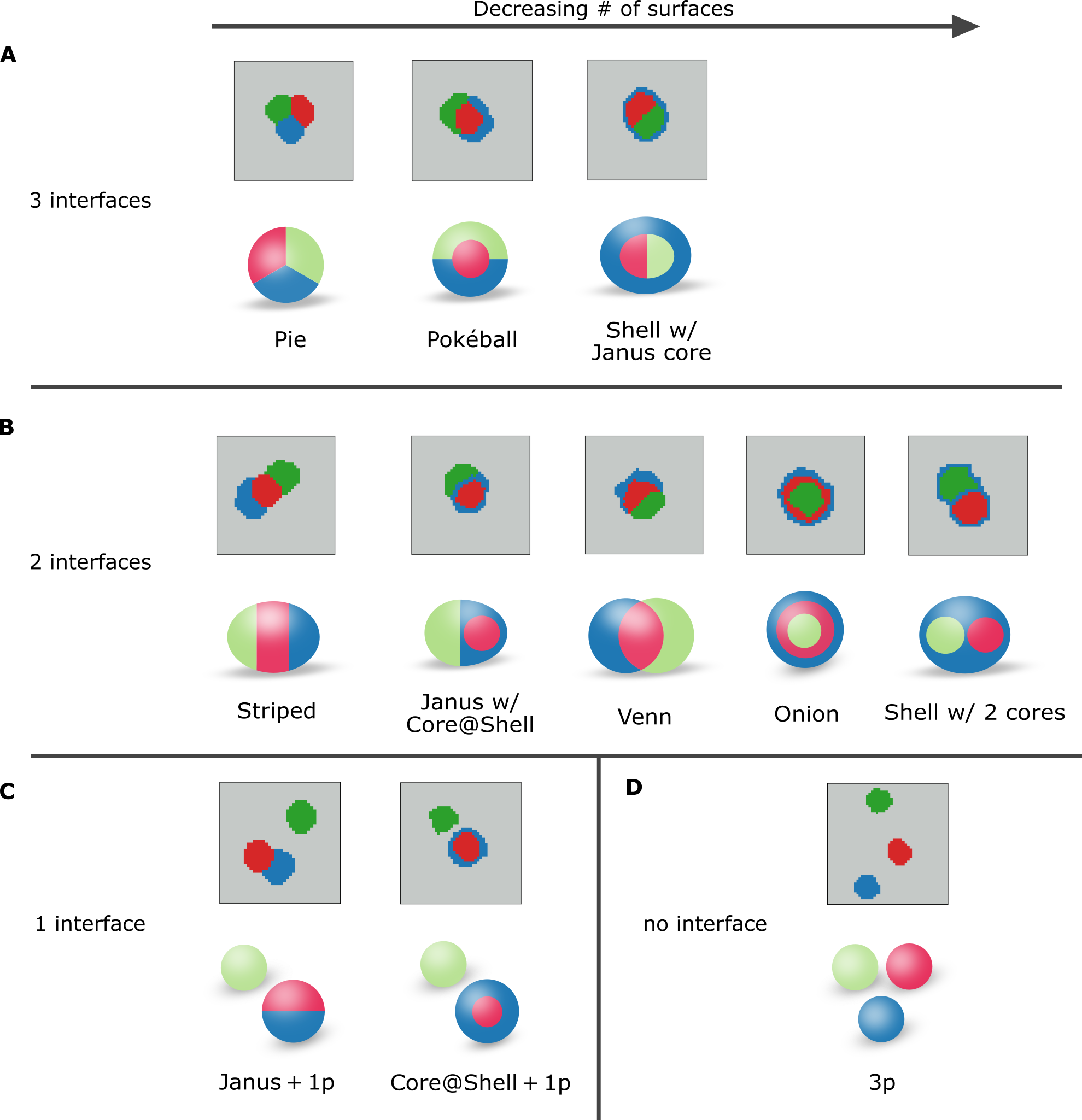}
 \caption{Snapshots of the triphase NP interface morphologies found via MCP as well as their corresponding schematic depiction. (A) The three-interface triphase morphologies, ordered by decreasing number of surfaces, are Pie, Pokéball, and Shell with Janus core. (B) The two-interface triphase morphologies in order of decreasing number of surfaces are Striped, Janus with Core@Shell and Venn, followed by Onion and Shell with 2 cores. (C) The single-interface triphase morphologies observed in order of decreasing number of surfaces are the Janus + 1p and the Core@Shell + 1p morphologies. (D) The triphase morphology with no interface is the 3p morphology. We note the number of interfaces and surfaces do not uniquely define a triphase morphology. See Table \ref{tab:triphase} for qualitative descriptions of each morphology.}
 \label{fig:triphase}
\end{figure*}

Another more informative view of these 11 distinct triphase morphologies is presented in Figure \ref{fig:triphase} by decreasing number of interfaces, with Figure \ref{fig:triphase}A illustrating the snapshots and schematic representations of three-interface triphase morphologies, followed by Figure \ref{fig:triphase}B, containing two-interface morphologies, and so on with Figure \ref{fig:triphase}C\&D. Within each sub-section, the morphologies are ordered by decreasing number of surfaces. For example, starting in Figure \ref{fig:triphase}A, the Pie morphology exhibits three surfaces, followed by the two-surface Pokéball morphology, then the single-surface Shell with Janus core morphology. Of the 11 distinct morphologies, we only find five experimentally reported (Pie,\cite{jacs} Striped,\cite{jacs} Onion,\cite{schaak2025} Venn,\cite{jacs} and Janus + 1p\cite{jacs}).

The triphase morphologies found agree with the predicted number of possible interfaces from Ref. \citenum{MohanPNP}, who predict the number of interfaces for an $n$-phase particle to fall between $n-1$ and $\nicefrac{n(n-1)}{2}$. This relation implies that triphase particles must be composed of a minimum of two interfaces and a maximum of three interfaces, which is true for all of the eight contiguous triphase morphologies found (Pie, Striped, Venn, Pok\'eball, Janus with Core@Shell, Onion, Shell with Janus core, and Shell with 2 cores). We additionally find that the number of surfaces must lie between $1$ and $3$ in the triphase case, or more generally between $1$ and $n$. The categorization of triphase morphologies by number of surfaces and number of interfaces is shown in Figure \ref{fig:triphase}.
The distribution of morphologies in Figure \ref{fig:triphase} illustrates that the number of interfaces does not uniquely define a NP interface morphology, nor does the number of surfaces in conjunction with the number of interfaces. In Figure \ref{fig:triphase}B, both the Onion and the Shell with 2 cores morphologies have one surface and two interfaces, yet are topologically distinct morphologies. Likewise, the Janus with Core@Shell and Venn morphologies each have two surfaces and two interfaces.

\subsection{Prediction of triphase morphologies}
We predict triphase NP morphologies by extending the same approach: we select phase-separating ternary combinations of isostructural elemental phases from the interfacial energy database in Ref. \citenum{interfacedatabase} along the close-packed direction, and use the matching DFT-calculated surface energies from Ref. \citenum{Tran2016DataCrystals} to sort the phases by decreasing surface energy into labels A, B, and C. We then calculate the ratios $\nicefrac{\sigma_{AB}}{\gamma_A},\nicefrac{\sigma_{BC}}{\gamma_A},\nicefrac{\sigma_{AC}}{\gamma_A}$ and $\nicefrac{\gamma_{B}}{\gamma_A},\nicefrac{\gamma_{C}}{\gamma_A}$ and use these as inputs into the MCP model to predict triphase NP morphologies, the results of which are summarized in Table \ref{tab:triphase_pred}, which lists the identities of Phases A, B, and C, followed by their surface and interfacial energies ratios and the MCP-predicted morphology. The values used to calculate these ratios are reproduced in Table~\ref{tab:triphase_pred_dft} of the Supplementary Information and predictions using different choices of experimental surface and DFT-calculated interfacial energies are presented in the Supplementary Information.

\begin{longtable}[h]{@{\extracolsep{\fill}}ccccccccc}
  \caption{\ MCP model predictions of triphase morphologies from existing surface and interfacial energy databases. We select all phase-separating ternary isostructural combinations of elemental metals exhibiting FCC, BCC, or HCP structures that lie on the convex hull and for which their ternary combination does not form any ordered compound on the convex hull. Elemental phases A, B, and C are ordered according to decreasing DFT-calculated surface energy. The ratios of surface and interfacial energies are calculated from the DFT energies in Refs. \citenum{Tran2016DataCrystals} and \citenum{interfacedatabase} along the close-packed direction and input into the MCP model. For \ce{Cr-Mo-Fe}, \ce{Mo-Fe-Li}, and \ce{W-Mo-Fe}, the ratios are calculated along the [100] direction (Table~\ref{tab:triphase_pred_dft}). In this table, the input ratios are provided $(\nicefrac{\gamma_{B}}{\gamma_{A}}, \nicefrac{\gamma_{C}}{\gamma_{A}},\nicefrac{\sigma_{AB}}{\gamma_{A}},\nicefrac{\sigma_{BC}}{\gamma_{A}},\nicefrac{\sigma_{AC}}{\gamma_{A}})$, followed by the MCP-predicted morphology. Additional details regarding the DFT values used to calculate the parameter ratios are found in Table~\ref{tab:triphase_pred_dft}.}
  \label{tab:triphase_pred}
  \small
  \\
    \hline
    Phase A & Phase B & Phase C & $\nicefrac{\gamma_B}{\gamma_A}$ & $\nicefrac{\gamma_C}{\gamma_A}$ & $\nicefrac{\sigma_{AB}}{\gamma_A}$ 
    & $\nicefrac{\sigma_{BC}}{\gamma_A}$ 
    & $\nicefrac{\sigma_{AC}}{\gamma_A}$ & Predicted Morphology \\
    \hline
    \endfirsthead
    \hline
    Phase A & Phase B & Phase C & $\nicefrac{\gamma_B}{\gamma_A}$ & $\nicefrac{\gamma_C}{\gamma_A}$ & $\nicefrac{\sigma_{AB}}{\gamma_A}$ & $\nicefrac{\sigma_{BC}}{\gamma_A}$ & $\nicefrac{\sigma_{AC}}{\gamma_A}$ & Predicted Morphology \\
    \hline
    \endhead
Co & Zn & Cd & 0.165 & 0.095 & 0.035 & 0.087 & 0.276 & Janus w/ Core@Shell \\
Cr & Fe & Li & 0.764 & 0.156 & 0.032 & 0.153 & 0.406 & Onion \\
Cr & Mo & Fe & 0.876 & 0.688 & 0.074 & 0.042 & 0.041 & Shell w/ Janus core \\
Cr & Mo & Li & 0.874 & 0.156 & 0.002 & 0.296 & 0.406 & Onion \\
Ir & Cu & Ag & 0.575 & 0.338 & 0.088 & 0.046 & 0.310 & Onion \\
Ir & Pt & Au & 0.648 & 0.325 & 0.022 & 0.029 & 0.294 & Onion \\
Mo & Fe & Li & 0.785 & 0.145 & 0.048 & 0.310 & 0.426 & Onion \\
Os & Co & Cd & 0.722 & 0.068 & 0.060 & 0.200 & 0.407 & Onion \\
Re & Co & Cd & 0.817 & 0.077 & 0.041 & 0.226 & 0.421 & Onion \\
Re & Co & Zn & 0.817 & 0.135 & 0.041 & 0.029 & 0.002 & Shell w/ 2 cores \\
Re & Ti & Mg & 0.762 & 0.211 & 0.077 & 0.082 & 0.109 & Shell w/ Janus core-like \\
Re & Ti & Hf & 0.762 & 0.662 & 0.077 & 0.003 & 0.053 & Shell w/ 2 cores \\
Re & Tl & Cd & 0.099 & 0.077 & 0.546 & 0.020 & 0.421 & Janus w/ Core@Shell \\
Re & Zn & Cd & 0.135 & 0.077 & 0.002 & 0.071 & 0.421 & Janus w/ Core@Shell \\
Rh & Cu & Ag & 0.661 & 0.389 & 0.050 & 0.053 & 0.208 & Onion \\
Ru & Co & Cd & 0.818 & 0.078 & 0.087 & 0.226 & 0.256 & Shell w/ Janus core \\
Tc & Co & Cd & 0.944 & 0.089 & 0.025 & 0.261 & 0.361 & Onion \\
Tc & Ti & Cd & 0.880 & 0.089 & 0.008 & 0.042 & 0.361 & Onion \\
Tc & Ti & Zr & 0.880 & 0.721 & 0.008 & 0.014 & 0.075 & Onion \\
Tc & Tl & Cd & 0.115 & 0.089 & 0.483 & 0.023 & 0.361 & Janus w/ Core@Shell \\
Tc & Zr & Mg & 0.721 & 0.244 & 0.075 & 0.018 & 0.007 & Shell w/ 2 cores \\
Ti & Hf & Mg & 0.869 & 0.277 & 0.004 & 0.052 & 0.108 & Onion \\
Ti & Tl & Cd & 0.130 & 0.102 & 0.166 & 0.027 & 0.048 & Janus w/ Core@Shell \\
Ti & Zr & Tl & 0.819 & 0.130 & 0.016 & 0.040 & 0.166 & Onion \\
W & Cr & Fe & 0.992 & 0.758 & 0.005 & 0.032 & 0.032 & Shell w/ 2 cores \\
W & Cr & Li & 0.992 & 0.154 & 0.005 & 0.402 & 0.393 & Shell w/ Janus core \\
W & Cr & Mo & 0.992 & 0.866 & 0.005 & 0.002 & 0.009 & Shell w/ 2 cores \\
W & Fe & Li & 0.758 & 0.154 & 0.032 & 0.152 & 0.393 & Onion \\
W & Mo & Fe & 0.805 & 0.632 & 0.001 & 0.039 & 0.027 & Onion \\
W & Mo & Li & 0.866 & 0.154 & 0.009 & 0.293 & 0.393 & Onion \\
W & Mo & Na & 0.866 & 0.068 & 0.009 & 0.442 & 0.578 & Onion \\
    \hline
\end{longtable}

The predicted morphologies fall into four categories: Onion, Janus with Core@Shell, Shell with Janus core, and Shell with 2 cores. Morphologies denoted with the suffix ``-like'' are intermediate morphologies in transition to their listed morphology -- we list their competing morphology in the Supplementary Information.
These four morphologies are extrapolations of the Core@Shell biphase morphology, likely reflecting the effects discussed earlier in the biphase discussion: namely, MCP minimizes the number of surfaces, owing to the trend that surface energies are larger than interfacial energies.
The triphase morphologies are a Core@Shell morphology with varying topologies in the core. The core in the NPs in Table \ref{tab:triphase_pred} is either a Core@Shell (forming the Onion morphology), a Janus core (forming the Shell with Janus core), or the 2p morphology (forming the Shell with 2 cores morphology). The interfacial energy ratios determine which morphology forms, much the same way that the biphase counterpart $\nicefrac{\sigma_{AB}}{\gamma_A}$  differentiates between the Core@Shell, Janus, and 2p morphologies. Thus, ground-state morphology predictions from DFT-calculated surface and interfacial energies tend toward Core@Shell in biphase NPs and a Core@Shell variant in triphase NPs. Predictions using ratios along the [100], [110], and average amongst all three lowest-index directions are listed in Table~\ref{tab:triphase_pred_dft}, with no significant changes in our conclusions. Surface passivation, which is not captured by DFT surface energies calculated in vacuum, decreases the difference between surface energies,\cite{polymersurfacepassivation} driving $\nicefrac{\gamma_B}{\gamma_A}$ and $\nicefrac{\gamma_C}{\gamma_A}$ toward 1. In the triphase stability map, only Pie and Striped morphologies are predicted at $\nicefrac{\gamma_B}{\gamma_A} = 1.0$ (Figure~\ref{fig:3p3d_stability_map}j), and almost all predicted morphologies are Pie or Striped at $\nicefrac{\gamma_B}{\gamma_A} = 0.9$ (Figure~\ref{fig:3p3d_stability_map}i), which may explain why these are the most frequently reported experimental triphase morphologies.

\subsection{Comparison of MCP triphase morphologies with experiment}
We applied the MCP model to experimentally reported triphase NPs (\ce{Co-Cu-Ag}, \ce{Co-Pd3Sn-Ag}, \ce{Co-Pd3Sn-Au}, \ce{Co-Pd3Sn-Cu}, and \ce{Ni-Cu-Ag}) fabricated via megalibrary synthesis \cite{MohanPNP, jacs, mirkincatalystdesign}. We input the elemental composition of each NP into OQMD to determine the number of phases and their identity. The equilibrium OQMD phases are categorized as Phase A, B, and C
in Table \ref{tab:triphase_exp}. As in the biphase NP, the NPs are modeled in a vacuum for simplicity, which allows for the use of DFT-calculated surface energies. We calculate the values of $\gamma_A, \gamma_B, \gamma_C, \sigma_{AB}, \sigma_{AC}, \sigma_{BC}$ (which are reproduced in Table~\ref{tab:mohan_3r_temp}) using the corresponding DFT-calculated surface and interfacial energies from databases in Refs. \citenum{Tran2016DataCrystals} and \citenum{interfacedatabase}. These parameters are input into MCP and the output morphology  visualized and categorized by the number of each unlike spin interaction, and reported in Table \ref{tab:triphase_exp}. The MCP output is compared to the experimentally-reported morphology listed in the final column.

MCP predicts the ground-state thermodynamic morphology of the \ce{Co-Cu-Ag} NP to be Onion, the \ce{Co-Pd3Sn-Ag} NP to be a Shell with Janus core, the \ce{Co-Pd3Sn-Au} NP to be a Shell with 2 cores morphology, the \ce{Co-Pd3Sn-Cu} NP to be Janus with Core@Shell-like, and \ce{Ni-Cu-Ag} to exhibit an Onion morphology. The reader will note that the MCP-predicted triphase morphologies are different from the experimentally observed morphologies listed in Table \ref{tab:triphase_exp}. In the \ce{Co-Cu-Ag} NP, multiple morphologies are experimentally reported and the MCP prediction is close to one report. For the four remaining NPs (\ce{Co-Pd3Sn-Ag}, \ce{Co-Pd3Sn-Au}, \ce{Co-Pd3Sn-Cu}, and \ce{Ni-Cu-Ag}), the MCP-predicted morphologies are one topological transition away from the experimentally-observed morphology: they differ by the addition or subtraction of only one surface or interface. We discuss each NP prediction below, as well as reasons why the MCP-predicted morphology may differ from the experimentally-reported one.

\begin{table*}
\small
  \caption{\ MCP comparison with experimentally reported triphase polyelemental NPs. In the first column, the stoichiometric experimental composition is listed when reported, otherwise the general composition of each domain is provided instead. Next, the identities of equilibrium OQMD phases A, B, and C are provided. The experimentally-reported phases are equivalent to the OQMD equilibrium phases in all but the \ce{Pd_xSn_y} phases, where the experimental phase varies widely. We use \ce{Pd3Sn} instead for simplicity, as in Ref. \citenum{MohanPNP}. The corresponding parameters $\gamma_A, \gamma_B,\gamma_C,\sigma_{AB},\sigma_{AC},\sigma_{BC}$ are from DFT-calculated surface and interfacial energies in Refs. \citenum{MohanPNP,jacs,Tran2016DataCrystals,interfacedatabase,np_Au_surf_en}, which are only reported for the FCC or L1\textsubscript{2} structure for these phases. These parameter ratios were input into MCP to calculate an output morphology that is reported in the next column and compared to the experimentally-reported morphology in the final column. Further calculation details, including the values of the input parameters, are found in the Supplementary Information Table~\ref{tab:mohan_3r_temp}. *Using DFT-calculated interfacial and surface energies. MCP predictions from experimental surface energies are given in Table~\ref{tab:mohan_3r_temp}. **In Refs. \citenum{jacs,Murat}, the authors do not report a morphology. In these two cases, we examine the figures provided to discern morphology.}
\label{tab:triphase_exp}
  \begin{tabular*}{\textwidth}{@{\extracolsep{\fill}}ccccccc}
    \hline
    \thead{Experimentally-Reported \\ Composition} & \thead{Phase\\A} & \thead{Phase\\B} & \thead{Phase\\C} & \thead{MCP-Predicted\\Morphology*} & \thead{Experimentally-Reported\\Morphology} \\
    \hline
    \makecell{ \ce{Ag_{{0.23}}Cu_{{0.47}}Co_{{0.30}}}\ \cite{MohanPNP}\\
        \ce{Ag_{{0.66}}Cu_{{0.27}}Co_{{0.07}}}\ \cite{Murat}\\
        \ce{Ag_{{0.43}}Cu_{{0.19}}Co_{{0.38}}}\ \cite{Chen2016PolyelementalLibraries} \\
        \ce{Ag-Cu-Co}\ \cite{MohanPNP, jacs}
        }
        & Co & Cu & Ag & Onion & \makecell{Striped,\cite{jacs, MohanPNP} \\ Pie\cite{Chen2016PolyelementalLibraries},\\Striped with\\Onion character,**\cite{Murat} \\Janus + 1p\cite{jacs}}  \\
    \ce{Ag-Co-PdSn} \cite{MohanPNP}
        & Co & \ce{Pd3Sn} & Ag & \makecell{Shell w/\\Janus core} & Pie \cite{MohanPNP} \\
    \makecell{
        \ce{Au_{{0.30}}Co_{{0.37}}Pd_{{0.19}}Sn_{{0.14}}}\ \cite{MohanPNP}\\
        \ce{Au_{{0.33}}Co_{{0.34}}Pd_{{0.18}}Sn_{{0.15}}}\ \cite{MohanPNP}\\
        \ce{Au_{{0.24}}Co_{{0.58}}(PdSn)_{{0.18}}}\ \cite{MohanPNP}\\
        \ce{Au_{{0.31}}Co_{{0.34}}Pd_{{0.21}}Sn_{{0.14}}}\ \cite{MohanPNP}\\
        \ce{Au_{{0.30}}Co_{{0.26}}Pd_{{0.30}}Sn_{{0.14}}}\ \cite{MohanPNP}\\
        \ce{Au_{{0.33}}Co_{{0.24}}Pd_{{0.26}}Sn_{{0.17}}}\ \cite{MohanPNP}\\
        \ce{Au_{{0.30}}Co_{{0.29}}Pd_{{0.30}}Sn_{{0.11}}}\ \cite{MohanPNP}\\
        \ce{Au_{{0.38}}Co_{{0.33}}Pd_{{0.23}}Sn_{{0.06}}}\ \cite{MohanPNP}
        }
        & Co & \ce{Pd3Sn} & Au & Shell w/ 2 cores
        & Pie \cite{MohanPNP} \\
    \ce{Co-Cu-PdSn(Cu)}\ \cite{MohanPNP}
         & \ce{Co} & \ce{Pd3Sn} & \ce{Cu} & \makecell{Janus w/  \\Core@Shell-like} & Striped \cite{MohanPNP} \\
    \ce{Ag-Cu-Ni}\ \cite{jacs}
        & \ce{Ni} & \ce{Cu} & \ce{Ag} & Onion & \makecell{Janus w/ \\ Core@Shell-like**} \\
    \hline
  \end{tabular*}
\end{table*}

As in the biphase case above, we attribute discrepancies between MCP-predicted morphologies and experimentally reported triphase morphologies to categories (I) and (II). The triphase case also introduces a third possibility: (III) the discovery of new triphase morphologies here may warrant re-evaluation of previously reported morphologies. For example, some transition triphase morphologies are observed experimentally -- such as the morphology transition between Janus with Core@Shell and Pokéball, listed in Table \ref{tab:triphase_exp} as Janus with Core@Shell-like, but have previously been categorized as Striped or Pie.\cite{jacs} 
Category (II) also carries extra weight in the triphase case: the greater structural complexity of three- and four-element NPs makes phase-identity and interfacial-energy approximations (IIA and IIC) more likely to introduce error.
For example, the \ce{Pd-Sn} domain in these NPs is approximated as the $L1_2$-structured \ce{Pd3Sn}, a simplification of the true, more variable \ce{Pd_xSn_y} stoichiometry.
For these reasons, there may be a discrepancy between the MCP-predicted NP morphology and its experimental counterpart. Each NP in the table from top to bottom is discussed below, beginning with the \ce{Co-Cu-Ag} NP.

The \ce{Co-Cu-Ag} NP is experimentally reported to exhibit either a Striped,\cite{jacs, MohanPNP} Pie, \cite{Chen2016PolyelementalLibraries} or Striped with Onion character\cite{Murat} morphology. The final report of a Janus + 1p morphology\cite{jacs} was annealed at a much lower temperature and for less time than the other reports, thus it is likely that this fabrication method is diffusion-limited and the NP did not reach the energetic ground state (I). The remaining reported NP morphologies were fabricated using megalibrary synthesis methods, with small variations in annealing time and temperature. One challenge faced in the megalibrary fabrication techniques is the diffusion of the components through the polymer solution,\cite{Smith2023MolecularCatalysts} which highlights the impact of kinetics on morphology formation.
Both the Striped and Pie morphologies are reached when the outer layer of the Onion morphology retracts and does not form a shell, implying the Onion morphology may be reached if diffusion were permitted to continue over an extended period of time.
In the final study, no morphology is reported but the images provided\cite{Murat} indicate a Striped morphology with Onion character -- a Striped morphology in process of transitioning to an Onion morphology, which aligns with the MCP-predicted morphology.

For the \ce{Co-Pd3Sn-Ag} NP, upon first glance the predicted Shell with Janus core morphology appears far from the reported Pie morphology. However, the only difference between the two is that one of the Pie domains has extended to cover the other two. The topological transition from Pie to Shell with Janus core suggests that it is possible the NP morphology is diffusion-limited, which prevents the formation of the ground-state morphology, falling into category (I). We additionally find support for (II) as there is a wide distribution of phases that could represent Phase B, the \ce{Pd_{x}Sn_{y}} phase. Of the available  surface and interfacial energies for \ce{Pd_{x}Sn_{y}} phases, DFT-calculated values are only available for $L1_2$-structured \ce{Pd3Sn}.\cite{MohanPNP}
Lastly, support for (III) comes from the analysis of the transition between the reported Pie and predicted Shell with Janus core morphologies: it is possible that a Pie morphology may be confused for a Shell with Janus core morphology depending on the angle of visualization.

In the \ce{Co-Pd3Sn-Au} NP, we find a few reasons for the discrepancy between the predicted Shell with 2 cores morphology and the observed Pie, all of which fall into category (II). We see the same issue as discussed above with the \ce{Pd_{x}Sn_{y}} phase. Closer examination of the NP compositions reveals that in all but one case, the reported \ce{Pd-Sn} phase is not \ce{Pd3Sn}. Additionally, the experimental surface energy of \ce{Au} is known to be approximately double that of the DFT-calculated surface energy.\cite{np_Au_surf_en}
Instead, we use the experimental surface energy of \ce{Au} given in Ref. \citenum{expsurfen} and extrapolate to temperatures matching experimental conditions ($T=773, 873, 1023 \ \si{\kelvin}$) to obtain a corrected MCP prediction, which is an Onion-like morphology: an Onion with mixing between the two outer layers.

The \ce{Co-Cu-PdSn(Cu)} NP is predicted as Janus with Core@Shell-like, meaning a Janus with Core@Shell in transition to a Pokéball morphology. The difference between the predicted Janus with Core@Shell-like morphology and the reported Striped morphology is that the core domain has extended to come into contact with the surface as well as another domain.
The transformation from one morphology to the other lends support to category (I), that the reported experimental morphology is diffusion-limited, which prevents the formation of the ground-state morphology. Category (II) also applies here because Phase B, the \ce{Pd_{x}Sn_{y}} phase, is approximated using the $L1_2$ \ce{Pd3Sn} structure in the DFT interfacial and surface energy calculations\cite{MohanPNP}.
To investigate this effect, we applied MCP using experimental surface energies for \ce{Cu} and \ce{Co} \cite{expsurfen} and predicted a Janus with Core@Shell morphology. This is an interesting case because, when the temperature dependence of the experimental surface energies is included, the ordering of the surface energies changes, so the identities of phases A, B, and C change from \ce{Co}, \ce{Pd3Sn}, and \ce{Cu} to \ce{Co}, \ce{Cu}, and \ce{Pd3Sn}. This causes the phases with surfaces exposed in the Janus with Core@Shell morphology to switch places, which does not change the type of interface present but does change the amount of each one present. There is some experimentally reported \ce{Cu} presence in Phase B, which is not taken into consideration in our approach.
Lastly, Janus with Core@Shell is a complex morphology that may be difficult to discern from Striped in a two-dimensional microscopy image, so this case also falls into category (III).

The \ce{Ni-Cu-Ag} NP is reported in Ref. \citenum{jacs} without a determined morphology. The images provided suggest a Janus with Core@Shell-like morphology. In contrast, MCP predicts an Onion morphology. The annealing temperatures and times in Ref. \citenum{jacs} are much lower than in the megalibrary synthesis methods from the other sources listed in the table, so the reported morphologies are unlikely to be equilibrium morphologies, falling into category (I). The predicted Onion morphology is again one topological transition away from the Janus with Core@Shell morphology: the outer, non-core-containing, Janus domain of the Janus with Core@Shell morphology can envelop the Core@Shell portion to form the Onion morphology, perhaps with a further annealing.

The predicted ground-state thermodynamically stable morphologies are more complex than the experimentally reported morphologies, which may imply that these more complex morphologies are less kinetically accessible since the as-implemented MCP model only captures energetically stable morphologies. Additionally, the more complex MCP-predicted morphologies may be easily confused when identified from a microscopy image, a two-dimensional projection of a three-dimensional NP. MCP can inform NP characterization, particularly the internal architecture of interfaces, as fabricated NPs become more complex. Although this study does not include kinetic or strain considerations, it may be possible for future studies to modify the MCP approach presented here to include these effects and study their relative importance. 
Despite the predictions not aligning concretely with experimentally reported morphologies, this work provides a tool for studying triphase NPs and comparing the impact of surface and interfacial energies on ground-state NP morphology predictions.

We conducted a study of the distribution of DFT-calculated surface and interfacial energies of elemental phases using the databases in Refs. \citenum{Tran2016DataCrystals} and \citenum{interfacedatabase}. The range of the ratios of interfacial to surface energies, or equivalently $\nicefrac{\sigma}{\gamma}$, as well as the ratios between surface energies, $\nicefrac{\gamma}{\gamma}$, are calculated from this dataset to screen for MCP-predicted NP morphologies. The Venn, Pie, Striped, Onion, and Janus + 1p morphologies are experimentally reported\cite{MohanPNP,jacs, schaak2025,Chen2016PolyelementalLibraries} and lie within range of DFT-calculated surface and interfacial energies, whereas the Pokéball, Shell with Janus core, Janus with Core@Shell, Shell with 2 cores, and Core@Shell + 1p do lie within range of DFT-calculated surface and interfacial energies, but are not yet experimentally reported. The remaining morphology, 3p, is neither experimentally reported nor within range of DFT-calculated surface and interfacial energies.
3p is unlikely to be experimentally reported for the same reason that 2p is unlikely to be observed in the biphase NP case: the interfacial energies between the three constituent phases must be larger than their individual surfaces. The remaining ten morphologies lie within range of DFT-calculated surface and interfacial energies, which implies they may occur at the proper combination of ratios.

\begin{figure*}[h!]
 \centering
 \includegraphics[width=0.7\textwidth]{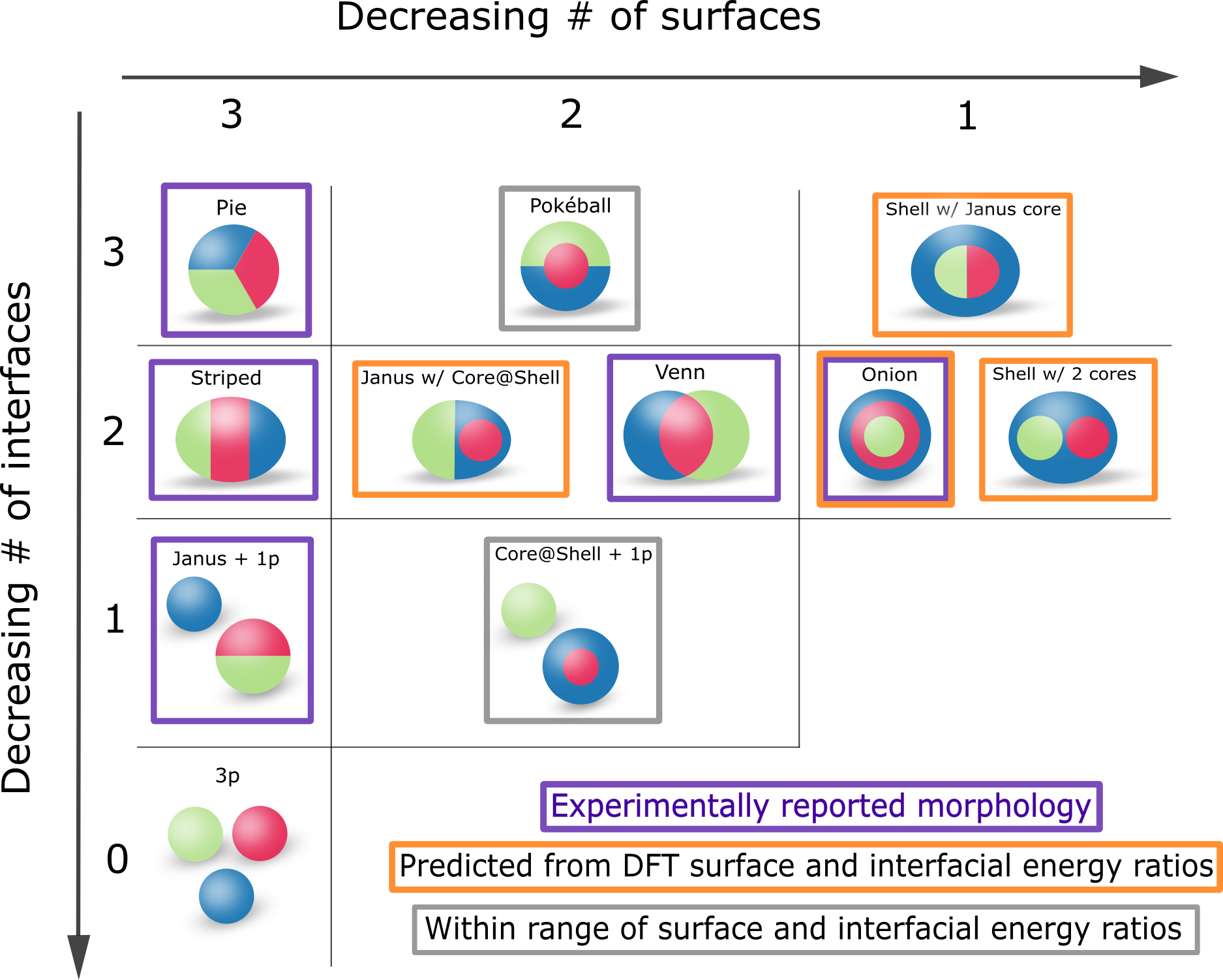}
 \caption{Idealized schematic representations of all triphase morphologies predicted, distributed by number of surfaces and number of interfaces. Those outlined with boxes are within the range of DFT-calculated surface and interfacial energy ratios from DFT surface energy\cite{Tran2016DataCrystals} and interfacial energy\cite{interfacedatabase} databases. The morphologies highlighted in purple are experimentally reported\cite{MohanPNP, jacs, schaak2025, Chen2016PolyelementalLibraries}. Morphologies highlighted by orange boxes are predicted by systems in the DFT surface energy\cite{Tran2016DataCrystals} and interfacial energy\cite{interfacedatabase} databases (individual systems reported in Table \ref{tab:triphase_pred}).
 }
 \label{fig:triphase_boxed}
\end{figure*}

\section*{Conclusions}
The field of nanocombinatorics is growing rapidly but currently lacks a predictive computational approach to determining internal interface morphology of multiphase nanoparticles. This work addresses that gap with our proposed Monte Carlo Potts model, to predict multiphase nanoparticle interface morphologies as a function of surface and interfacial energies.
MCP provides a discretized, flexible representation of multiphase NPs that requires no prior enumeration of candidate morphologies, in contrast to the idealized-geometry approach. It applies to any number of phases, any stoichiometric ratio, and any environment, and recovers non-idealized transition morphologies -- such as the Dumbbell and Lens-type biphase architectures -- that analytically-derived phase boundaries from idealized geometry cannot capture. This systematic search across surface and interfacial energy ratios makes it possible to construct stability maps for bi- and triphase NPs, which can then be used to predict morphologies of other phase-separating elemental combinations. The approach is extensible to $n$-phase NPs beyond $n = 3$, and its discretized lattice additionally provides a three-dimensional view of internal interface architecture that is difficult to capture in the two-dimensional snapshots typical of experimental microscopy.

Applying MCP to biphase and triphase NPs surfaced three biphase morphologies and 11 triphase morphologies.
These morphologies follow the ``number of interfaces in an $n$-phase NP'' rule proposed in Ref. \citenum{MohanPNP} (between $n-1$ and $\nicefrac{n(n-1)}{2}$) and an analogous rule for the number of surfaces (between $1$ and $n$), but we show that neither the interface count, nor the surface count, nor the two together, uniquely determine a morphology. The ground-state architecture depends on the relative surface and interfacial energies of the component phases, as captured in the stability maps. Applying DFT-calculated energies to elemental phase combinations, MCP predictions tend toward Core@Shell-like morphologies, consistent with the hypothesis that DFT surface energies calculated in vacuum favor Core@Shell formation.\cite{polymersurfacepassivation} Comparison with experimentally reported NPs shows agreement in the majority of biphase systems, and in the triphase case, the discrepancies are systematic and interpretable -- most MCP predictions differ from experiment by only a single topological transition, consistent with kinetically limited rather than incorrect ground-state predictions.

The current model has limitations that point to clear next steps. It does not include kinetic effects, coherency strain, or temperature dependence, all of which likely contribute to the discrepancies observed between predicted and experimentally reported morphologies. Input approximations -- including the $T = 0~\si{\kelvin}$, bulk, fully-ordered-phase assumptions implemented within OQMD, and the use of vacuum-calculated DFT surface energies -- may cumulatively push predictions away from their true energetic ground state.
Additionally, the NP total energy equation does not factor in NP size-dependence or contact with the substrate. Potential avenues for improvement of the model include incorporating kinetic effects via a kinetic MC algorithm, or by increasing resolution via a further discretization of the Potts model into an atomistic simulation.

\section{Methods}
\subsection{Biphase NP morphologies}
\label{sec:method_biphase}
For MCP calculation of the 2D cases, we used a $50 \times 50$ square lattice with periodic boundary conditions and, by extension, a $50 \times 50\times 50$ cubic lattice for 3D cases. We initialized a random distribution of the 3 spins ($q=0,1,2$, representing the environment, phase A, and phase B, respectively) on the underlying lattice.
We tailored the number of spins representing the environment ($q=0$) to allow the formation of separate particles. If the amount of $q=0$ spins is too low, the system is artificially constrained in such a way that the A and B phases must be in contact.
The volume ratio of $\mathrm{Environment}$ spins on the 2D square lattice is $70\%$ and $90\%$ on the 3D cubic lattice. 
All results discussed assume equal amounts of A and B unless otherwise noted.
We applied a spin-conserved Metropolis MC algorithm with a simulated anneal to the random initial lattice using three dimensionless temperature steps of $kT=0.25,\ 0.125,\ 0.05$. As a check on the three-step anneal, we also repeated the calculations with a finer schedule
in which $kT$ was lowered from 0.25 to 0.04, each step set to $90\%$ of the previous value, and obtained
the same morphologies.

In 2D, each temperature step was held for 10,000 MC timesteps and 600 MC timesteps in 3D.
We fine-tuned the two simulation parameters of temperature and number of equilibration steps to allow the simulation to reach a converged energy state before a decrease in temperature while minimizing computational resources.
The initial temperature step is adjusted such that the acceptance probability of an uphill spin swap is $\sim 80\%$ and the number of MC timesteps is dictated by the minimum number required for the simulation total energy to not change beyond $0.01\%$ in the last $10\%$ of the timesteps.
For example, in 2D simulations, the simulation energy convergence at that temperature occurs when the energy does not change more than $0.01\%$ during the last 1,000 MC timesteps at that temperature step.
We repeat each run five times and select the one with the lowest energy as the representative morphology.

We systematically applied the MCP method to sample input ratios to calculate the effect of interfacial and surface energies on biphase NP morphologies.
We iterate in a grid search fashion over the two values in $J_{q_{m}q_{n}} = \left(\begin{smallmatrix}
  0 & 1 & \nicefrac{\gamma_B}{\gamma_A} \\
  1 & 0 & \nicefrac{\sigma_{AB}}{\gamma_A} \\
  \nicefrac{\gamma_B}{\gamma_A} & \nicefrac{\sigma_{AB}}{\gamma_A} & 0 \end{smallmatrix}\right)$ to calculate a stability map by visualization and categorization of the MCP output. We perform the grid search over $\nicefrac{\gamma_B}{\gamma_A}$ and $\nicefrac{\sigma_{AB}}{\gamma_A}$ in the range of 0.005 to 1 in increments of 0.0025, and extend the upper range to 2 for $\nicefrac{\sigma_{AB}}{\gamma_A}$.

We use a point-by-point iteration over the final lattice to count the number of \ce{A-Environment}, \ce{B-Environment}, and \ce{A-B} nearest neighbor spin interactions, which provides a numerical value on the amount of \ce{A} surface, \ce{B} surface, and \ce{A-B} interface observed in the output lattice. We categorize the output morphology from these values according to the definitions of each morphology (see also Table \ref{tab:biphase_morphs} for the surface and interface counts).
In short, any morphology with zero \ce{A-Environment} or \ce{B-Environment} interactions is a Core@Shell.
A morphology with no \ce{A-B} interactions is 2p.
A morphology with any number of all three interactions is a Janus morphology.
Thus, a categorical \ce{A-B} biphase NP interface morphology is mapped to a pair of $\nicefrac{\gamma_B}{\gamma_A}$ and $\nicefrac{\sigma_{AB}}{\gamma_A}$ values, forming a ground-state NP morphology stability map.

\subsection{Triphase NP morphologies}
\label{sec:method_triphase}
Following the construction of the biphase NP morphology stability map, we applied the same procedure to triphase particles. Triphase NPs are composed of sample phases A, B, and C, represented by four spins, $q=0,1,2,3$, representing the environment followed by each of the three phases. The four spins form six unlike interactions: \ce{A-Environment}, \ce{B-Environment}, \ce{C-Environment}, \ce{A-B}, \ce{A-C}, and \ce{B-C}. These interactions are described by the respective surface and interfacial energies of the phases, $\gamma_A$, $\gamma_B$, $\gamma_C$, $\sigma_{AB}$, $\sigma_{AC}$, and $\sigma_{BC}$. We conducted a comprehensive search of this reduced parameter space, as described earlier in the biphase section, to calculate a triphase morphology stability map. We conduct the grid search over the range of 0.1 to 1 in increments of 0.1 for each ratio. Our ordering of phases by decreasing surface energy allows us to remove some combinations of ratios without loss of generality, so for $m$ increments, we can map out the grid using $\nicefrac{m(m+1)}{2}\times\nicefrac{m(m+1)}{2}\times m$ points. As before, to allow separation of A, B, and C into three separate particles, a minimum amount of environment $q=0$ spins must be present in the lattice system.
The volume ratio of $\mathrm{Environment}$ spins on the 2D square lattice is $50\%$ and $90\%$ on the 3D cubic lattice. 
All results discussed use equal numbers of spins of A, B, and C.
As in the biphase case, we apply the same counting procedure to \ce{A-Environment}, \ce{B-Environment}, \ce{C-Environment}, \ce{A-B}, \ce{A-C}, and \ce{B-C} interactions and take into account the connectivity of each phase to distinguish topologically distinct triphase morphologies.

These approaches can extend to $n$-phase systems using an $(n+1) \times (n+1)$ interaction energy matrix populated with $\nicefrac{n(n+1)}{2}$ non-zero terms. The non-zero terms used to model the system comprise $n$ surface energy terms $(\gamma_{1},\gamma_2,...,\gamma_{n-1},\gamma_n)$ and $\nicefrac{n(n-1)}{2}$ interfacial energy terms between unlike phases. These terms populate the interaction energy matrix $J_{q_mq_n}$, which serves as input into MCP and allows for the solution of eq.~\ref{eqn:ham}, the discretized form of eq.~\ref{eqn:np_toten}.

\section{Conflicts of interest}
There are no conflicts to declare.

\section{Data availability}
The data supporting this article have been included as part of the Supplementary Information.

\begin{acknowledgement}
E.S. acknowledges support by the National Science Foundation Graduate Research Fellowship award NSF DGE-1842165. The authors acknowledge financial support of this research by the International Institute of Nanotechnology at Northwestern University. This research was supported in part through the computational resources provided by the Quest high performance computing facility at Northwestern University which is jointly supported by the Office of the Provost, the Office for Research, and Northwestern University Information Technology. 

\end{acknowledgement}

\bibliography{references}

\clearpage
\appendix
\input{si}

\end{document}

%% file: si.tex
\setcounter{table}{0}
\setcounter{figure}{0}
\renewcommand{\thetable}{S\arabic{table}}
\renewcommand{\thefigure}{S\arabic{figure}}
\renewcommand{\theHtable}{S\arabic{table}}   
\renewcommand{\theHfigure}{S\arabic{figure}}

\begin{figure}[h!]
\centering
\includegraphics[width=0.7\textwidth]{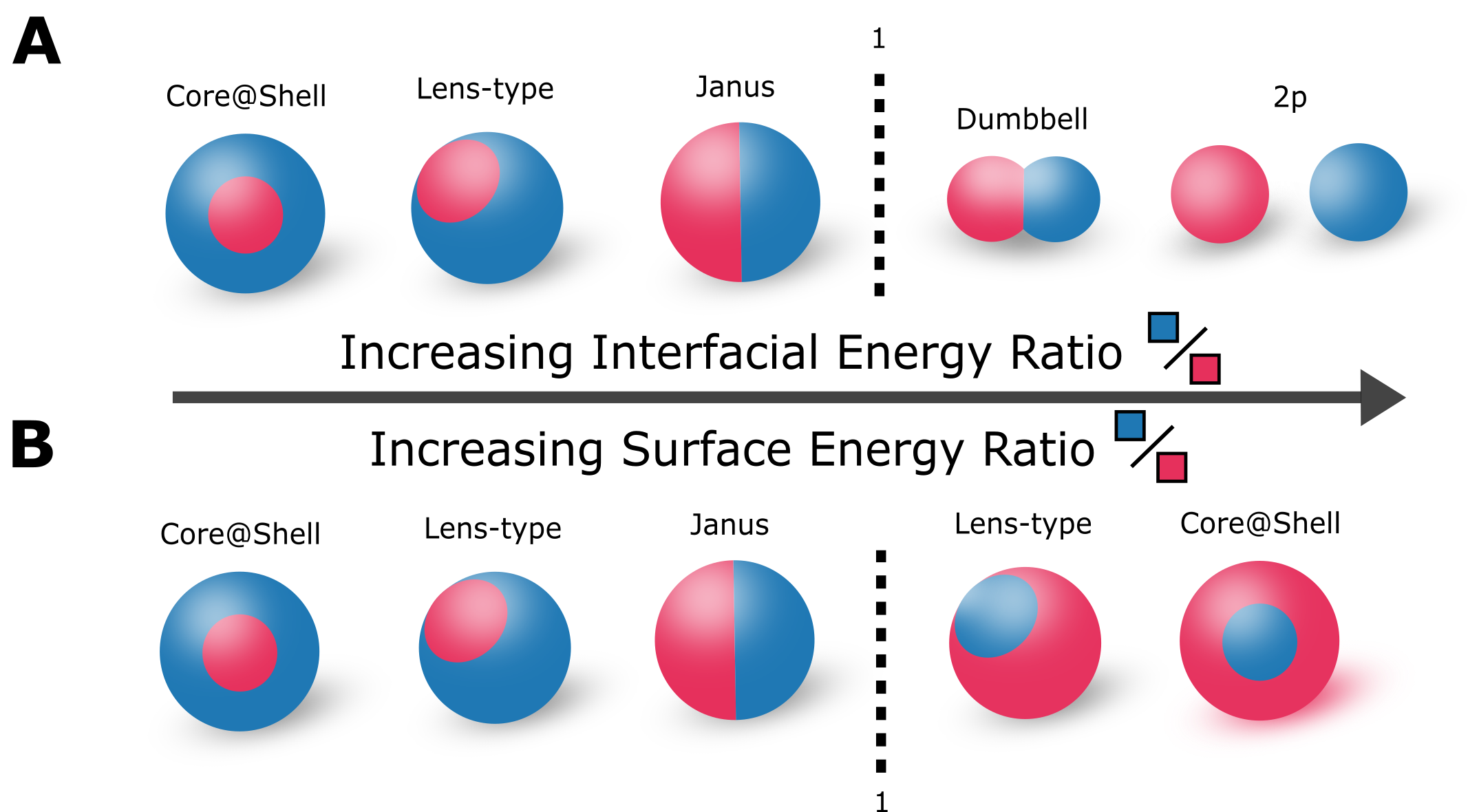}
  \caption{Schematic of biphase NP morphologies transition across the Potts-calculated biphase stability map. As the ratio of interfacial energy relative to the largest surface energy ( $\nicefrac{\sigma_{AB}}{\gamma_A}$ ) increases in (A), the ground-state NP morphology predicted by MCP evolves from Core@Shell, to Lens-type, and then to an idealized Janus in an effort to minimize interfacial area. Once the ratio surpasses 1, meaning $\sigma_{AB} > {\gamma_A}$, the predicted ground-state NP morphology is now Dumbbell, which decreases the shared interfacial area and increases the surface area, and eventually the interface is removed altogether, forming 2 separate particles (2p). In (B), the increase in surface energy ratio ($\nicefrac{\gamma_B}{\gamma_A}$) between the two phases changes the predicted ground-state NP morphology. Beginning at the same point as before, with the A@B Core@Shell morphology, the increase in surface energy of B relative to A results in the formation of the Lens-type morphology to reduce the amount of B surface area. Further increase in $\gamma_B$ results in the idealized Janus morphology, where the surface area of B has further decreased from the surface area of B in the Lens-type morphology. Once the ratio $\nicefrac{\gamma_B}{\gamma_A}$ surpasses 1, a Lens-type morphology forms once more, with the aim to decrease the surface area of B, since the surface energy of B now surpasses that of A. Continuing to increase the $\nicefrac{\gamma_B}{\gamma_A}$ ratio results in the formation of a B@A Core@Shell NP, where there is no more B surface present. }
  \label{fig:2r_transition_linear}
\end{figure}

Supplementary Information: additional calculation details, including surface and interfacial energies used in MCP predictions, further comparison of the use of DFT and experimental data in MCP predictions, and tables of compiled published experimental multiphase nanoparticle data
\begin{itemize}
    \item Table \ref{tab:biphase_pred}. Input parameter details for biphase nanoparticle interface morphology compared to experimental references
    \item Table \ref{tab:mohan_3r_temp}. Input parameter details for triphase nanoparticle interface morphology compared to experimental references, with consideration for temperature
    \item Figure \ref{fig:2r_transition_linear}. Schematic illustration of biphase NP morphology transitions along the biphase stability map
    \item Table \ref{tab:triphase_pred_dft}. Input parameter details for triphase nanoparticle interface morphology predictions using DFT-calculated surface and interfacial energies
    \item Table \ref{tab:multiphase_1}. Compilation of published experimental single-phase nanoparticle data
    \item Table \ref{tab:multiphase_2}. Compilation of published experimental biphase nanoparticle data
    \item Table \ref{tab:multiphase_3}. Compilation of published experimental triphase nanoparticle data
    \item Table \ref{tab:multiphase_4}. Compilation of published experimental tetraphase nanoparticle data
    \item Figure \ref{fig:3p2d_strict_stability_map}. Triphase NP morphology stability map calculated on a 2D lattice
    \item Figure \ref{fig:3p3d_stability_map}. Triphase NP morphology stability map calculated on a 3D lattice
    \item Figure \ref{fig:3p3d_loose_stability_map}. Triphase NP morphology stability map calculated on a 3D lattice with a looser domain-contact threshold (2.0)
\end{itemize}

\section{Supplementary text}

\input{table_biphase_pred}

\begin{landscape}\SItablefont\renewcommand{\arraystretch}{0.9}
\begin{xltabular}{\linewidth}{@{} *{11}{c} @{}}
    \caption{Input parameters with temperature effect on surface energy values. The DFT-calculated surface energy is given,\cite{Tran2016DataCrystals} followed by experimentally-reported surface energies extrapolated to the stated temperature.\cite{expt_surface_energy} Interfacial energy values reported are DFT-calculated values from Ref.~\citenum{MohanPNP}, except for \ce{Ni-Ag} from Ref.~\citenum{interfacedatabase}. *Experimental surface energies are not available, so the DFT-calculated value from Ref.~\citenum{MohanPNP} is used.}
\label{tab:mohan_3r_temp}    \\
\toprule
\makecell{Temperature for\\$\gamma$ extrapolation\\{}[\unit{\kelvin}]} & \makecell{Phase\\{}A} & \makecell{Phase\\B} & \makecell{Phase\\C} & \makecell{$\gamma_{A}$\\{}[\unit{\joule\per\meter\squared}]} & \makecell{$\gamma_B$\\{}[\unit{\joule\per\meter\squared}]} & \makecell{$\gamma_C$\\{}[\unit{\joule\per\meter\squared}]} & \makecell{$\sigma_{AB}$\\{}[\unit{\joule\per\meter\squared}]} & \makecell{$\sigma_{AC}$\\{}[\unit{\joule\per\meter\squared}]} & \makecell{$\sigma_{BC}$\\{}[\unit{\joule\per\meter\squared}]} & \makecell{MCP-predicted\\Morphology} \\
\endfirsthead
\midrule
DFT & \ce{Co} & \ce{Pd3Sn} & \ce{Au} & 2.042 & 1.330* & 0.742 & 0.769 & 0.513 & 0.016 & Shell w/ 2 cores \\
0 & \ce{Co} & \ce{Au} & \ce{Pd3Sn} & 2.519 & 1.351 & 1.330* & 0.513 & 0.769 & 0.016 & Onion \\
773.15 & \ce{Co} & \ce{Pd3Sn} & \ce{Au} & 2.248 & 1.330* & 1.235 & 0.769 & 0.513 & 0.016 & Onion-like \\
873.15 & \ce{Co} & \ce{Pd3Sn} & \ce{Au} & 2.213 & 1.330* & 1.220 & 0.769 & 0.513 & 0.016 & Onion-like \\
1023.15 & \ce{Co} & \ce{Pd3Sn} & \ce{Au} & 2.161 & 1.330* & 1.197 & 0.769 & 0.513 & 0.016 & Onion-like \\
DFT & \ce{Co} & \ce{Cu} & \ce{Ag} & 2.042 & 1.314 & 0.773 & 0.208 & 0.833 & 0.369 & Onion \\
0 & \ce{Co} & \ce{Cu} & \ce{Ag} & 2.519 & 1.700 & 1.197 & 0.208 & 0.833 & 0.369 & Onion \\
773.15 & \ce{Co} & \ce{Cu} & \ce{Ag} & 2.248 & 1.484 & 1.027 & 0.208 & 0.833 & 0.369 & Onion \\
873.15 & \ce{Co} & \ce{Cu} & \ce{Ag} & 2.213 & 1.456 & 1.005 & 0.208 & 0.833 & 0.369 & Onion \\
1023.15 & \ce{Co} & \ce{Cu} & \ce{Ag} & 2.161 & 1.414 & 0.972 & 0.208 & 0.833 & 0.369 & Onion \\
DFT & \ce{Co} & \ce{Pd3Sn} & \ce{Cu} & 2.042 & 1.330* & 1.314 & 0.769 & 0.208 & 0.336 & Janus w/ Core@Shell-like \\
0 & \ce{Co} & \ce{Cu} & \ce{Pd3Sn} & 2.519 & 1.700 & 1.330* & 0.208 & 0.769 & 0.336 & Janus w/ Core@Shell-like \\
773.15 & \ce{Co} & \ce{Cu} & \ce{Pd3Sn} & 2.248 & 1.484 & 1.330* & 0.208 & 0.769 & 0.336 & Janus w/ Core@Shell \\
873.15 & \ce{Co} & \ce{Cu} & \ce{Pd3Sn} & 2.213 & 1.456 & 1.330* & 0.208 & 0.769 & 0.336 & Janus w/ Core@Shell \\
1023.15 & \ce{Co} & \ce{Cu} & \ce{Pd3Sn} & 2.161 & 1.414 & 1.330* & 0.208 & 0.769 & 0.336 & Janus w/ Core@Shell \\
DFT & \ce{Co} & \ce{Pd3Sn} & \ce{Ag} & 2.042 & 1.330* & 0.773 & 0.769 & 0.833 & 0.192 & Shell w/ Janus core \\
DFT & \ce{Ni} & \ce{Cu} & \ce{Ag} & 1.924 & 1.314 & 0.773 & 0.112 & 0.212 & 0.369 & Onion \\
\bottomrule
\end{xltabular}
\end{landscape}

\input{table_triphase_pred_dft_updated}

\input{table_single_phase}
\clearpage
\input{table_biphase}
\clearpage
\input{table_triphase}
\clearpage
\input{table_tetraphase}
\clearpage
\input{triphase_2d_stability_map}
\clearpage
\input{triphase_3d_stability_map}
\clearpage
\input{triphase_3d_loose_stability_map}

%% file: table_biphase_pred.tex
\begingroup\SItablefont
\begin{longtable}{@{\extracolsep{\fill}}ccccccc}
  \caption{\ This table provides further details regarding the input parameters for the prediction of biphase morphologies from DFT-calculated interfacial energies,\cite{interfacedatabase} as well as the DFT-calculated\cite{Tran2016DataCrystals} and experimental\cite{expt_surface_energy} surface energies. Experimental surface energies are difficult to measure along specific crystallographic directions, so they can be interpreted as an average over all directions.\cite{expt_surface_energy} The experimental surface energies are extrapolated to 0~K, and \lq\lq---\rq\rq{} indicates that no experimental value is available. The DFT-calculated interfacial energies are reported along the close-packed direction, in other words, [111] for FCC structures, [110] for BCC, and [0001] for HCP. Predictions are made for all phase-separating isostructural binary combinations of elemental metals. Phases A and B are ordered according to decreasing DFT-calculated surface energy. Co is paired with FCC metals (Ag, Au, Cu) using the DFT-calculated surface energy of FCC Co, and with HCP metals (Cd, Os, Re, Ru, Tc, Zn) using that of HCP Co. }
  \label{tab:biphase_pred}
  \\
    \hline
    Phase A & Phase B & $\gamma_A^{DFT}$ [\unit{\joule\per\meter\squared}] & $\gamma_A^{exp, 0K}$ [\unit{\joule\per\meter\squared}] & $\gamma_B^{DFT}$ [\unit{\joule\per\meter\squared}] & $\gamma_B^{exp, 0K}$ [\unit{\joule\per\meter\squared}] & $\sigma_{AB}^{DFT}$ [\unit{\joule\per\meter\squared}] \\
    \hline
    \endfirsthead
    \hline
    Phase A & Phase B & $\gamma_A^{DFT}$ [\unit{\joule\per\meter\squared}] & $\gamma_A^{exp, 0K}$ [\unit{\joule\per\meter\squared}] & $\gamma_B^{DFT}$ [\unit{\joule\per\meter\squared}] & $\gamma_B^{exp, 0K}$ [\unit{\joule\per\meter\squared}] & $\sigma_{AB}^{DFT}$ [\unit{\joule\per\meter\squared}] \\
    \hline
    \endhead
Co (fcc) & Ag & 2.042 & 2.519 & 0.773 & 1.197 & 0.505  \\
Cu & Ag & 1.314 & 1.700 & 0.773 & 1.197 & 0.106  \\
Ir & Ag & 2.284 & 2.929 & 0.773 & 1.197 & 0.707  \\
Ni & Ag & 1.924 & 2.365 & 0.773 & 1.197 & 0.212  \\
Ag & Pb & 0.773 & 1.197 & 0.253 & 0.523 & 0.006  \\
Rh & Ag & 1.989 & 2.611 & 0.773 & 1.197 & 0.413  \\
Al & Pb & 0.795 & 1.283 & 0.253 & 0.523 & 0.317  \\
Co (fcc) & Au & 2.042 & 2.519 & 0.742 & 1.351 & 0.187  \\
Ir & Au & 2.284 & 2.929 & 0.742 & 1.351 & 0.671  \\
Pt & Au & 1.479 & 2.373 & 0.742 & 1.351 & 0.066  \\
Rh & Au & 1.989 & 2.611 & 0.742 & 1.351 & 0.243  \\
Ca & Sr & 0.461 & 0.474 & 0.342 & 0.379 & 0.005  \\
Co (fcc) & Cu & 2.042 & 2.519 & 1.314 & 1.700 & 0.357  \\
Ir & Cu & 2.284 & 2.929 & 1.314 & 1.700 & 0.201  \\
Ni & Cu & 1.924 & 2.365 & 1.314 & 1.700 & 0.0004  \\
Rh & Cu & 1.989 & 2.611 & 1.314 & 1.700 & 0.100  \\
Ir & Pd & 2.284 & 2.929 & 1.338 & 1.991 & 0.086  \\
Ir & Pt & 2.284 & 2.929 & 1.479 & 2.373 & 0.050  \\
Rh & Pd & 1.989 & 2.611 & 1.338 & 1.991 & 0.090  \\
Ba & Cs & 0.312 & 0.345 & 0.060 & 0.704 & 0.056  \\
Ba & K & 0.312 & 0.345 & 0.108 & 0.134 & 0.077  \\
Ba & Rb & 0.312 & 0.345 & 0.082 & 0.107 & 0.082  \\
Cr & Fe & 3.201 & 2.885 & 2.447 & 2.623 & 0.104  \\
Cr & Li & 3.201 & 2.885 & 0.498 & 0.468 & 1.299  \\
Cr & Mo & 3.201 & 2.885 & 2.797 & 2.080 & 0.005  \\
W & Cr & 3.228 & 2.310 & 3.201 & 2.885 & 0.017  \\
Fe & Li & 2.447 & 2.623 & 0.498 & 0.468 & 0.491  \\
W & Fe & 3.228 & 2.310 & 2.447 & 2.623 & 0.103  \\
Mo & Li & 2.797 & 2.080 & 0.498 & 0.468 & 0.946  \\
Nb & Li & 2.074 & 2.485 & 0.498 & 0.468 & 0.631  \\
Ta & Li & 2.342 & 2.050 & 0.498 & 0.468 & 0.868  \\
V & Li & 2.421 & 1.860 & 0.498 & 0.468 & 0.882  \\
W & Li & 3.228 & 2.310 & 0.498 & 0.468 & 1.270  \\
Mo & Na & 2.797 & 2.080 & 0.218 & 0.230 & 1.427  \\
W & Mo & 3.228 & 2.310 & 2.797 & 2.080 & 0.029  \\
Nb & Na & 2.074 & 2.485 & 0.218 & 0.230 & 1.032  \\
Ta & Na & 2.342 & 2.050 & 0.218 & 0.230 & 1.143  \\
W & Na & 3.228 & 2.310 & 0.218 & 0.230 & 1.865  \\
V & Nb & 2.421 & 1.860 & 2.074 & 2.485 & 0.037  \\
V & Ta & 2.421 & 1.860 & 2.342 & 2.050 & 0.016  \\
Be & Zn & 1.790 & --- & 0.348 & --- & 0.625  \\
Co (hcp) & Cd & 2.108 & --- & 0.200 & --- & 0.583  \\
Os & Cd & 2.920 & --- & 0.200 & --- & 1.189  \\
Re & Cd & 2.581 & --- & 0.200 & --- & 1.087  \\
Ru & Cd & 2.577 & --- & 0.200 & --- & 0.661  \\
Tc & Cd & 2.234 & --- & 0.200 & --- & 0.806  \\
Ti & Cd & 1.966 & --- & 0.200 & --- & 0.093  \\
Tl & Cd & 0.256 & --- & 0.200 & --- & 0.052  \\
Zn & Cd & 0.348 & --- & 0.200 & --- & 0.184  \\
Os & Co (hcp) & 2.920 & --- & 2.108 & --- & 0.176  \\
Re & Co (hcp) & 2.581 & --- & 2.108 & --- & 0.106  \\
Ru & Co (hcp) & 2.577 & --- & 2.108 & --- & 0.223  \\
Tc & Co (hcp) & 2.234 & --- & 2.108 & --- & 0.057  \\
Co (hcp) & Zn & 2.108 & --- & 0.348 & --- & 0.075  \\
Hf & Mg & 1.709 & --- & 0.545 & --- & 0.103  \\
Re & Hf & 2.581 & --- & 1.709 & --- & 0.138  \\
Ti & Hf & 1.966 & --- & 1.709 & --- & 0.008  \\
Hf & Tl & 1.709 & --- & 0.256 & --- & 0.189  \\
Hf & Y & 1.709 & --- & 1.004 & --- & 0.062  \\
Re & Mg & 2.581 & --- & 0.545 & --- & 0.281  \\
Tc & Mg & 2.234 & --- & 0.545 & --- & 0.015  \\
Ti & Mg & 1.966 & --- & 0.545 & --- & 0.211  \\
Zr & Mg & 1.610 & --- & 0.545 & --- & 0.041  \\
Os & Tl & 2.920 & --- & 0.256 & --- & 1.646  \\
Re & Ti & 2.581 & --- & 1.966 & --- & 0.200  \\
Re & Tl & 2.581 & --- & 0.256 & --- & 1.410  \\
Re & Zn & 2.581 & --- & 0.348 & --- & 0.005  \\
Re & Zr & 2.581 & --- & 1.610 & --- & 0.385  \\
Ti & Sc & 1.966 & --- & 1.267 & --- & 0.024  \\
Tc & Ti & 2.234 & --- & 1.966 & --- & 0.019  \\
Tc & Tl & 2.234 & --- & 0.256 & --- & 1.079  \\
Tc & Zr & 2.234 & --- & 1.610 & --- & 0.167  \\
Ti & Tl & 1.966 & --- & 0.256 & --- & 0.327  \\
Ti & Y & 1.966 & --- & 1.004 & --- & 0.201  \\
Ti & Zr & 1.966 & --- & 1.610 & --- & 0.032  \\
Zr & Tl & 1.610 & --- & 0.256 & --- & 0.079  \\
Zr & Y & 1.610 & --- & 1.004 & --- & 0.036  \\
    \hline
\end{longtable}
\endgroup

%% file: table_triphase_pred_dft_updated.tex
\begin{landscape}
\begin{longtable}{@{\extracolsep{\fill}}ccccccccc}
  \caption{\ This table reproduces the input parameters for the prediction of triphase morphologies from DFT-calculated interfacial energies\cite{interfacedatabase} and surface energies.\cite{Tran2016DataCrystals} Predictions are made for all phase-separating isostructural ternary combinations of elemental metals. Phases A, B, and C are ordered according to decreasing DFT-calculated surface energy. Next, we provide the orientation selected for the surface and interfacial energies. All surface and interfacial energies are normalized by the surface energy of Phase A, $\gamma_A^{DFT}$, and are therefore dimensionless. }
  \label{tab:triphase_pred_dft}
  \small
  \\
    \hline
    \makecell{Phase\\A} & \makecell{Phase\\B} & \makecell{Phase\\C} & \makecell{[hkl]} & \makecell{$\gamma_B^{DFT}/\gamma_A^{DFT}$} & \makecell{$\gamma_C^{DFT}/\gamma_A^{DFT}$} & \makecell{$\sigma_{AB}^{DFT}/\gamma_A^{DFT}$} & \makecell{$\sigma_{AC}^{DFT}/\gamma_A^{DFT}$} & \makecell{$\sigma_{BC}^{DFT}/\gamma_A^{DFT}$} \\
    \hline
    \endfirsthead
    \hline
    \makecell{Phase\\A} & \makecell{Phase\\B} & \makecell{Phase\\C} & \makecell{[hkl]} & \makecell{$\gamma_B^{DFT}/\gamma_A^{DFT}$} & \makecell{$\gamma_C^{DFT}/\gamma_A^{DFT}$} & \makecell{$\sigma_{AB}^{DFT}/\gamma_A^{DFT}$} & \makecell{$\sigma_{AC}^{DFT}/\gamma_A^{DFT}$} & \makecell{$\sigma_{BC}^{DFT}/\gamma_A^{DFT}$} \\
    \hline
    \endhead
Ba & K & Rb & [100] & 0.378 & 0.294 & 0.331 & 0.369 & 0.001 \\
Co & Cu & Ag & [100] & 0.592 & 0.327 & 0.159 & 0.339 & 0.063 \\
Co & Cu & Ag & [110] & 0.645 & 0.358 & 0.198 & 0.352 & 0.085 \\
Co & Cu & Ag & [111] & 0.644 & 0.378 & 0.175 & 0.247 & 0.052 \\
Co & Rh & Ag & [100] & 0.938 & 0.327 & 0.013 & 0.339 & 0.170 \\
Co & Rh & Au & [100] & 0.938 & 0.347 & 0.013 & 0.159 & 0.133 \\
Co & Rh & Cu & [100] & 0.938 & 0.592 & 0.013 & 0.159 & 0.060 \\
Co & Zn & Cd & [0001] & 0.165 & 0.095 & 0.035 & 0.276 & 0.087 \\
Cr & Fe & Li & [100] & 0.688 & 0.127 & 0.041 & 0.525 & 0.271 \\
Cr & Fe & Li & [110] & 0.764 & 0.156 & 0.032 & 0.406 & 0.153 \\
Cr & Fe & Li & [111] & 0.795 & 0.158 & 0.044 & 0.433 & 0.189 \\
Cr & Mo & Fe & [100] & 0.876 & 0.688 & 0.074 & 0.041 & 0.042 \\
Cr & Mo & Li & [100] & 0.876 & 0.127 & 0.074 & 0.525 & 0.373 \\
Cr & Mo & Li & [110] & 0.874 & 0.156 & 0.002 & 0.406 & 0.296 \\
Cr & Mo & Li & [111] & 0.862 & 0.158 & 0.073 & 0.433 & 0.334 \\
Cr & Nb & Li & [100] & 0.626 & 0.127 & 0.083 & 0.525 & 0.313 \\
Cr & Nb & Li & [111] & 0.681 & 0.158 & 0.114 & 0.433 & 0.257 \\
Cr & Ta & Li & [100] & 0.680 & 0.127 & 0.053 & 0.525 & 0.323 \\
Cr & Ta & Li & [111] & 0.786 & 0.158 & 0.109 & 0.433 & 0.323 \\
Ir & Co & Ag & [100] & 0.880 & 0.288 & 0.011 & 0.283 & 0.298 \\
Ir & Co & Au & [100] & 0.880 & 0.305 & 0.011 & 0.294 & 0.140 \\
Ir & Co & Cu & [100] & 0.880 & 0.521 & 0.011 & 0.096 & 0.140 \\
Ir & Cu & Ag & [100] & 0.521 & 0.288 & 0.096 & 0.283 & 0.056 \\
Ir & Cu & Ag & [110] & 0.563 & 0.312 & 0.095 & 0.333 & 0.074 \\
Ir & Cu & Ag & [111] & 0.575 & 0.338 & 0.088 & 0.310 & 0.046 \\
Ir & Pt & Ag & [110] & 0.606 & 0.312 & 0.036 & 0.333 & 0.015 \\
Ir & Pt & Au & [100] & 0.654 & 0.305 & 0.024 & 0.294 & 0.034 \\
Ir & Pt & Au & [110] & 0.606 & 0.298 & 0.036 & 0.307 & 0.052 \\
Ir & Pt & Au & [111] & 0.648 & 0.325 & 0.022 & 0.294 & 0.029 \\
Mo & Fe & Li & [100] & 0.785 & 0.145 & 0.048 & 0.426 & 0.310 \\
Mo & Li & Na & [100] & 0.145 & 0.068 & 0.426 & 0.662 & 0.042 \\
Mo & Li & Na & [111] & 0.184 & 0.084 & 0.388 & 0.581 & 0.009 \\
Nb & Li & Na & [100] & 0.202 & 0.096 & 0.499 & 0.501 & 0.059 \\
Nb & Li & Na & [111] & 0.232 & 0.107 & 0.378 & 0.529 & 0.011 \\
Ni & Cu & Ag & [100] & 0.665 & 0.367 & 0.022 & 0.233 & 0.071 \\
Ni & Cu & Ag & [110] & 0.683 & 0.379 & 0.029 & 0.252 & 0.090 \\
Ni & Cu & Ag & [111] & 0.683 & 0.402 & 0.000 & 0.110 & 0.055 \\
Os & Co & Cd & [0001] & 0.722 & 0.068 & 0.060 & 0.407 & 0.200 \\
Os & Tl & Cd & [0001] & 0.088 & 0.068 & 0.564 & 0.407 & 0.018 \\
Re & Co & Cd & [0001] & 0.817 & 0.077 & 0.041 & 0.421 & 0.226 \\
Re & Co & Zn & [0001] & 0.817 & 0.135 & 0.041 & 0.002 & 0.029 \\
Re & Hf & Mg & [0001] & 0.662 & 0.211 & 0.053 & 0.109 & 0.040 \\
Re & Hf & Tl & [0001] & 0.662 & 0.099 & 0.053 & 0.546 & 0.073 \\
Re & Ti & Cd & [0001] & 0.762 & 0.077 & 0.077 & 0.421 & 0.036 \\
Re & Ti & Hf & [0001] & 0.762 & 0.662 & 0.077 & 0.053 & 0.003 \\
Re & Ti & Mg & [0001] & 0.762 & 0.211 & 0.077 & 0.109 & 0.082 \\
Re & Ti & Tl & [0001] & 0.762 & 0.099 & 0.077 & 0.546 & 0.127 \\
Re & Ti & Zr & [0001] & 0.762 & 0.624 & 0.077 & 0.149 & 0.013 \\
Re & Tl & Cd & [0001] & 0.099 & 0.077 & 0.546 & 0.421 & 0.020 \\
Re & Zn & Cd & [0001] & 0.135 & 0.077 & 0.002 & 0.421 & 0.071 \\
Re & Zr & Mg & [0001] & 0.624 & 0.211 & 0.149 & 0.109 & 0.016 \\
Re & Zr & Tl & [0001] & 0.624 & 0.099 & 0.149 & 0.546 & 0.030 \\
Rh & Cu & Ag & [100] & 0.631 & 0.348 & 0.064 & 0.181 & 0.068 \\
Rh & Cu & Ag & [110] & 0.670 & 0.372 & 0.055 & 0.236 & 0.088 \\
Rh & Cu & Ag & [111] & 0.661 & 0.389 & 0.050 & 0.208 & 0.053 \\
Rh & Ni & Ag & [100] & 0.950 & 0.348 & 0.024 & 0.181 & 0.221 \\
Rh & Ni & Au & [100] & 0.950 & 0.370 & 0.024 & 0.142 & 0.080 \\
Rh & Ni & Cu & [100] & 0.950 & 0.631 & 0.024 & 0.064 & 0.021 \\
Ru & Co & Cd & [0001] & 0.818 & 0.078 & 0.087 & 0.256 & 0.226 \\
Ta & Li & Na & [100] & 0.186 & 0.088 & 0.475 & 0.646 & 0.055 \\
Ta & Li & Na & [111] & 0.201 & 0.093 & 0.412 & 0.578 & 0.010 \\
Tc & Co & Cd & [0001] & 0.944 & 0.089 & 0.025 & 0.361 & 0.261 \\
Tc & Ti & Cd & [0001] & 0.880 & 0.089 & 0.008 & 0.361 & 0.042 \\
Tc & Ti & Mg & [0001] & 0.880 & 0.244 & 0.008 & 0.007 & 0.095 \\
Tc & Ti & Tl & [0001] & 0.880 & 0.115 & 0.008 & 0.483 & 0.146 \\
Tc & Ti & Zr & [0001] & 0.880 & 0.721 & 0.008 & 0.075 & 0.014 \\
Tc & Tl & Cd & [0001] & 0.115 & 0.089 & 0.483 & 0.361 & 0.023 \\
Tc & Zr & Mg & [0001] & 0.721 & 0.244 & 0.075 & 0.007 & 0.018 \\
Tc & Zr & Tl & [0001] & 0.721 & 0.115 & 0.075 & 0.483 & 0.035 \\
Ti & Hf & Mg & [0001] & 0.869 & 0.277 & 0.004 & 0.108 & 0.052 \\
Ti & Hf & Tl & [0001] & 0.869 & 0.130 & 0.004 & 0.166 & 0.096 \\
Ti & Hf & Y & [0001] & 0.869 & 0.511 & 0.004 & 0.102 & 0.031 \\
Ti & Tl & Cd & [0001] & 0.130 & 0.102 & 0.166 & 0.048 & 0.027 \\
Ti & Zr & Mg & [0001] & 0.819 & 0.277 & 0.016 & 0.108 & 0.021 \\
Ti & Zr & Tl & [0001] & 0.819 & 0.130 & 0.016 & 0.166 & 0.040 \\
Ti & Zr & Y & [0001] & 0.819 & 0.511 & 0.016 & 0.102 & 0.018 \\
V & Nb & Li & [110] & 0.856 & 0.206 & 0.015 & 0.364 & 0.261 \\
V & Nb & Li & [111] & 0.866 & 0.201 & 0.072 & 0.436 & 0.327 \\
V & Ta & Li & [110] & 0.967 & 0.206 & 0.007 & 0.364 & 0.358 \\
V & Ta & Li & [111] & 1.000 & 0.201 & 0.078 & 0.436 & 0.411 \\
W & Cr & Fe & [100] & 0.919 & 0.632 & 0.082 & 0.027 & 0.038 \\
W & Cr & Fe & [110] & 0.992 & 0.758 & 0.005 & 0.032 & 0.032 \\
W & Cr & Li & [100] & 0.919 & 0.116 & 0.082 & 0.507 & 0.482 \\
W & Cr & Li & [110] & 0.992 & 0.154 & 0.005 & 0.393 & 0.402 \\
W & Cr & Li & [111] & 0.991 & 0.157 & 0.090 & 0.434 & 0.429 \\
W & Cr & Mo & [100] & 0.919 & 0.805 & 0.082 & 0.001 & 0.068 \\
W & Cr & Mo & [110] & 0.992 & 0.866 & 0.005 & 0.009 & 0.002 \\
W & Fe & Li & [100] & 0.632 & 0.116 & 0.027 & 0.507 & 0.249 \\
W & Fe & Li & [110] & 0.758 & 0.154 & 0.032 & 0.393 & 0.152 \\
W & Li & Na & [100] & 0.116 & 0.055 & 0.507 & 0.723 & 0.034 \\
W & Li & Na & [111] & 0.157 & 0.072 & 0.434 & 0.616 & 0.008 \\
W & Mo & Fe & [100] & 0.805 & 0.632 & 0.001 & 0.027 & 0.039 \\
W & Mo & Li & [100] & 0.805 & 0.116 & 0.001 & 0.507 & 0.343 \\
W & Mo & Li & [110] & 0.866 & 0.154 & 0.009 & 0.393 & 0.293 \\
W & Mo & Na & [100] & 0.805 & 0.055 & 0.001 & 0.723 & 0.533 \\
W & Mo & Na & [110] & 0.866 & 0.068 & 0.009 & 0.578 & 0.442 \\
    \hline
\end{longtable}
\end{landscape}

%% file: table_single_phase.tex
\begingroup\SItablefont\setlength{\tabcolsep}{3pt}\renewcommand{\arraystretch}{1.15}
\begin{longtable}{>{\raggedright\arraybackslash}p{\dimexpr 0.1912\linewidth-2\tabcolsep\relax}>{\raggedright\arraybackslash}p{\dimexpr 0.2500\linewidth-2\tabcolsep\relax}>{\raggedright\arraybackslash}p{\dimexpr 0.2794\linewidth-2\tabcolsep\relax}>{\raggedright\arraybackslash}p{\dimexpr 0.2794\linewidth-2\tabcolsep\relax}}
\caption{Experimentally reported single-phase (solid-solution) multicomponent nanoparticles.}
\label{tab:multiphase_1}\\
\toprule
\textbf{Composition} & \textbf{Expt.\ morphology} & \textbf{Expt.\ reference} & \textbf{Expt.\ notes} \\
\midrule
\endfirsthead
\multicolumn{4}{l}{\itshape Table~\ref{tab:multiphase_1} (continued)}\\
\toprule
\textbf{Composition} & \textbf{Expt.\ morphology} & \textbf{Expt.\ reference} & \textbf{Expt.\ notes} \\
\midrule
\endhead
\midrule
\multicolumn{4}{r}{\itshape Continued on next page}\\
\endfoot
\bottomrule
\endlastfoot
Pt-Fe-Ni & single-phase NP & \doi{10.1126/science.1252553} &  \\
\midrule
AuAgCu &  & \doi{10.1126/science.aaf8402} & Au 70\%, $\sim$40nm diameter \\
\midrule
PtAuCu & random (alloy) & \doi{10.1073/pnas.1800884115} & -- \\
\midrule
PtCu & random (alloy) & \doi{10.1073/pnas.1800884115} & -- \\
\midrule
PtNi & random (alloy) & \doi{10.1073/pnas.1800884115} & -- \\
\midrule
Pt-Ni & solid solution (single-phase) & \doi{10.1016/j.calphad.2018.11.013} & randomly mixed core structure \\
\midrule
PdAu &  & \doi{10.1073/pnas.1800884115} &  \\
\midrule
PdNi &  & \doi{10.1073/pnas.1800884115} &  \\
\midrule
Pt$_{47}$\allowbreak{}Ni$_{53}$\allowbreak{} &  & \doi{10.1073/pnas.1800884115} &  \\
\midrule
RhNi &  & \doi{10.1073/pnas.1800884115} &  \\
\midrule
RhCo &  & \doi{10.1073/pnas.1800884115} &  \\
\midrule
PtPd &  & \doi{10.1073/pnas.1800884115} &  \\
\midrule
RhPt &  & \doi{10.1073/pnas.1800884115} &  \\
\midrule
Pt$_{49}$\allowbreak{}Cu$_{51}$\allowbreak{} &  & \doi{10.1073/pnas.1800884115} &  \\
\midrule
PdCu &  & \doi{10.1073/pnas.1800884115} &  \\
\midrule
PdCo &  & \doi{10.1073/pnas.1800884115} &  \\
\midrule
PtCo &  & \doi{10.1073/pnas.1800884115} &  \\
\midrule
PtBi &  & \doi{10.1073/pnas.1800884115} &  \\
\midrule
Pt$_{0.69}$\allowbreak{}Sb$_{0.31}$\allowbreak{} &  & \doi{10.1073/pnas.1800884115} &  \\
\midrule
Pt$_{0.85}$\allowbreak{}Sb$_{0.15}$\allowbreak{} &  & \doi{10.1073/pnas.1800884115} &  \\
\midrule
Pt$_{30}$\allowbreak{}Au$_{36}$\allowbreak{}Cu$_{34}$\allowbreak{} &  & \doi{10.1073/pnas.1800884115} & $\sim$18 nm \\
\midrule
PtCu &  & \doi{10.1073/pnas.1800884115} &  \\
\midrule
PtNi &  & \doi{10.1073/pnas.1800884115} &  \\
\midrule
Sn-Ag-Cu & amorphous outer layer, mixed core with Ag$_{3}$\allowbreak{}Sn precipitates & \doi{10.1039/C5NR00462D} & $\sim$18 nm \\
\midrule
PdCuPtNiCo & HEA & \doi{10.1021/acsnano.2c07787} & $\sim$10 nm \\
\midrule
CuPdIr & solid solution (single-phase) & \doi{10.1016/j.ijhydene.2021.09.192} &  \\
\midrule
CuAgIr & solid solution (single-phase) & \doi{10.1016/j.ijhydene.2021.09.192} &  \\
\midrule
CuAgPd & solid solution (single-phase) & \doi{10.1016/j.ijhydene.2021.09.192} &  \\
\end{longtable}
\endgroup

%% file: table_biphase.tex
\begin{landscape}
\begingroup\SItablefont\setlength{\tabcolsep}{3pt}\renewcommand{\arraystretch}{1.15}
\begin{longtable}{>{\raggedright\arraybackslash}p{\dimexpr 0.1080\linewidth-2\tabcolsep\relax}>{\raggedright\arraybackslash}p{\dimexpr 0.1000\linewidth-2\tabcolsep\relax}>{\raggedright\arraybackslash}p{\dimexpr 0.1070\linewidth-2\tabcolsep\relax}>{\raggedright\arraybackslash}p{\dimexpr 0.1036\linewidth-2\tabcolsep\relax}>{\raggedright\arraybackslash}p{\dimexpr 0.1036\linewidth-2\tabcolsep\relax}>{\raggedright\arraybackslash}p{\dimexpr 0.1347\linewidth-2\tabcolsep\relax}>{\raggedright\arraybackslash}p{\dimexpr 0.1254\linewidth-2\tabcolsep\relax}>{\raggedright\arraybackslash}p{\dimexpr 0.1144\linewidth-2\tabcolsep\relax}>{\raggedright\arraybackslash}p{\dimexpr 0.1033\linewidth-2\tabcolsep\relax}}
\caption{Experimentally reported biphase nanoparticle morphologies compared with Monte Carlo Potts (MCP) predictions.}
\label{tab:multiphase_2}\\
\toprule
\textbf{Composition} & \textbf{Phase A} & \textbf{Phase B} & \textbf{Expt.\ morphology} & \textbf{MCP predicted morphology} & \textbf{Expt.\ reference} & \textbf{Expt.\ notes} & \textbf{Related comp.\ references} & \textbf{Comp.\ notes} \\
\midrule
\endfirsthead
\multicolumn{9}{l}{\itshape Table~\ref{tab:multiphase_2} (continued)}\\
\toprule
\textbf{Composition} & \textbf{Phase A} & \textbf{Phase B} & \textbf{Expt.\ morphology} & \textbf{MCP predicted morphology} & \textbf{Expt.\ reference} & \textbf{Expt.\ notes} & \textbf{Related comp.\ references} & \textbf{Comp.\ notes} \\
\midrule
\endhead
\midrule
\multicolumn{9}{r}{\itshape Continued on next page}\\
\endfoot
\bottomrule
\endlastfoot
 & AgCuAu & AuAg & Janus & -- & \doi{10.1126/sciadv.abj5505} &  &  &  \\
\midrule
Not reported & Au & Ag & solid solution & -- & \doi{10.1021/jacs.1c11118} & Core@Shell (Au@Ag) domains mix after annealing at 500~\textdegree{}C for another 12 h; 30-50 nm & \doi{10.1021/jp510725a}; \doi{10.1021/jp200642d} & no Janus at large scale \\
\midrule
Not reported & Au & Cu & Core@Shell (Au@Cu) & solid solution & \doi{10.1021/jacs.1c11118} & Authors note similar heating causes evolution from Core@Shell to alloy structure &  &  \\
\midrule
 & Au & Cu & Core@Shell (Cu@Au) & -- & \doi{10.1039/C6NR03560D} & surface reconstruction for thin shells leads to new shapes &  &  \\
\addlinespace[1pt]
Not reported &  &  & Core@Shell (Au@Cu) & alloy & \doi{10.1021/jacs.1c11118} & Authors note similar heating causes evolution from Core@Shell to alloy structure &  &  \\
\midrule
 & Au (fcc) & Fe (bcc) & Janus & -- & \doi{10.1039/D1NR00383F} & solid solution at room temperature; phase-separation to Janus when heated to 700~\textdegree{}C &  &  \\
\midrule
Au$_{0.57}$\allowbreak{}Pd$_{0.27}$\allowbreak{}Sn$_{0.16}$\allowbreak{} & Au & Pd$_{1.7}$\allowbreak{}Sn & Janus & -- & \doi{10.1126/science.aav4302} &  &  &  \\
\midrule
Au$_{0.57}$\allowbreak{}Pd$_{0.27}$\allowbreak{}Sn$_{0.16}$\allowbreak{} & Au & PdSn & Janus & -- & \doi{10.1126/science.aav4302} &  &  &  \\
\midrule
 & Au & Pd$_{3}$\allowbreak{}Sn &  & Core@Shell &  &  &  &  \\
\midrule
 & Au & Pt & Core@Shell & Core@Shell (Pt@Au) & \doi{10.1038/nmat2156} &  &  &  \\
\midrule
 & Au & PtNi & Janus & -- & \doi{10.1073/pnas.1800884115} &  &  &  \\
\midrule
 & AuAg & AuCu & Janus & -- & \doi{10.1126/science.aaf8402} & Au 30\%, Ag$_{x}$\allowbreak{} Cu$_{y}$\allowbreak{}; $\sim$40 nm &  &  \\
\midrule
Au$_{0.21}$\allowbreak{}Ag$_{0.32}$\allowbreak{}Pd$_{0.32}$\allowbreak{}Sn$_{0.15}$\allowbreak{} & AuAg & PdSn & Janus & -- & \doi{10.1126/science.aav4302} &  &  &  \\
\midrule
Co$_{0.17}$\allowbreak{}Ni$_{0.25}$\allowbreak{}Cu$_{0.1}$\allowbreak{}Pd$_{0.15}$\allowbreak{}Ag$_{0.15}$\allowbreak{}Au$_{0.15}$\allowbreak{} & AuAgCuPd & (Cu)CoNi(Pd) & Janus & -- & \doi{10.1126/sciadv.abj5505} &  &  &  \\
\midrule
 & AuAgCuPd & CoNi & Janus & -- & \doi{10.1126/sciadv.abj5505} &  &  &  \\
\midrule
 & AuCu & AuAg & Janus & -- & \doi{10.1126/sciadv.abj5505} &  &  &  \\
\midrule
 & AuCu & CoNi & Janus & -- & \doi{10.1126/science.aaf8402} & 10 - 15 nm &  &  \\
\midrule
AuCu-PdSn & AuCu & PdSn & Janus & -- & \doi{10.1126/science.aav4302} &  &  &  \\
\midrule
Co$_{2}$\allowbreak{}Ni$_{4}$\allowbreak{}CuPdAu$_{2}$\allowbreak{} & AuCuPd & CoNi & Janus & -- & \doi{10.1126/sciadv.abj5505} &  &  &  \\
\midrule
 & AuNi & Pt & Core@Shell (AuNi@Pt) & -- & \doi{10.1021/acscatal.5b01656} &  &  &  \\
\midrule
Au$_{0.32}$\allowbreak{}Co$_{0.34}$\allowbreak{}Pd$_{0.34}$\allowbreak{} & AuPd & Co &  & -- & \doi{10.1126/science.aaf8402} &  &  &  \\
\addlinespace[1pt]
 &  &  & Janus & -- & \doi{10.1126/science.aav4302} &  &  &  \\
\midrule
Not reported & Co & Ag & Janus/2p & Core@Shell & \doi{10.1021/jacs.1c11118} & 160~\textdegree{}C for 4h leads to 2p; 500~\textdegree{}C anneal for 12h leads to Janus; annealing at 230~\textdegree{}C for 6 h also leads to Janus & \doi{10.1039/d2fd00114d}: predicts quasi-Janus; \doi{10.1039/C6NR03560D}: surface reconstruction leads to new shapes; \doi{10.1088/0957-4484/27/11/115702} &  \\
\addlinespace[1pt]
 &  &  & alloy &  & \url{https://pubmed.ncbi.nlm.nih.gov/36349781/} & depends on substrate &  &  \\
\midrule
 & Co & Au & Core@Shell & Core@Shell (Co@Au), next closest is Janus & \doi{10.1021/cr100449n} & Redox reaction to avoid diffusion of shell into core & \doi{10.1039/C6NR03560D} & Core@Shell (Co@Au); surface reconstruction for thin shells \\
\addlinespace[1pt]
Not reported &  &  & Janus &  & \doi{10.1021/jacs.1c11118} & Annealing at 230~\textdegree{}C for 6 h & \doi{10.1021/jp302260b} & Changes with decreasing temperature: solid solution at high temp, then Onion-like, then Janus \\
\midrule
Not reported & Co & Ni & Janus & -- & \doi{10.1021/jacs.1c11118} & 500~\textdegree{}C anneal for 12h leads to Janus; 185~\textdegree{}C for 6 h &  &  \\
\addlinespace[1pt]
Not reported &  &  & 2p &  &  & Annealing at 230~\textdegree{}C for 6h &  &  \\
\midrule
Co$_{0.50}$\allowbreak{}Pd$_{0.32}$\allowbreak{}Sn$_{0.18}$\allowbreak{} & Co & Pd$_{1.7}$\allowbreak{}Sn & Janus & -- & \doi{10.1126/science.aav4302} &  &  &  \\
\midrule
-- & Co & Pd$_{3}$\allowbreak{}Sn & -- & Janus & -- &  &  &  \\
\midrule
Co$_{0.66}$\allowbreak{}Pd$_{0.17}$\allowbreak{}Sn$_{0.17}$\allowbreak{} & Co & PdSn & Janus & -- & \doi{10.1126/science.aav4302} &  &  &  \\
\midrule
Co-PdSn & Co & CuPdSn & Janus & -- & \doi{10.1126/science.aav4302} &  &  &  \\
\midrule
 & Co & Pt & alloyed; L1$_0$ structure & -- & \doi{10.1039/B705912B} & controlled by deposition rate &  &  \\
\addlinespace[1pt]
 &  &  & Core@Shell (Co@Pt; CoPt@Pt) &  & \doi{10.1039/C6NR03560D}; \doi{10.1038/nmat3458} & surface reconstruction for thin shells leads to new shapes &  &  \\
\midrule
 & Co & Ni & Janus/2p & -- & \doi{10.1021/jacs.1c11118} & 230~\textdegree{}C for 6h leads to 2p; 500~\textdegree{}C anneal for 12h leads to Janus &  &  \\
\midrule
Co-PdSn & Co & CuPdSn & Janus & -- & \doi{10.1126/science.aav4302} &  &  &  \\
\midrule
 & Co & Pd$_{3}$\allowbreak{}Sn & Janus & Janus & \doi{10.1126/science.aav4302} &  &  &  \\
\midrule
Co$_{0.38}$\allowbreak{}Ni$_{0.23}$\allowbreak{}Pd$_{0.18}$\allowbreak{}Sn$_{0.21}$\allowbreak{} & CoNi & PdSn & Janus & -- & \doi{10.1126/science.aav4302} & ``For example, CoNiPdSn NPs with two different compositions show either a bi-phase structure (Fig. S27B) or a tri-phase structure (Fig. 5B)'' &  &  \\
\addlinespace[1pt]
(tri-phase composition is Co$_{0.13}$\allowbreak{}Ni$_{0.35}$\allowbreak{}Pd$_{0.26}$\allowbreak{}Sn$_{0.26}$\allowbreak{}) &  &  & (Elodie's note: looks more like Striped to me) &  &  &  &  &  \\
\midrule
CoNi-PdSn & CoNi & PdSn & Janus & -- & \doi{10.1126/science.aav4302} &  &  &  \\
\midrule
 & Cu & Ag & Janus, 2p & Core@Shell (Cu@Ag) & \doi{10.1021/jacs.1c11118} & 2p small NP, then heated to form Janus; Janus when heated 160 for 4h, ``lenticular'' shape &  &  \\
\addlinespace[1pt]
 &  &  & Core@Shell (Cu@Ag) &  & \doi{10.1039/C6NR03560D} & surface reconstruction for thin shells leads to new shapes &  &  \\
\addlinespace[1pt]
 &  &  & Core@Shell (Cu@Ag) &  & \doi{10.1103/PhysRevB.73.113402} & surface reconstruction for thin shells leads to new shapes &  &  \\
\addlinespace[1pt]
 &  &  & Janus & Janus or Core@Shell depending on surface \& interfacial energy method used & \doi{10.1126/sciadv.adj6129} &  &  &  \\
\addlinespace[1pt]
 &  &  & Core@Shell (Cu@Ag), 2p, Janus &  & \doi{10.1039/B705912B} & morphology changes depending on size of Cu cores &  &  \\
\addlinespace[1pt]
 &  &  & solid solution &  & \doi{10.1016/j.calphad.2018.11.013} & The randomly mixed core of the Ag-Cu nanoparticles (22\,mol\% Cu) -- choice of precursor is important and decides if Core@Shell or random alloy & \doi{10.1039/C2NR11954D} & DFT to study transition from CS to Janus \\
\midrule
 & Cu & Co & Janus, 2p & Janus & \doi{10.1021/jacs.1c11118} & Heat to form Janus: Janus obtained after annealing at 230~\textdegree{}C for 6 h ($\sim$40 nm); 2p obtained after heating at 185~\textdegree{}C for 6 h ($\sim$20nm Co \& $\sim$15 nm Cu) &  &  \\
\midrule
Cu$_{0.92}$\allowbreak{}Pd$_{0.08}$\allowbreak{}-Cu$_{0.2}$\allowbreak{}(PdSn)$_{0.8}$\allowbreak{}; Cu$_{0.46}$\allowbreak{}Pd$_{0.39}$\allowbreak{}Sn$_{0.15}$\allowbreak{} & CuPdSn & Cu(Pd) & Janus & -- & \doi{10.1126/science.aav4302} &  &  &  \\
\midrule
CuNi-PdSn & CuNi & PdSn & Janus & -- & \doi{10.1126/science.aav4302} &  &  &  \\
\midrule
Not reported & Cu & Ag & Janus & Core@Shell (Cu@Ag) & \doi{10.1021/jacs.1c11118} & Annealing at 160~\textdegree{}C for 4 h ($\sim$25 nm) &  &  \\
\addlinespace[1pt]
 &  &  &  &  & \doi{10.1016/j.calphad.2018.11.013} & 2p small NP, then heated to form Janus; ``lenticular'' shape &  &  \\
\addlinespace[1pt]
 &  &  &  &  & \doi{10.1039/C6NR03560D} & surface reconstruction for thin shells leads to new shapes &  &  \\
\addlinespace[1pt]
 &  &  & Core@Shell (Cu@Ag) &  & \doi{10.1103/PhysRevB.73.113402} &  &  &  \\
\midrule
 & Cu & Au & Core@Shell (Cu@Au) & -- & \doi{10.1039/C6NR03560D} & surface reconstruction for thin shells leads to new shapes &  &  \\
\midrule
 & Cu & Co & Janus & Core@Shell (Co@Cu) in 2D; Janus in 3D & \doi{10.1021/jacs.1c11118} & Annealing at 230~\textdegree{}C for 6 h ($\sim$40 nm) &  &  \\
\addlinespace[1pt]
 &  &  & 2p &  &  & Annealing at 185~\textdegree{}C for 6 h, $\sim$20nm Co \& $\sim$15 nm Cu &  &  \\
\midrule
 & Cu & Ni & Core@Shell (Cu@Ni) & Core@Shell & \doi{10.1016/j.calphad.2018.11.013} &  &  &  \\
\addlinespace[1pt]
 &  &  &  &  & \doi{10.1021/jacs.1c11118} & 230~\textdegree{}C for 6h; depends on order of deposition & \doi{10.1021/jp302260b} & Changes with temperature: solid solution @ high temp, then Onion-like, then Janus @ low temp \\
\midrule
 & Cu & Ni & Core@Shell (Cu@Ni) & Core@Shell & \doi{10.1016/j.calphad.2018.11.013} &  &  &  \\
\addlinespace[1pt]
 &  &  & Core@Shell (Cu@Ni) &  & \doi{10.1021/jacs.1c11118} & 230~\textdegree{}C for 6h; depends on order of deposition &  &  \\
\midrule
Cu$_{0.88}$\allowbreak{}Ir$_{0.12}$\allowbreak{}, Cu$_{0.79}$\allowbreak{}Ir$_{0.21}$\allowbreak{} & Cu & Ir & Core@Shell &  & \doi{10.1021/jacs.2c02538} & on &  &  \\
\midrule
Cu$_{0.88}$\allowbreak{}Rh$_{0.12}$\allowbreak{}, Cu$_{0.79}$\allowbreak{}Rh$_{0.21}$\allowbreak{} & Cu & Rh & Core@Shell &  & \doi{10.1021/jacs.2c02538} & reaction temperature of 260~\textdegree{}C; ``If the temperature is increased too slowly, the reaction will yield a higher dispersion of particle diameters and lead to the formation of very large bimetallic crystals.'' &  &  \\
\midrule
Cu$_{0.88}$\allowbreak{}Ru$_{0.12}$\allowbreak{}, Cu$_{0.94}$\allowbreak{}Ru$_{0.06}$\allowbreak{}, Cu$_{0.79}$\allowbreak{}Ru$_{0.21}$\allowbreak{} & Cu & Ru & Core@Shell &  & \doi{10.1021/jacs.2c02538} & reaction temperature of 270~\textdegree{}C; ``If the temperature is increased too slowly, the reaction will yield a higher dispersion of particle diameters and lead to the formation of very large bimetallic crystals.'' &  &  \\
\midrule
 & Ga & In & Janus &  & \url{https://www.sciencedirect.com/science/article/pii/S2590238519300050} &  &  &  \\
\midrule
Au$_{3}$\allowbreak{}Ir$_{7}$\allowbreak{}, Au$_{5}$\allowbreak{}Ir$_{5}$\allowbreak{}, Au$_{7}$\allowbreak{}Ir$_{3}$\allowbreak{} & Ir & Au & Core@Shell (Ir@Au) &  & \url{https://www.sciencedirect.com/science/article/pii/S016943321831938X} & Laser irradiation for 30 minutes; caption description indicates Lens &  &  \\
\midrule
 & Ir & Pt & Core@Shell & Core@Shell & \doi{10.1038/nmat2156} &  &  &  \\
\midrule
 & Mn & Ni & cubic Core@Shell & -- & \doi{10.1021/cr040090g} &  &  &  \\
\midrule
 & Ni & Ag & Janus & Core@Shell (Ni@Ag) & \doi{10.1126/sciadv.adj6129} &  & \doi{10.1016/j.calphad.2018.11.013} says core@Shell &  \\
\addlinespace[1pt]
 &  &  & 2p &  & \doi{10.1021/jacs.1c11118} & Annealing at 230~\textdegree{}C for 6 h. &  &  \\
\addlinespace[1pt]
 &  &  & Core@Shell (Ni@Ag) &  & \doi{10.1039/C6NR03560D} & surface reconstruction for thin shells leads to new shapes &  &  \\
\addlinespace[1pt]
Ag$_{3}$\allowbreak{}Ni, AgNi, AgNi$_{3}$\allowbreak{} &  &  &  &  & \doi{10.1103/PhysRevB.65.165422} & small (<7 nm); Ag/Ni interface is relatively loose; NPs in a dielectric matrix & \doi{10.1039/C6NR03560D} & surface reconstruction for thin shells leads to new shapes; Ni@Ag \\
\addlinespace[1pt]
AgNi &  &  & Core@Shell (Ag@Ni) &  & \doi{10.1016/j.calphad.2018.11.013} & 23-31 nm depending on composition &  &  \\
\midrule
Not reported & Ni & Au & Janus & -- & \doi{10.1021/jacs.1c11118} & Annealed at 500~\textdegree{}C for 6h, then 700~\textdegree{}C for 2 hours to see Janus; $\sim$50 nm diameter & \doi{10.1016/j.calphad.2017.05.002} & phase diagram study of changing particle size \\
\addlinespace[1pt]
Not reported &  &  & Janus $\rightarrow$ Core@Shell &  & \doi{10.1039/C4NR03559C} & some Au mixed into Ni; forms Core@Shell after oxidation (Onion w/ outer oxidized layer) &  &  \\
\addlinespace[1pt]
 &  &  & Janus &  & \doi{10.1126/sciadv.adj6129} & annealing at 230~\textdegree{}C for 6 h &  &  \\
\addlinespace[1pt]
Not reported &  &  &  &  & \doi{10.1016/j.calphad.2018.11.013} & interesting for both experimental and theoretical studies if the size of alloy particles decreases to nanoscale and the discontinuous precipitation is not possible. &  &  \\
\midrule
Au$_{0.24}$\allowbreak{}Ag$_{0.24}$\allowbreak{}Ni$_{0.52}$\allowbreak{} & Ni & AuAg & Janus & -- & \doi{10.1126/science.aaf8402} & likely AuAg alloy (111) and Ni (111); $\sim$30 nm, $\sim$20 nm, $\sim$35 nm'' &  &  \\
\addlinespace[1pt]
Not reported &  &  &  &  & \doi{10.1021/jacs.1c11118} & Annealed at 500~\textdegree{}C for 4 h; Ag deposited on Ni/Au Janus seeds &  &  \\
\midrule
 & Ni & AuPt & Core@Shell (Ni@AuPt) & -- & \doi{10.1021/acscatal.5b01656} &  &  &  \\
\midrule
 & Ni & Cu & Core@Shell (Ni@Cu) & Core@Shell & \doi{10.1021/jacs.1c11118} &  &  &  \\
\addlinespace[1pt]
 &  &  &  &  & \doi{10.1016/j.calphad.2018.11.013} &  &  &  \\
\midrule
 & Ni & Rh & Core@Shell (Ni@Rh) & &  &  & \doi{10.1039/C6NR03560D} & surface reconstruction for thin shells leads to new shapes \\
\midrule
 & Ni & Pt & Core@Shell (Ni@Pt) & -- & \doi{10.1021/acscatal.5b01656} &  & \doi{10.1063/1.1828033} &  \\
\addlinespace[1pt]
 &  &  & Core@Shell (Ni@Pt) &  & \doi{10.1039/C6NR03560D} & surface reconstruction for thin shells leads to new shapes &  &  \\
\addlinespace[1pt]
 &  &  & solid solution (single-phase); &  & \doi{10.1016/j.calphad.2018.11.013} &  &  &  \\
\addlinespace[1pt]
 &  &  & Janus &  &  &  &  &  \\
\addlinespace[1pt]
 &  &  & alloy (single-phase) &  & \doi{10.1073/pnas.1800884115} &  &  &  \\
\midrule
 & Ni & Pd & Core@Shell (Ni@Pd) & -- & \doi{10.1039/B513587G} & clusters: <=6nm &  &  \\
\addlinespace[1pt]
 &  &  &  &  & \doi{10.1039/C6NR03560D} & surface reconstruction for thin shells leads to new shapes &  &  \\
\midrule
 & Ni & PdSn & Janus & -- & \doi{10.1126/science.aav4302} &  &  &  \\
\midrule
 & Ni & Pt & Core@Shell (Ni@Pt) & -- & \doi{10.1021/acscatal.5b01656} &  &  &  \\
\addlinespace[1pt]
 &  &  & solid solution (single-phase) &  & \doi{10.1016/j.calphad.2018.11.013} & randomly mixed core structure &  &  \\
\addlinespace[1pt]
 &  &  & random (alloy) &  & \doi{10.1073/pnas.1800884115} &  &  &  \\
\midrule
Ni$_{0.6}$\allowbreak{}Sn$_{0.4}$\allowbreak{}-Ni$_{0.08}$\allowbreak{}(PdSn)$_{0.92}$\allowbreak{} & NiSn & (Ni)PdSn & Janus & -- & \doi{10.1126/science.aav4302} &  &  &  \\
\midrule
Cu$_{0.33}$\allowbreak{}Ni$_{0.32}$\allowbreak{}Pd$_{0.16}$\allowbreak{}Sn$_{0.19}$\allowbreak{} & NiCu & PdSn(NiCu) & Janus & -- & \doi{10.1126/science.aav4302} &  &  &  \\
\addlinespace[1pt]
Co$_{0.38}$\allowbreak{}Ni$_{0.23}$\allowbreak{}Pd$_{0.18}$\allowbreak{}Sn$_{0.21}$\allowbreak{} &  &  & Janus & -- & \doi{10.1126/science.aav4302} &  &  &  \\
\midrule
Ni$_{0.32}$\allowbreak{}Pd$_{0.37}$\allowbreak{}Sn$_{0.31}$\allowbreak{} & NiPd & PdSn & Janus & -- & \doi{10.1126/science.aav4302} &  &  &  \\
\midrule
 & Rh & Au & Random (alloy) & Core@Shell (Rh@Au) & \doi{10.1038/s41565-024-01626-0} &  &  &  \\
\addlinespace[1pt]
 &  &  & Core@Shell (Rh@Au \& Au@Rh) &  & \doi{10.1021/la902674u} & can depend on the support &  &  \\
\addlinespace[1pt]
 &  &  & Core@Shell (Au@Rh) &  & \doi{10.1016/J.VACUUM.2011.07.014} &  &  &  \\
\addlinespace[1pt]
 &  &  & Janus &  & \doi{10.1021/ACS.NANOLETT.1C02225} &  &  &  \\
\addlinespace[1pt]
 &  &  & Core@Shell (Rh@Au) & Core@Shell & \doi{10.1039/C6NR03560D} & surface reconstruction for thin shells leads to new shapes &  &  \\
\midrule
Co$_{0.75}$\allowbreak{}Rh$_{0.25}$\allowbreak{} & Rh (fcc) & Co & Core@Shell & Core@Shell (for FCC [100]/FCC [100]) & \doi{10.1103/PhysRevB.69.235416} & <7 nm; surface segregation; Co is fcc at smaller sizes, then hcp at 4-6 nm; ``The Co-rich Co$_{3}$\allowbreak{}Rh$_{1}$\allowbreak{} [has] a large majority of hcp grains, while the Rh-rich sample (Co$_{0.39}$\allowbreak{}Rh$_{0.61}$\allowbreak{}) is mainly fcc, although its composition is closer to Co$_{2}$\allowbreak{}Rh$_{3}$\allowbreak{} than to Co$_{1}$\allowbreak{}Rh$_{3}$\allowbreak{}...The two samples with compositions close to Co$_{1}$\allowbreak{}Rh$_{1}$\allowbreak{} ( Co$_{0.53}$\allowbreak{}Rh$_{0.47}$\allowbreak{} and Co$_{0.45}$\allowbreak{}Rh$_{0.55}$\allowbreak{}) are mainly fcc. However, many regions of these samples exhibit structural distorsions, probably due to the presence of numerous faults, which prevents us from unambiguous determination of the structure'' &  &  \\
\midrule
 & Ru & Pt & Core@Shell & -- & \doi{10.1038/nmat2156} &  & \url{https://pubmed.ncbi.nlm.nih.gov/19731934/} & ``highly distorted hcp Ru cores that are primarily in the metallic state but show little order beyond 8 A. In contrast, the 1-2 monolayer thick Pt shells are relatively crystalline but are slightly distorted (compressed) relative to bulk fcc P'' \\
\midrule
Au$_{1}$\allowbreak{}Pd$_{4}$\allowbreak{} & Pd & Au & Core@Shell & -- & \doi{10.1039/A805753B} & $\sim$2nm &  &  \\
\midrule
 & Pd & Pt & Core@Shell & -- & \doi{10.1038/nmat2156} &  &  &  \\
\addlinespace[1pt]
 &  &  & mixes &  & \doi{10.1021/cr040090g} &  &  &  \\
\midrule
Pd$_{3}$\allowbreak{}Ni, Pd$_{4}$\allowbreak{}Ni, Pd$_{6}$\allowbreak{}Ni, Pd$_{8}$\allowbreak{}Ni & Pd & Ni & Core@Shell (Pd@Ni) & -- & \doi{10.1126/sciadv.aap8817} &  &  &  \\
\midrule
Ag$_{0.55}$\allowbreak{}Pd$_{0.26}$\allowbreak{}Sn$_{0.19}$\allowbreak{} & PdSn & Ag & Janus & -- & \doi{10.1126/science.aav4302} &  &  &  \\
\midrule
 & Sn & Ag &  & -- & \doi{10.1016/j.calphad.2018.11.013} &  &  &  \\
\midrule
 & Pd & Pt & Core@Shell & -- & \doi{10.1039/A805753B} & PVP-stabilized; particle size depends on composition; $<5$ \si{\nano\meter} &  &  \\
\midrule
 & Pd$_{3}$\allowbreak{}Sn & Ag & -- & Core@Shell & -- &  &  &  \\
\midrule
 & PdSn & Cu & Janus & -- & \doi{10.1126/science.aav4302} &  &  &  \\
\midrule
 & Pd$_{3}$\allowbreak{}Sn & Cu &  & Janus &  &  &  &  \\
\midrule
Cu$_{0.46}$\allowbreak{}Pd$_{0.39}$\allowbreak{}Sn$_{0.15}$\allowbreak{} & Pd$_{2.6}$\allowbreak{}Sn & Cu & Janus & -- & \doi{10.1126/science.aav4302} &  &  &  \\
\midrule
Cu$_{0.92}$\allowbreak{}Pd$_{0.08}$\allowbreak{}-Cu$_{0.2}$\allowbreak{}(PdSn)$_{0.8}$\allowbreak{} & CuPdSn & Cu(Pd) & Janus & -- & \doi{10.1126/science.aav4302} &  &  &  \\
\midrule
 & PdAuCoPd & PdAuCoPd & Janus & -- &  &  &  &  \\
\midrule
 & PdCu & Ir & Core@Shell & -- & \doi{10.1002/ange.202016199} &  &  &  \\
\midrule
Pt$_{46}$\allowbreak{}Au$_{54}$\allowbreak{} & Pt & Au & Janus & Core@Shell (Pt@Au) & \doi{10.1073/pnas.1800884115} & 18nm; annealing: under Ar gas flow, ramp to 150~\textdegree{}C within 1 h, hold at 150~\textdegree{}C for 48 h, cool down to 25~\textdegree{}C within 1 h, switch the atmosphere to H$_{2}$\allowbreak{}, ramp to 500~\textdegree{}C within 2 h, hold at 500~\textdegree{}C for 12 h, cool down to 25~\textdegree{}C within 2 h &  &  \\
\addlinespace[1pt]
 &  &  & Core@Shell (Au@Pt) &  & \doi{10.1039/A805753B} & electrochem &  &  \\
\addlinespace[1pt]
 &  &  &  &  & \doi{10.1002/ppsc.201700401} & redox; sequential synthesis &  &  \\
\midrule
 & Pt & Rh & Core@Shell & -- & \doi{10.1021/cr040090g} & some surface segregation of core to shell &  &  \\
\midrule
 & Pt & Ru & Core@Shell & -- &  & some surface segregation of core to shell &  &  \\
\midrule
 & Pt & Pd & Core@Shell (Pt@Pd) & -- & \doi{10.1021/cr040090g} & bulk mixes; strong surface segregation of Pd &  &  \\
\midrule
 & PtAuCu & -- & random (alloy) & -- & \doi{10.1073/pnas.1800884115} &  &  &  \\
\midrule
 & PtCu & -- & random (alloy) & -- & \doi{10.1073/pnas.1800884115} &  &  &  \\
\midrule
Pt$_{32}$\allowbreak{}Au$_{36}$\allowbreak{}Ni$_{32}$\allowbreak{} & PtNi & Au & Janus & -- & \doi{10.1073/pnas.1800884115} &  &  &  \\
\end{longtable}
\endgroup
\end{landscape}

%% file: table_triphase.tex
\begin{landscape}
\begingroup\SItablefont\setlength{\tabcolsep}{3pt}\renewcommand{\arraystretch}{1.15}
\begin{longtable}{>{\raggedright\arraybackslash}p{\dimexpr 0.1702\linewidth-2\tabcolsep\relax}>{\raggedright\arraybackslash}p{\dimexpr 0.0800\linewidth-2\tabcolsep\relax}>{\raggedright\arraybackslash}p{\dimexpr 0.0800\linewidth-2\tabcolsep\relax}>{\raggedright\arraybackslash}p{\dimexpr 0.0800\linewidth-2\tabcolsep\relax}>{\raggedright\arraybackslash}p{\dimexpr 0.1170\linewidth-2\tabcolsep\relax}>{\raggedright\arraybackslash}p{\dimexpr 0.0957\linewidth-2\tabcolsep\relax}>{\raggedright\arraybackslash}p{\dimexpr 0.1489\linewidth-2\tabcolsep\relax}>{\raggedright\arraybackslash}p{\dimexpr 0.2281\linewidth-2\tabcolsep\relax}}
\caption{Experimentally reported triphase nanoparticle morphologies compared with Monte Carlo Potts (MCP) predictions. An asterisk (*) marks morphologies discerned by eye. Phases are listed as reported in each experimental study and are not ordered by surface energy.}
\label{tab:multiphase_3}\\
\toprule
\textbf{Composition} & \textbf{Phase A} & \textbf{Phase B} & \textbf{Phase C} & \textbf{Expt.\ morphology$^{*}$} & \textbf{MCP predicted morphology} & \textbf{Expt.\ reference} & \textbf{Expt.\ notes} \\
\midrule
\endfirsthead
\multicolumn{8}{l}{\itshape Table~\ref{tab:multiphase_3} (continued)}\\
\toprule
\textbf{Composition} & \textbf{Phase A} & \textbf{Phase B} & \textbf{Phase C} & \textbf{Expt.\ morphology$^{*}$} & \textbf{MCP predicted morphology} & \textbf{Expt.\ reference} & \textbf{Expt.\ notes} \\
\midrule
\endhead
\midrule
\multicolumn{8}{r}{\itshape Continued on next page}\\
\endfoot
\bottomrule
\endlastfoot
Ag$_{0.23}$\allowbreak{}Cu$_{0.47}$\allowbreak{}Co$_{0.30}$\allowbreak{} & Ag & Cu & Co & Striped & Onion & \doi{10.1126/science.aav4302} &  \\
\addlinespace[1pt]
Co$_{0.07}$\allowbreak{}Cu$_{0.27}$\allowbreak{}Ag$_{0.66}$\allowbreak{} &  &  &  & Striped (in transition to Onion) &  & \doi{10.1126/sciadv.abj5505} & $\sim$30 nm \\
\addlinespace[1pt]
Not listed &  &  &  & Pie &  & \doi{10.1126/sciadv.adj6129} & $\sim$75 nm; Co deposited on Ag/Cu Janus seed \\
\addlinespace[1pt]
Not listed &  &  &  & Striped &  & \doi{10.1021/jacs.1c11118} & Annealed at 600~\textdegree{}C, $\sim$55 nm in diameter \\
\addlinespace[1pt]
Not listed &  &  &  & Janus + 1p &  & \doi{10.1021/jacs.1c11118} & Annealed at 185~\textdegree{}C \\
\addlinespace[1pt]
Ag$_{0.43}$\allowbreak{}Cu$_{0.19}$\allowbreak{}Co$_{0.38}$\allowbreak{} &  &  &  & Pie &  & \doi{10.1126/science.aaf8402} &  \\
\midrule
Co$_{0.13}$\allowbreak{}Ni$_{0.35}$\allowbreak{}Pd$_{0.26}$\allowbreak{}Sn$_{0.26}$\allowbreak{} & NiSn & PdSn & CoNi & Pie & -- & \doi{10.1126/science.aav4302} &  \\
\midrule
Au$_{0.37}$\allowbreak{}Co$_{0.18}$\allowbreak{}Ni$_{0.36}$\allowbreak{}Sn$_{0.09}$\allowbreak{} & NiSn & CoNi & Au & Pie & -- & \doi{10.1126/science.aav4302} &  \\
\midrule
Au$_{0.29}$\allowbreak{}Co$_{0.17}$\allowbreak{}Ni$_{0.19}$\allowbreak{}Pd$_{0.20}$\allowbreak{}Sn$_{0.15}$\allowbreak{} & Au & PdSn & CoNi & Pie & -- & \doi{10.1126/science.aav4302} &  \\
\midrule
Au$_{0.25}$\allowbreak{}Ni$_{0.24}$\allowbreak{}Pd$_{0.20}$\allowbreak{}Sn$_{0.31}$\allowbreak{}; Au-NiSn-PdSn & Au & PdSn & NiSn & Pie & -- & \doi{10.1126/science.aav4302} &  \\
\midrule
Ag-NiSn-PdSn & Ag & PdSn & NiSn & Pie & -- & \doi{10.1126/science.aav4302} &  \\
\midrule
Au$_{0.30}$\allowbreak{}Ag$_{0.19}$\allowbreak{}Cu$_{0.29}$\allowbreak{}Co$_{0.22}$\allowbreak{} & AuAg & AuCu & Co & Striped & -- & \doi{10.1126/science.aav4302} & $\sim$45 nm diameter \\
\midrule
Au$_{0.12}$\allowbreak{}Ag$_{0.30}$\allowbreak{}Cu$_{0.28}$\allowbreak{}Pd$_{0.20}$\allowbreak{}Sn$_{0.10}$\allowbreak{} & AuAg & AuCu & PdSn(Cu) & Pie & -- & \doi{10.1126/science.aav4302} & $\sim$45 nm diameter \\
\midrule
Au$_{0.06}$\allowbreak{}Ag$_{0.10}$\allowbreak{}Co$_{0.24}$\allowbreak{}Pd$_{0.39}$\allowbreak{}Sn$_{0.21}$\allowbreak{} & AuAg & Co & PdSn & Pie & -- & \doi{10.1126/science.aav4302} & $\sim$45 nm diameter \\
\midrule
Au$_{0.13}$\allowbreak{}Cu$_{0.27}$\allowbreak{}Co$_{0.30}$\allowbreak{}Pd$_{0.15}$\allowbreak{}Sn$_{0.15}$\allowbreak{} & AuCu & PdSn(Cu) & Co & Pie & -- & \doi{10.1126/science.aav4302} & $\sim$45 nm diameter \\
\midrule
Co$_{0.34}$\allowbreak{}Cu$_{0.29}$\allowbreak{}Pd$_{0.21}$\allowbreak{}Sn$_{0.16}$\allowbreak{} & Co & Cu & PdSn(Cu) & Striped & -- & \doi{10.1126/science.aav4302} &  \\
\midrule
Ag$_{0.30}$\allowbreak{}Cu$_{0.30}$\allowbreak{}Pd$_{0.27}$\allowbreak{}Sn$_{0.13}$\allowbreak{} & Ag & Cu & PdSn(Cu) & Pie & -- & \doi{10.1126/science.aav4302} &  \\
\midrule
Ag$_{0.33}$\allowbreak{}Co$_{0.23}$\allowbreak{}Pd$_{0.27}$\allowbreak{}Sn$_{0.17}$\allowbreak{} & Ag & Co & Pd$_{2}$\allowbreak{}Sn & Pie & -- & \doi{10.1126/science.aav4302} &  \\
\midrule
Ag$_{0.43}$\allowbreak{}Co$_{0.23}$\allowbreak{}Pd$_{0.23}$\allowbreak{}Sn$_{0.11}$\allowbreak{} & Ag & Co & PdSn & Pie & -- & \doi{10.1126/science.aav4302} &  \\
\midrule
Ag-Cu-PdSn & Ag & Cu & PdSn & Pie (3/4); Striped (1/4) & -- & \doi{10.1126/science.aav4302} &  \\
\midrule
Au$_{0.30}$\allowbreak{}Co$_{0.37}$\allowbreak{}Pd$_{0.19}$\allowbreak{}Sn$_{0.14}$\allowbreak{} & Au & Co & Pd$_{1}$\allowbreak{}Sn$_{1}$\allowbreak{} & Pie & -- & \doi{10.1126/science.aav4302} & When annealed for longer times (5 days) at a higher temperature (750~\textdegree{}C), the two-interface NPs evolved into three-interface structures; three-interface NPs remained relatively unchanged; evaporative loss of Sn; 15 to 60 nm \\
\addlinespace[1pt]
Au$_{0.33}$\allowbreak{}Co$_{0.34}$\allowbreak{}Pd$_{0.18}$\allowbreak{}Sn$_{0.15}$\allowbreak{} &  &  &  &  &  &  &  \\
\addlinespace[1pt]
Au$_{0.24}$\allowbreak{}Co$_{0.58}$\allowbreak{}(PdSn)$_{0.18}$\allowbreak{} &  &  &  &  &  &  &  \\
\addlinespace[1pt]
Au$_{0.31}$\allowbreak{}Co$_{0.34}$\allowbreak{}Pd$_{0.21}$\allowbreak{}Sn$_{0.14}$\allowbreak{} & Au & Co & Pd$_{1.5}$\allowbreak{}Sn & Pie & -- &  &  \\
\addlinespace[1pt]
Au$_{0.30}$\allowbreak{}Co$_{0.26}$\allowbreak{}Pd$_{0.30}$\allowbreak{}Sn$_{0.14}$\allowbreak{} & Au & Co & Pd$_{2}$\allowbreak{}Sn & Pie & -- &  &  \\
\addlinespace[1pt]
Au$_{0.33}$\allowbreak{}Co$_{0.24}$\allowbreak{}Pd$_{0.26}$\allowbreak{}Sn$_{0.17}$\allowbreak{} & Au & Co & Pd$_{3}$\allowbreak{}Sn & Pie & Shell w/ 2 cores &  &  \\
\addlinespace[1pt]
Au$_{0.30}$\allowbreak{}Co$_{0.29}$\allowbreak{}Pd$_{0.30}$\allowbreak{}Sn$_{0.11}$\allowbreak{} &  &  &  &  &  &  &  \\
\addlinespace[1pt]
Au$_{0.38}$\allowbreak{}Co$_{0.33}$\allowbreak{}Pd$_{0.23}$\allowbreak{}Sn$_{0.06}$\allowbreak{} & Au & Co & Pd$_{4}$\allowbreak{}Sn & Pie & -- &  &  \\
\midrule
Ag-AuCu-Ni & Ag & AuCu & Ni & Venn* &  & \doi{10.1021/jacs.1c11118} &  \\
\midrule
Au-Ag-Ni & Ni & Au & Ag & Venn (Au@Ag-Ni) &  & \doi{10.1021/jacs.1c11118} & 160~\textdegree{}C for 4h; Ag deposition only on Au surface, Cu deposition has not been triggered yet. Au-Ni Janus is formed with Ag deposited on top \\
\midrule
Au-Cu-Ni & Ni & Au & Cu & Venn (Au@Cu-Ni) &  & \doi{10.1021/jacs.1c11118} &  \\
\midrule
AuCu-CuNi-PdSn & CuNi & AuCu & PdSn & Pie &  & \doi{10.1126/science.aav4302} &  \\
\midrule
AuAg-NiSn-PdSn & NiSn & AuAg & PdSn & Pie &  & \doi{10.1126/science.aav4302} &  \\
\midrule
Au-CoNi-PdSn & CoNi & Au & PdSn & Pie &  & \doi{10.1126/science.aav4302} &  \\
\midrule
Ag-CuNi-PdSn & CuNi & Ag & PdSn & Pie &  & \doi{10.1126/science.aav4302} &  \\
\midrule
Ag-CoNi-PdSn & CoNi & Ag & PdSn & Pie &  & \doi{10.1126/science.aav4302} &  \\
\midrule
CoNi-CuNi-PdSn & CoNi & CuNi & PdSn & Striped &  & \doi{10.1126/science.aav4302} &  \\
\midrule
AuAgCu-CoNi-PdSn & CoNi & AuAgCu & PdSn & Pie &  & \doi{10.1126/science.aav4302} &  \\
\midrule
Co$_{0.13}$\allowbreak{}Ni$_{0.35}$\allowbreak{}Pd$_{0.26}$\allowbreak{}Sn$_{0.26}$\allowbreak{}; biphase is Co$_{0.38}$\allowbreak{}Ni$_{0.23}$\allowbreak{}Pd$_{0.18}$\allowbreak{}Sn$_{0.21}$\allowbreak{} & Co & Ni & PdSn & Pie; Striped* &  & \doi{10.1126/science.aav4302} & CoNiPdSn NPs with two different compositions show either a bi-phase structure or a tri-phase structure \\
\midrule
Au$_{0.25}$\allowbreak{}Ni$_{0.24}$\allowbreak{}Pd$_{0.20}$\allowbreak{}Sn$_{0.31}$\allowbreak{} & Au & PdSn & NiSn & Pie &  & \doi{10.1126/science.aav4302} &  \\
\midrule
Au$_{0.10}$\allowbreak{}Ag$_{x}$\allowbreak{}Cu$_{y}$\allowbreak{} & Ag & AuAg & AuCu & Striped &  & \doi{10.1126/science.aaf8402} &  \\
\midrule
Au$_{0.50}$\allowbreak{}Ag$_{x}$\allowbreak{}Cu$_{y}$\allowbreak{} & AuAg & AuAgCo & AuCu & Striped &  & \doi{10.1126/science.aaf8402} &  \\
\midrule
Au$_{0.19}$\allowbreak{}Ag$_{0.24}$\allowbreak{}Cu$_{0.28}$\allowbreak{}Co$_{0.14}$\allowbreak{}Ni$_{0.15}$\allowbreak{} & AuAg & AuCu & CoNi & Striped &  & \doi{10.1126/science.aaf8402} &  \\
\midrule
Au$_{0.25}$\allowbreak{}Ni$_{0.24}$\allowbreak{}Pd$_{0.20}$\allowbreak{}Sn$_{0.31}$\allowbreak{} & Au & Ni & PdSn(Ni) & Pie &  & \doi{10.1126/science.aav4302} &  \\
\midrule
Au$_{0.24}$\allowbreak{}Cu$_{0.38}$\allowbreak{}Pd$_{0.25}$\allowbreak{}Sn$_{0.13}$\allowbreak{} & Au(Cu) & Cu & Pd$_{2}$\allowbreak{}Sn & Striped &  & \doi{10.1126/science.aav4302} &  \\
\midrule
Ag$_{0.22}$\allowbreak{}Ni$_{0.24}$\allowbreak{}Pd$_{0.21}$\allowbreak{}Sn$_{0.33}$\allowbreak{} & Ag & Pd(Ni) & PdSn(Ni) & Pie &  & \doi{10.1126/science.aav4302} &  \\
\midrule
Au-Co-Ni & Au & Co & Ni & Striped &  & \doi{10.1021/jacs.1c11118} &  \\
\midrule
Ag-Cu-Ni & Ag & Cu & Ni & Janus w/ Core@Shell* & Onion & \doi{10.1021/jacs.1c11118} &  \\
\end{longtable}
\endgroup
\end{landscape}

%% file: table_tetraphase.tex
\begin{landscape}
\begingroup\SItablefont\setlength{\tabcolsep}{3pt}\renewcommand{\arraystretch}{1.15}
\begin{longtable}{>{\raggedright\arraybackslash}p{\dimexpr 0.1750\linewidth-2\tabcolsep\relax}>{\raggedright\arraybackslash}p{\dimexpr 0.0800\linewidth-2\tabcolsep\relax}>{\raggedright\arraybackslash}p{\dimexpr 0.0800\linewidth-2\tabcolsep\relax}>{\raggedright\arraybackslash}p{\dimexpr 0.0800\linewidth-2\tabcolsep\relax}>{\raggedright\arraybackslash}p{\dimexpr 0.0800\linewidth-2\tabcolsep\relax}>{\raggedright\arraybackslash}p{\dimexpr 0.2070\linewidth-2\tabcolsep\relax}>{\raggedright\arraybackslash}p{\dimexpr 0.1489\linewidth-2\tabcolsep\relax}>{\raggedright\arraybackslash}p{\dimexpr 0.1489\linewidth-2\tabcolsep\relax}}
\caption{Experimentally reported tetraphase nanoparticle morphologies.}
\label{tab:multiphase_4}\\
\toprule
\textbf{Composition} & \textbf{Phase A} & \textbf{Phase B} & \textbf{Phase C} & \textbf{Phase D} & \textbf{Expt.\ morphology} & \textbf{Expt.\ reference} & \textbf{Expt.\ notes} \\
\midrule
\endfirsthead
\multicolumn{8}{l}{\itshape Table~\ref{tab:multiphase_4} (continued)}\\
\toprule
\textbf{Composition} & \textbf{Phase A} & \textbf{Phase B} & \textbf{Phase C} & \textbf{Phase D} & \textbf{Expt.\ morphology} & \textbf{Expt.\ reference} & \textbf{Expt.\ notes} \\
\midrule
\endhead
\midrule
\multicolumn{8}{r}{\itshape Continued on next page}\\
\endfoot
\bottomrule
\endlastfoot
 & Ag & Cu & Au & Ni & Janus w/ inner Janus: Ag shell around Au-Ni w/ Cu shell & \doi{10.1021/jacs.1c11118} & annealed at 160~\textdegree{}C for 4h followed by 185~\textdegree{}C for 6h \\
\midrule
 & Ag & Cu & Au & Ni & Striped w/ split: Ag + Cu half-shells around Au-Ni & \doi{10.1021/jacs.1c11118} & annealed at 185~\textdegree{}C for 6h \\
\midrule
CoPdAgCu & Co & PdSn(Cu) & AuCu & AuAg & striped w/ split; 4-phase with 5 interfaces & \doi{10.1126/science.aav4302} &  \\
\midrule
Au-CoNi-NiSn-PdSn & Au & Ni & PdSn & Ni &  &  &  \\
\midrule
Au$_{0.18}$\allowbreak{}Co$_{0.14}$\allowbreak{}Ni$_{0.33}$\allowbreak{}Pd$_{0.18}$\allowbreak{}Sn$_{0.17}$\allowbreak{}; Au$_{0.17}$\allowbreak{}Co$_{0.11}$\allowbreak{}Ni$_{0.33}$\allowbreak{}Pd$_{0.24}$\allowbreak{}Sn$_{0.15}$\allowbreak{}; Au$_{0.18}$\allowbreak{}Co$_{0.09}$\allowbreak{}Ni$_{0.33}$\allowbreak{}Pd$_{0.25}$\allowbreak{}Sn$_{0.15}$\allowbreak{}; Au$_{0.20}$\allowbreak{}Co$_{0.12}$\allowbreak{}Ni$_{0.32}$\allowbreak{}Pd$_{0.20}$\allowbreak{}Sn$_{0.16}$\allowbreak{}; Au$_{0.20}$\allowbreak{}Co$_{0.11}$\allowbreak{}Ni$_{0.30}$\allowbreak{}Pd$_{0.21}$\allowbreak{}Sn$_{0.18}$\allowbreak{} & Au & CoNi & NiSn & PdSn & pie w/ core (most common); then Striped followed by the 4-interface, 3-surface morphology & \doi{10.1126/science.aav4302} &  \\
\midrule
AuAg-AuCu-NiSn-PdSn and AuAg-AuCu-Co-PdSn &  &  &  &  & heterotetramers with five interfaces & \doi{10.1126/science.aav4302} &  \\
\midrule
Ag-Cu-CoNi-PdSn &  &  &  &  & heterotetramer with four interfaces & \doi{10.1126/science.aav4302} &  \\
\midrule
Ag-Cu-Ni-PdSn & Cu & Ag & PdSn & Ni & heterotetramer with four interfaces & \doi{10.1126/science.aav4302} &  \\
\midrule
Au$_{0.17}$\allowbreak{}Ag$_{0.12}$\allowbreak{}Cu$_{0.22}$\allowbreak{}Ni$_{0.14}$\allowbreak{}Pd$_{0.19}$\allowbreak{}Sn$_{0.16}$\allowbreak{} &  &  &  &  & heterotetramer with four interfaces & \doi{10.1126/science.aav4302} &  \\
\midrule
Au$_{0.17}$\allowbreak{}Ag$_{0.15}$\allowbreak{}Cu$_{0.17}$\allowbreak{}Co$_{0.23}$\allowbreak{}Pd$_{0.18}$\allowbreak{}Sn$_{0.10}$\allowbreak{} & AuAg & AuCu & Co & PdSn & striped w/ split; 4-phase with 5 interfaces & \doi{10.1126/science.aav4302} &  \\
\addlinespace[1pt]
Au$_{0.18}$\allowbreak{}Ag$_{0.16}$\allowbreak{}Cu$_{0.20}$\allowbreak{}Co$_{0.23}$\allowbreak{}Pd$_{0.13}$\allowbreak{}Sn$_{0.10}$\allowbreak{} &  &  &  &  &  &  &  \\
\addlinespace[1pt]
Au$_{0.11}$\allowbreak{}Ag$_{0.15}$\allowbreak{}Cu$_{0.17}$\allowbreak{}Co$_{0.28}$\allowbreak{}Pd$_{0.19}$\allowbreak{}Sn$_{0.10}$\allowbreak{} &  &  &  &  &  &  &  \\
\midrule
Ag$_{0.32}$\allowbreak{}Cu$_{0.20}$\allowbreak{}Co$_{0.21}$\allowbreak{}Pd$_{0.15}$\allowbreak{}Sn$_{0.12}$\allowbreak{} & Co & Cu & Ag & PdSn &  & \doi{10.1126/science.aav4302} &  \\
\end{longtable}
\endgroup
\end{landscape}

%% file: triphase_2d_stability_map.tex
\newcommand{\figcompcap}{Stability map for triphase NP morphologies calculated from MCP on a 2D lattice.}

\setcounter{panelctr}{0}

\begin{center}
\captionof{figure}{\figcompcap}\label{fig:3p2d_strict_stability_map}

\panel{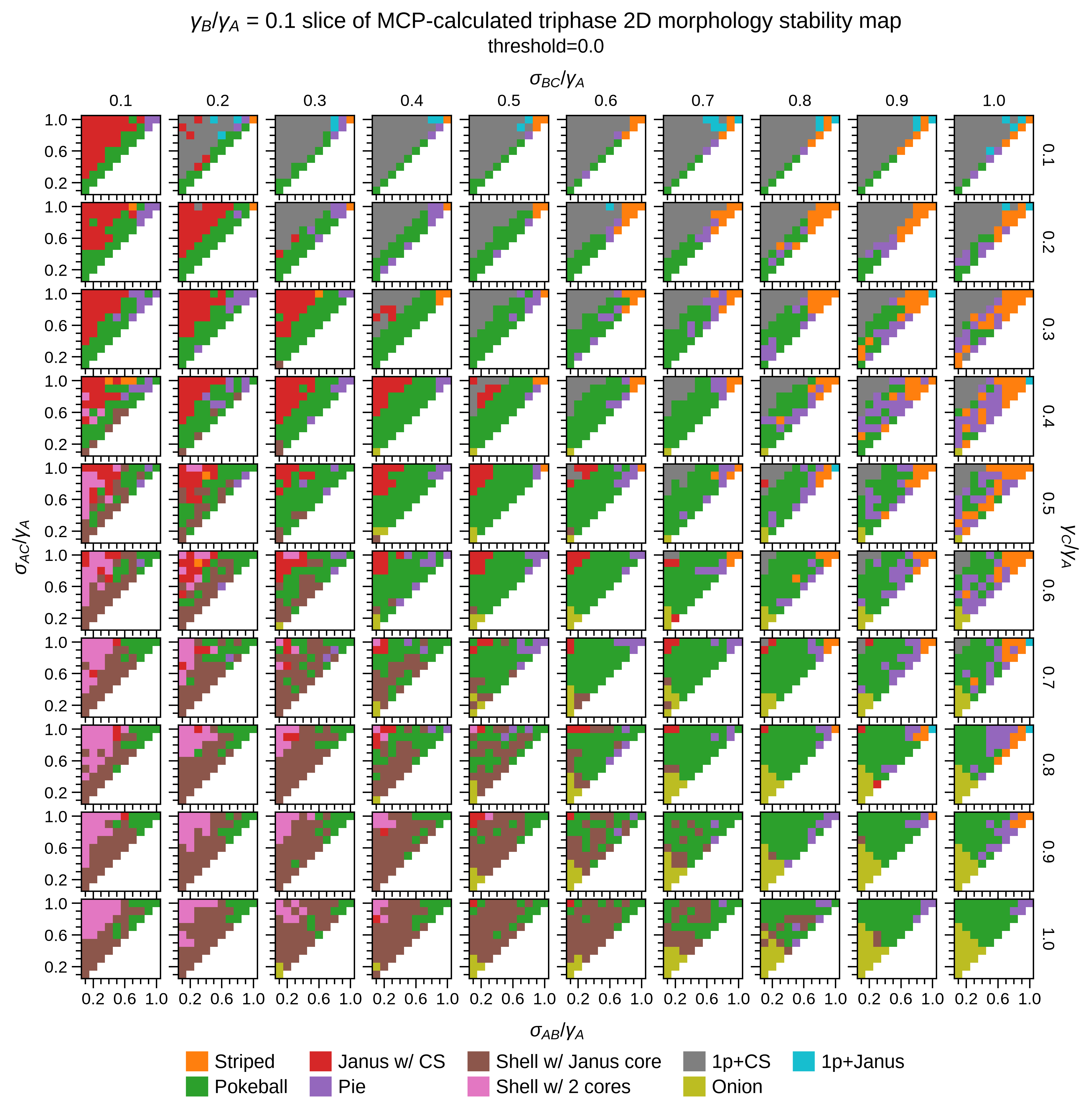}{fig:3p2d_strict_01}\\[8pt]
\panel{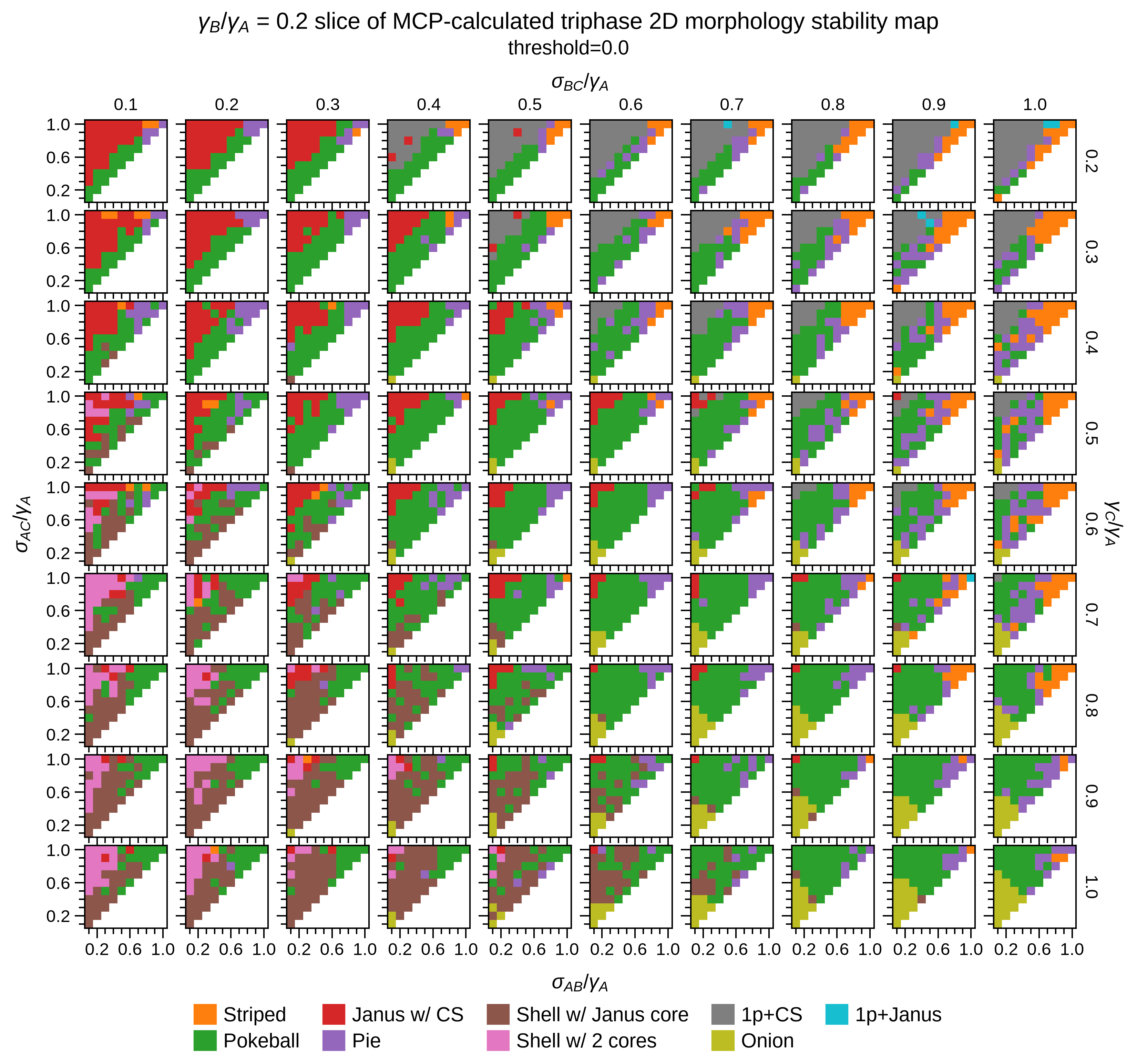}{fig:3p2d_strict_02}\\[8pt]
\panel{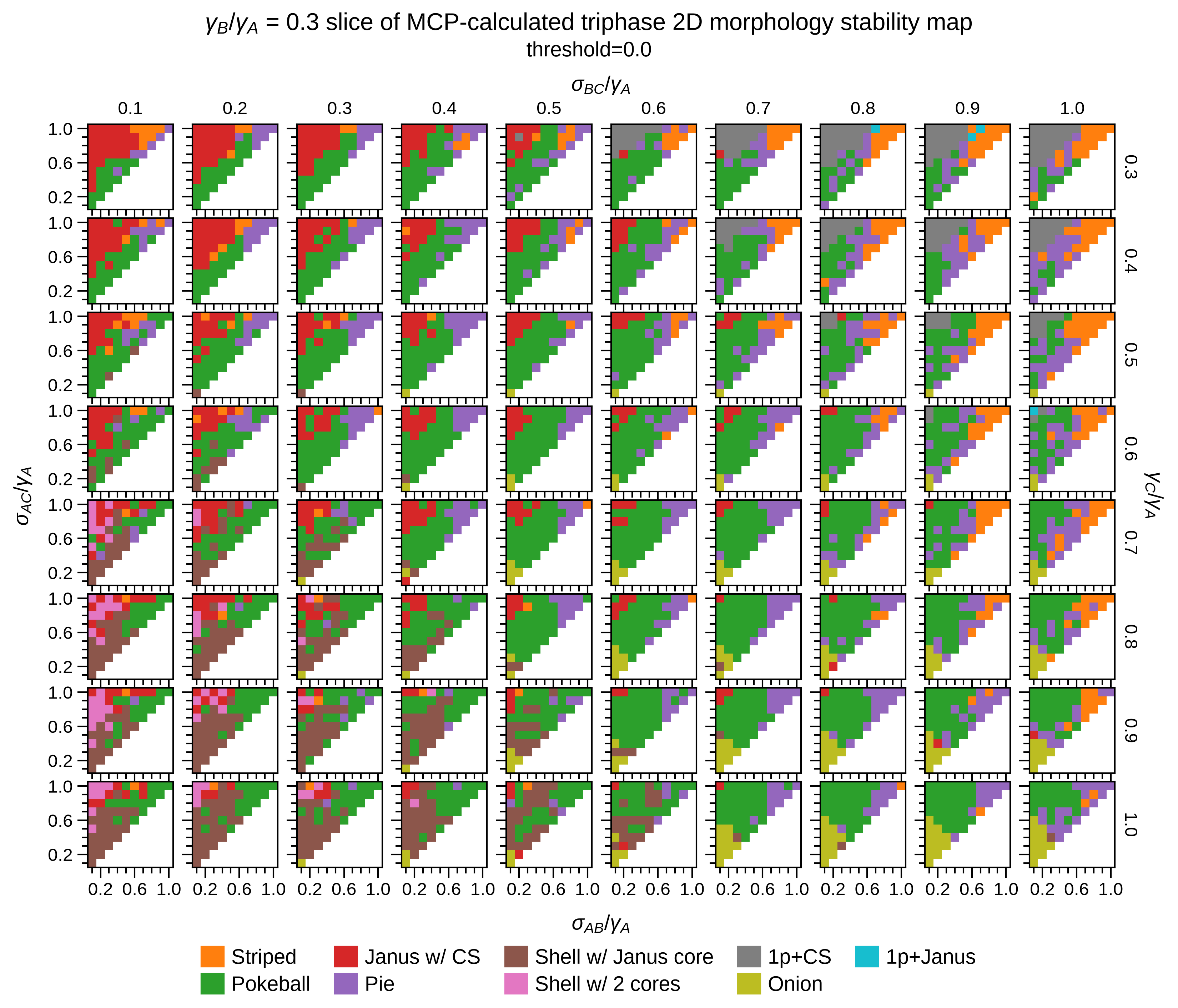}{fig:3p2d_strict_03}\\[8pt]
\panel{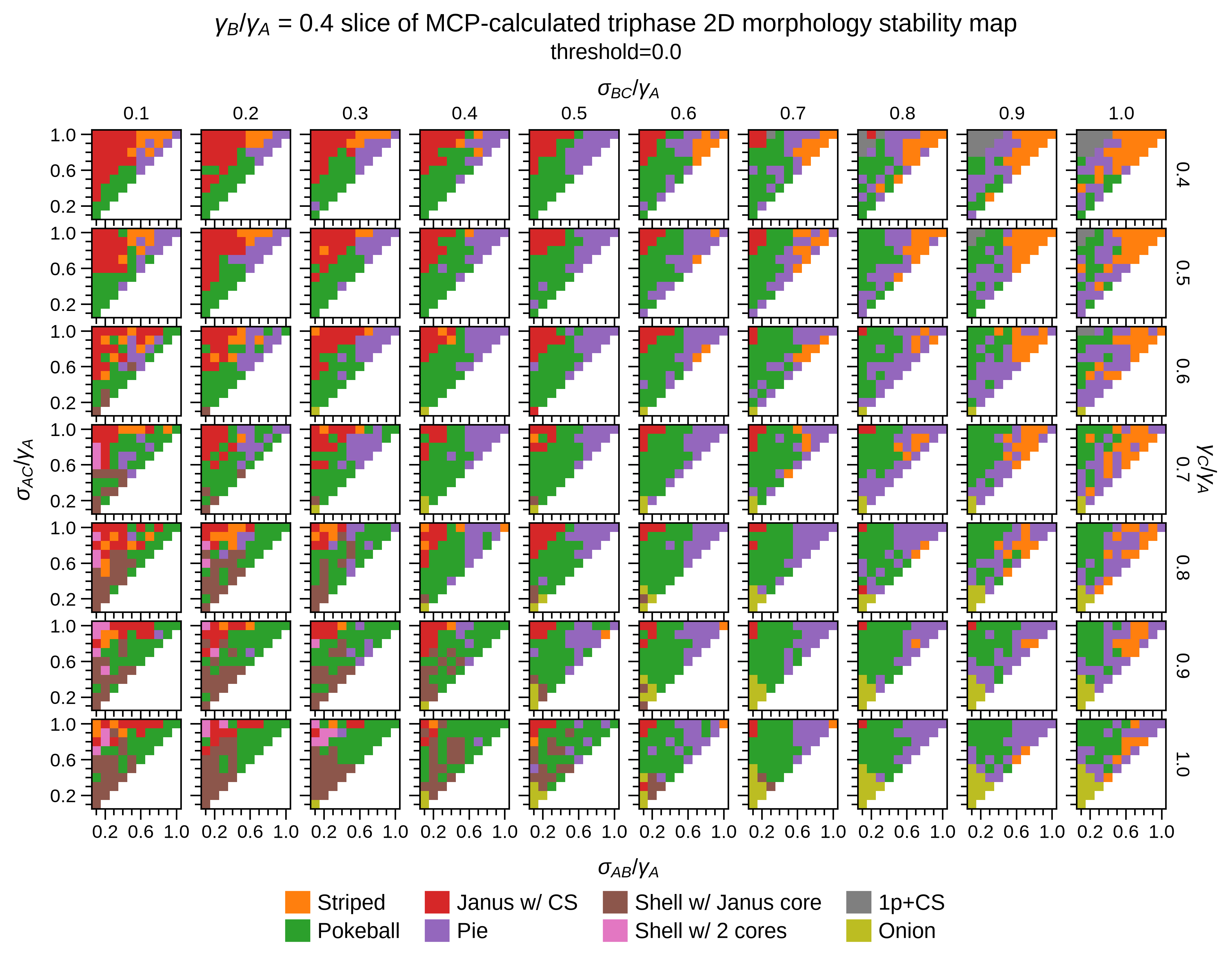}{fig:3p2d_strict_04}\\[8pt]
\panel{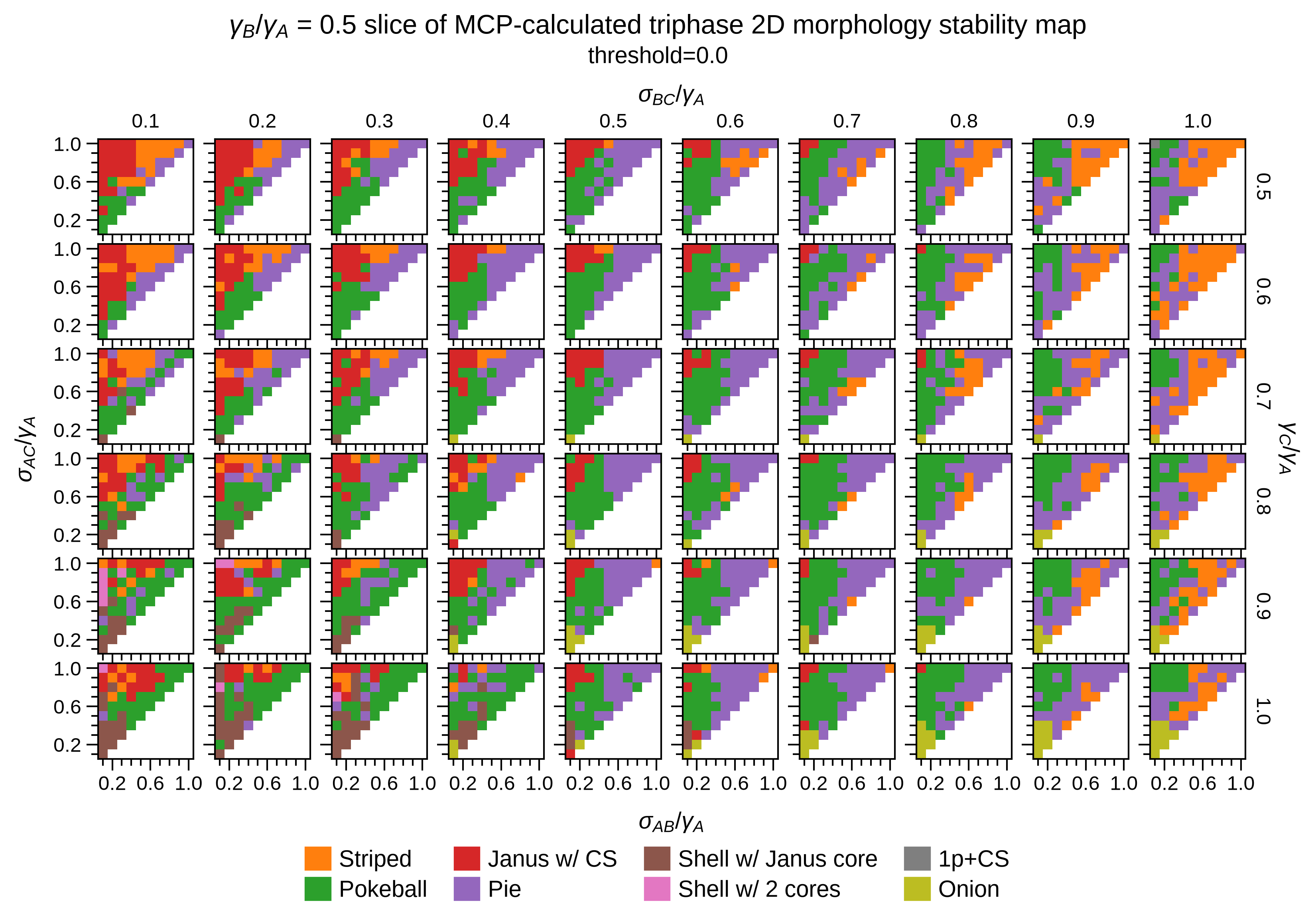}{fig:3p2d_strict_05}\\[8pt]
\panel{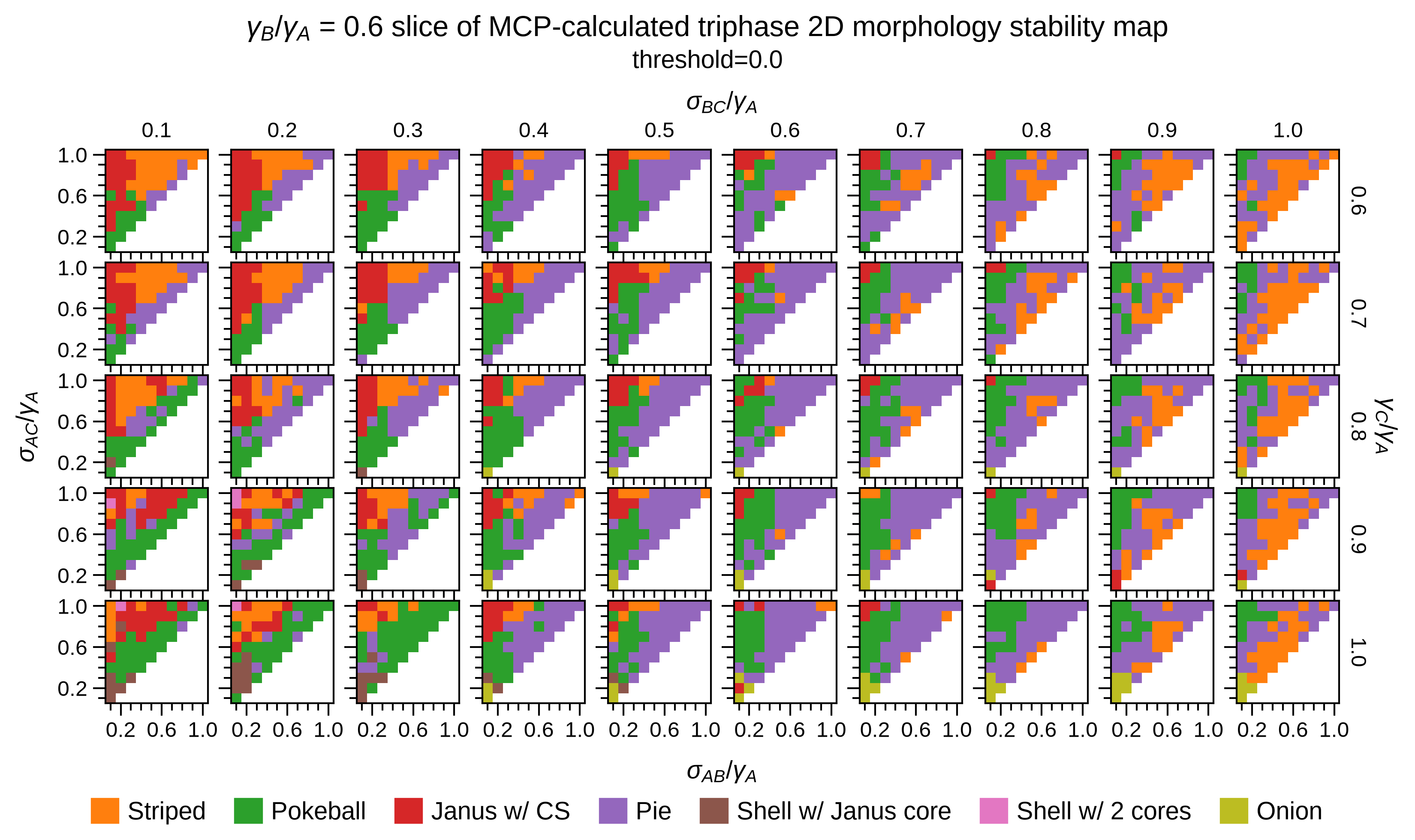}{fig:3p2d_strict_06}\\[8pt]
\panel{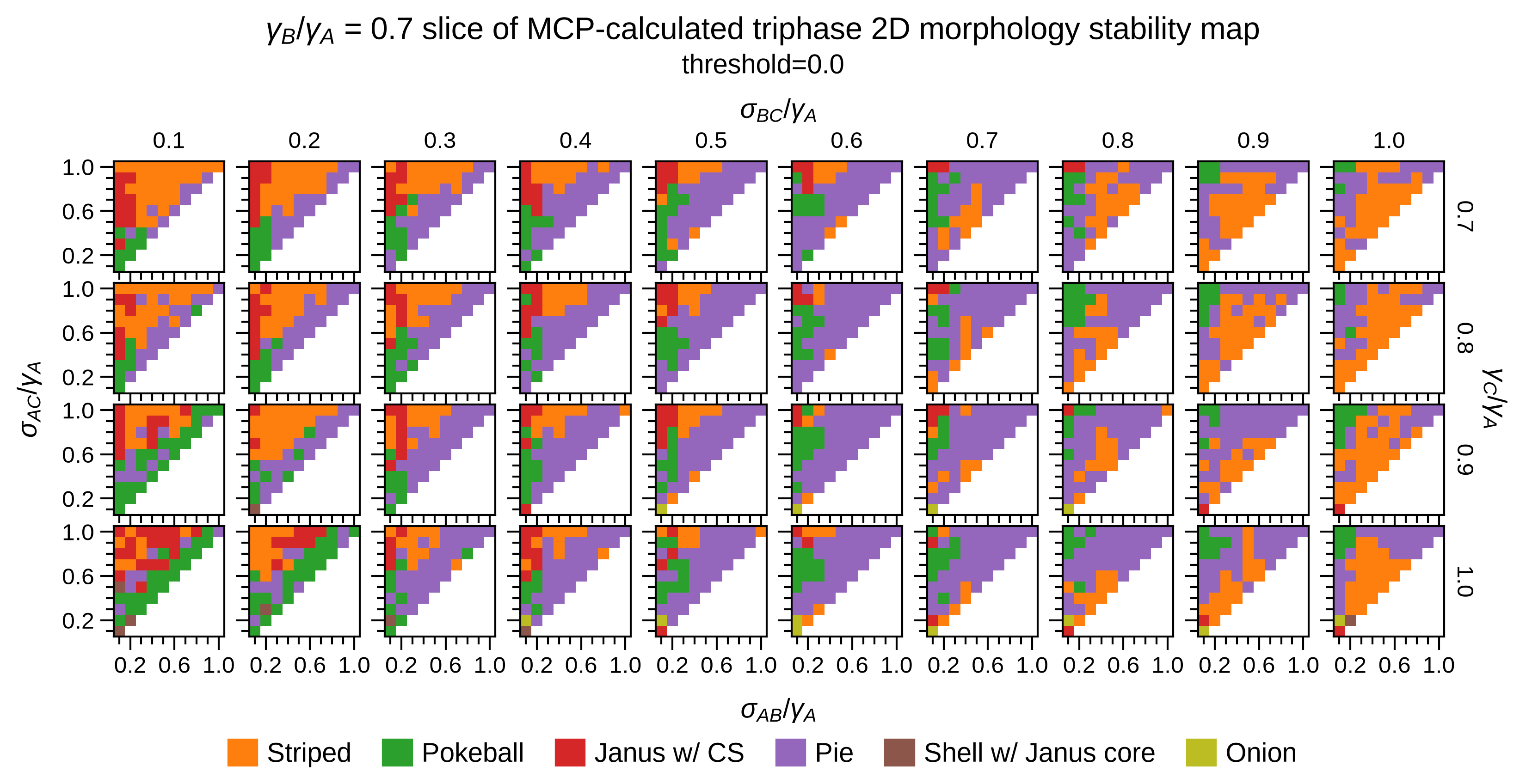}{fig:3p2d_strict_07}\\[8pt]
\panel{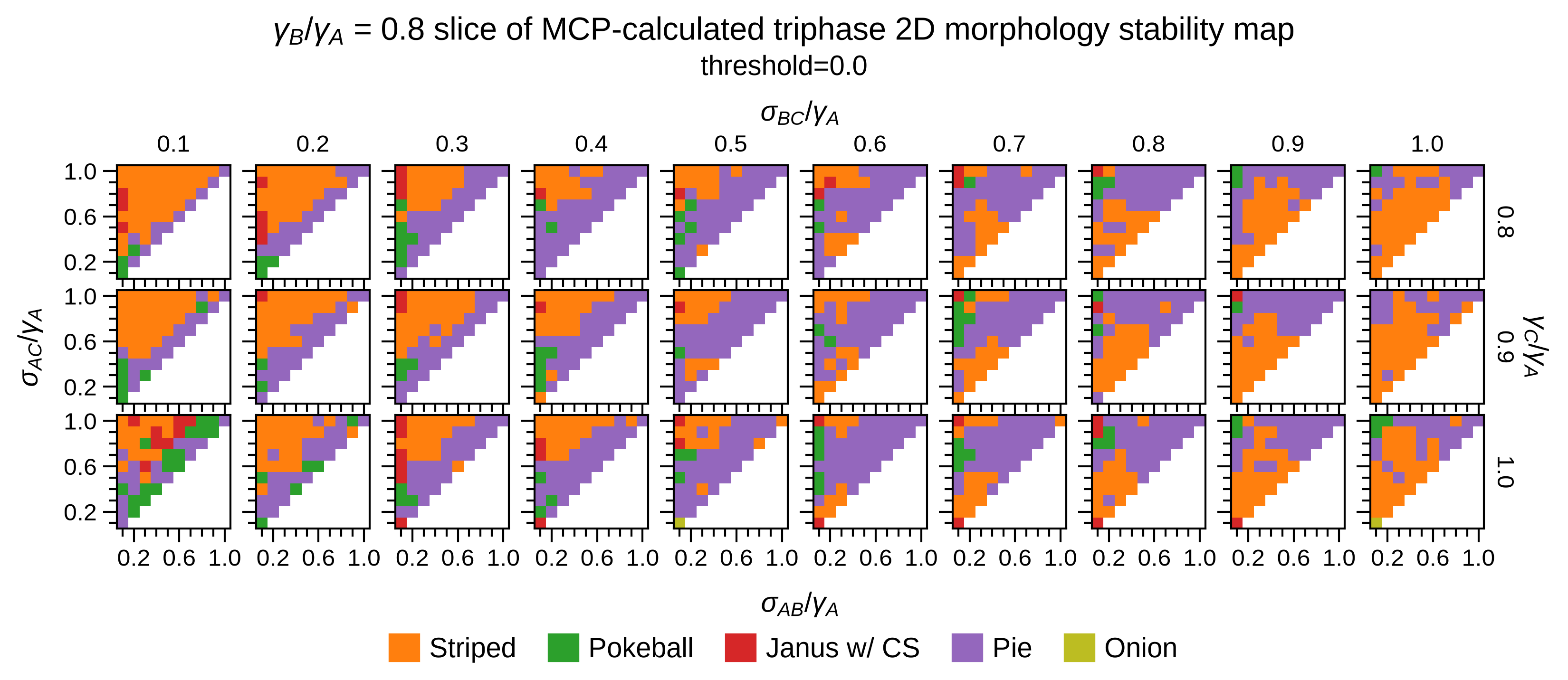}{fig:3p2d_strict_08}\\[8pt]
\panel{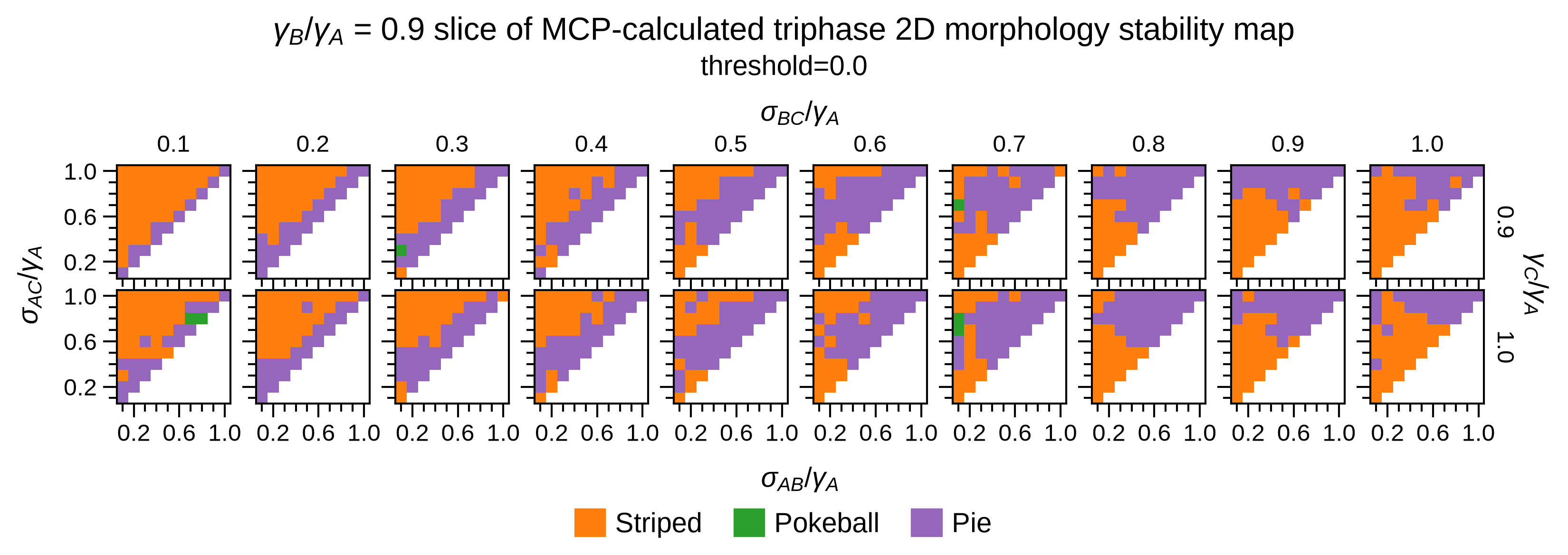}{fig:3p2d_strict_09}\\[8pt]
\panel{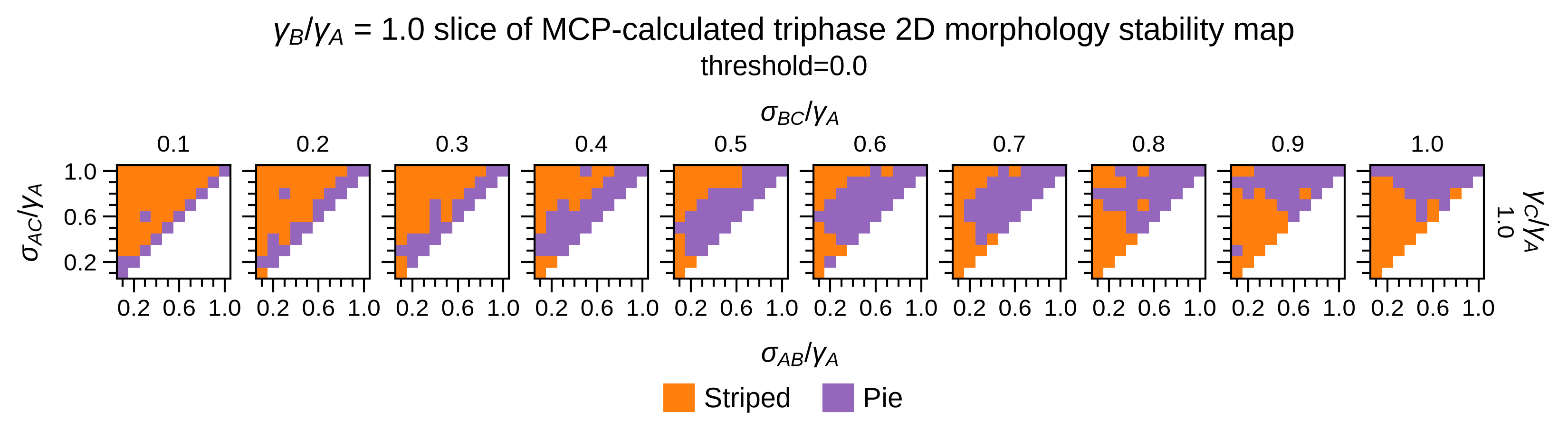}{fig:3p2d_strict_10}
\end{center}

%% file: triphase_3d_stability_map.tex
\newcommand{\figcompcapthreed}{Stability map for triphase NP morphologies calculated from MCP on a 3D lattice using a domain-contact threshold of 0.0: two domains are considered to share an interface if they have at least one NN or NNN interaction between them. A looser threshold is shown in Figure~\ref{fig:3p3d_loose_stability_map}.}

\setcounter{panelctr}{0}

\begin{center}
\captionof{figure}{\figcompcapthreed}\label{fig:3p3d_stability_map}

\panel{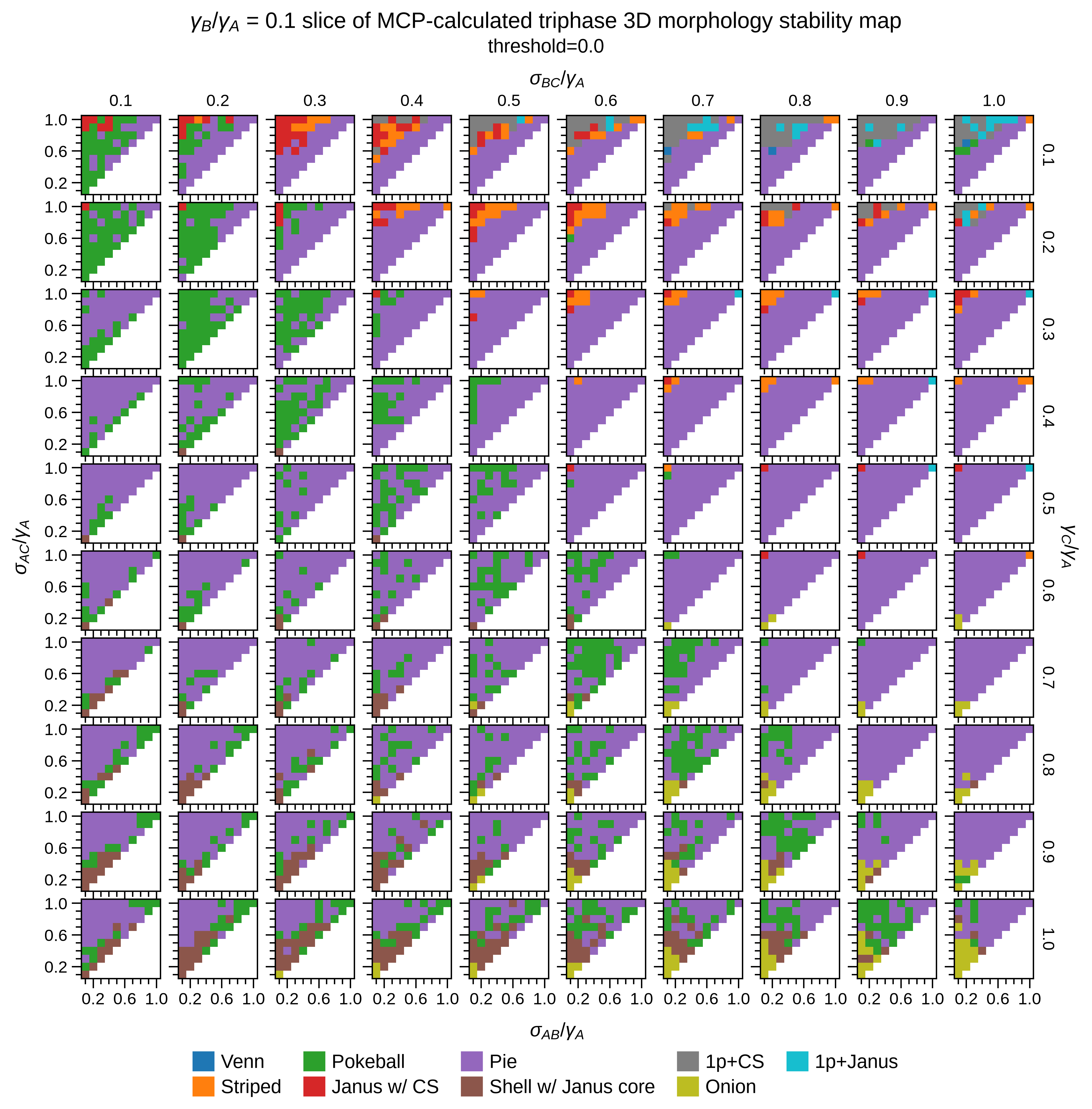}{fig:3p3d_01}\\[8pt]
\panel{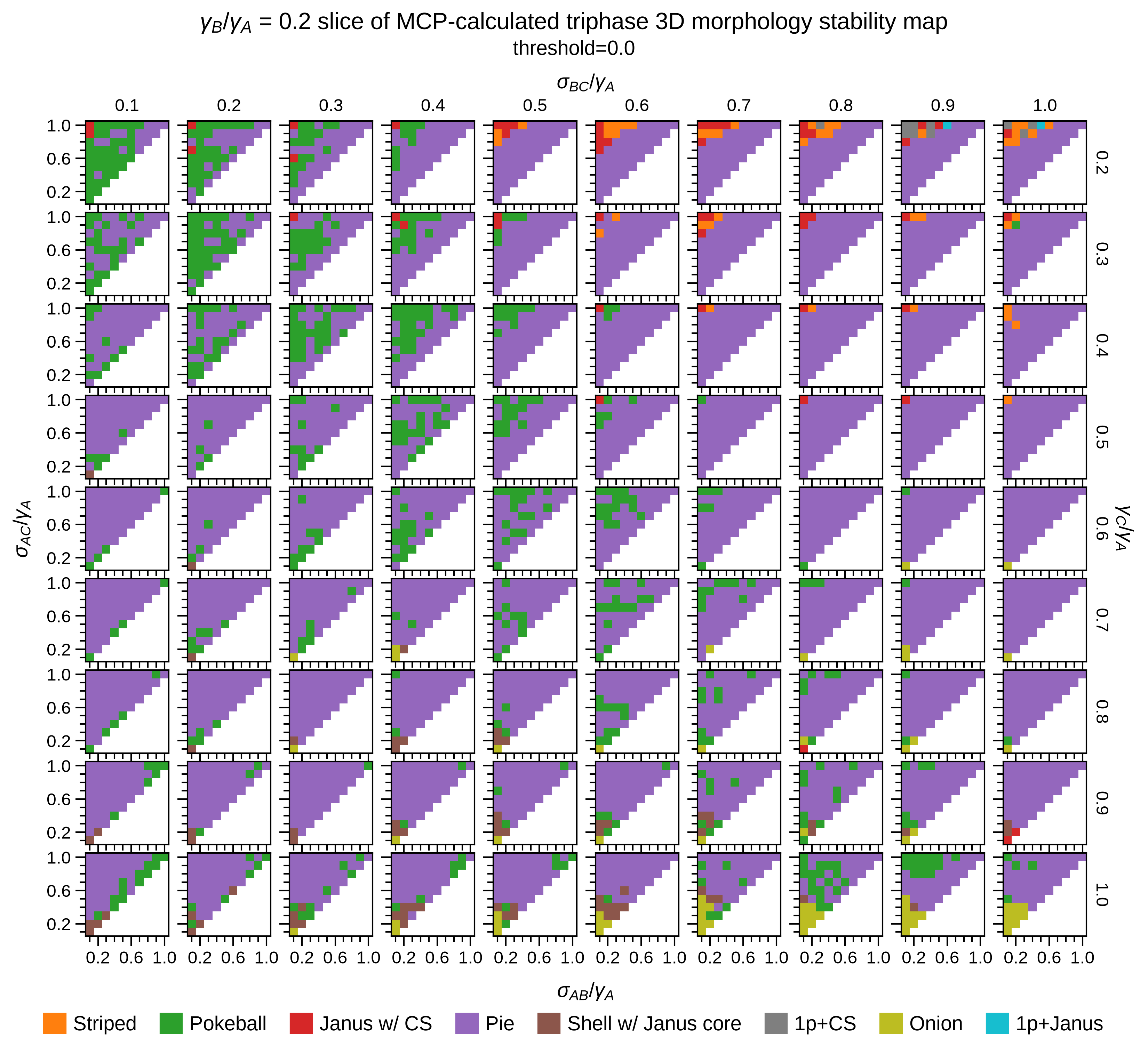}{fig:3p3d_02}\\[8pt]
\panel{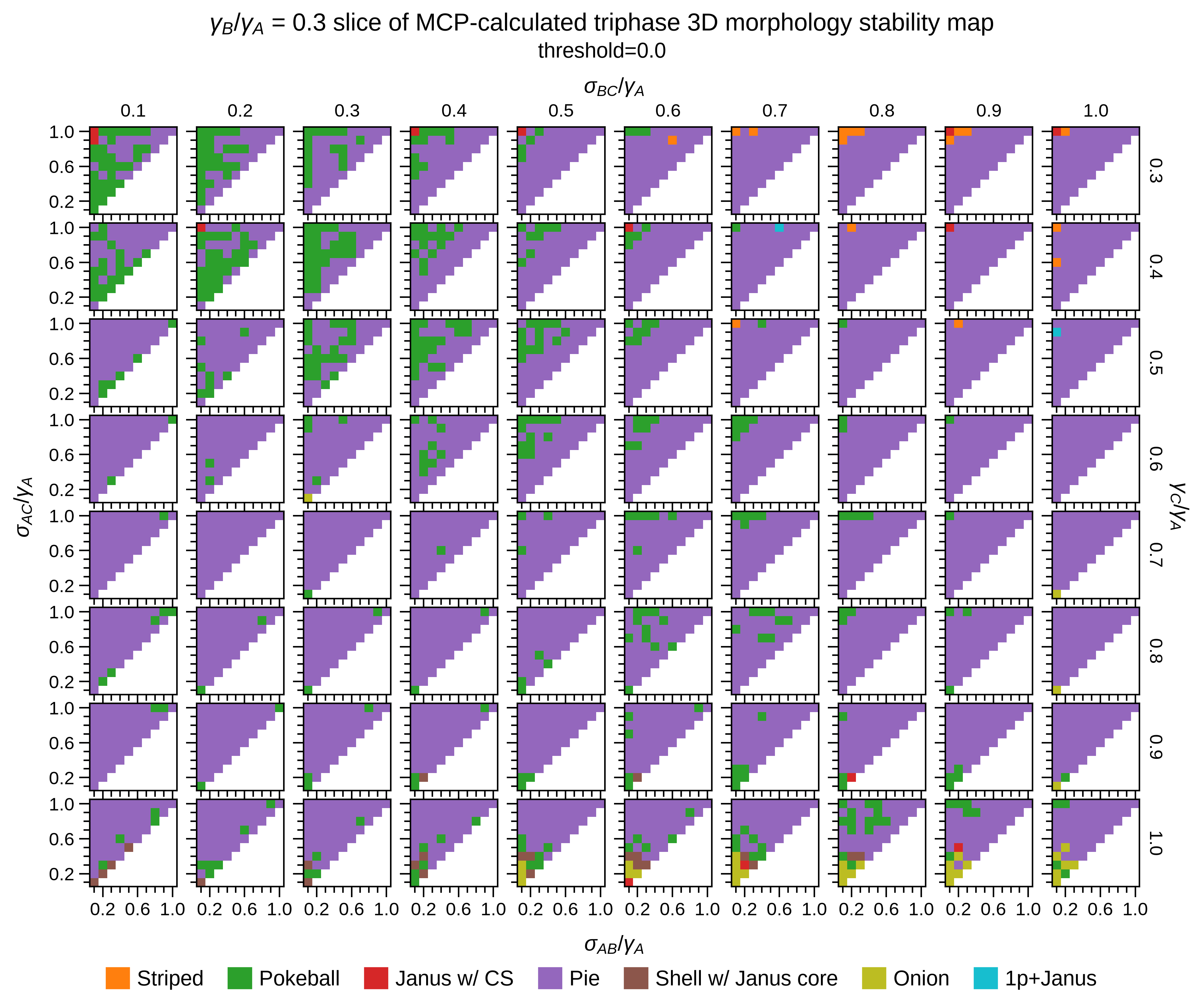}{fig:3p3d_03}\\[8pt]
\panel{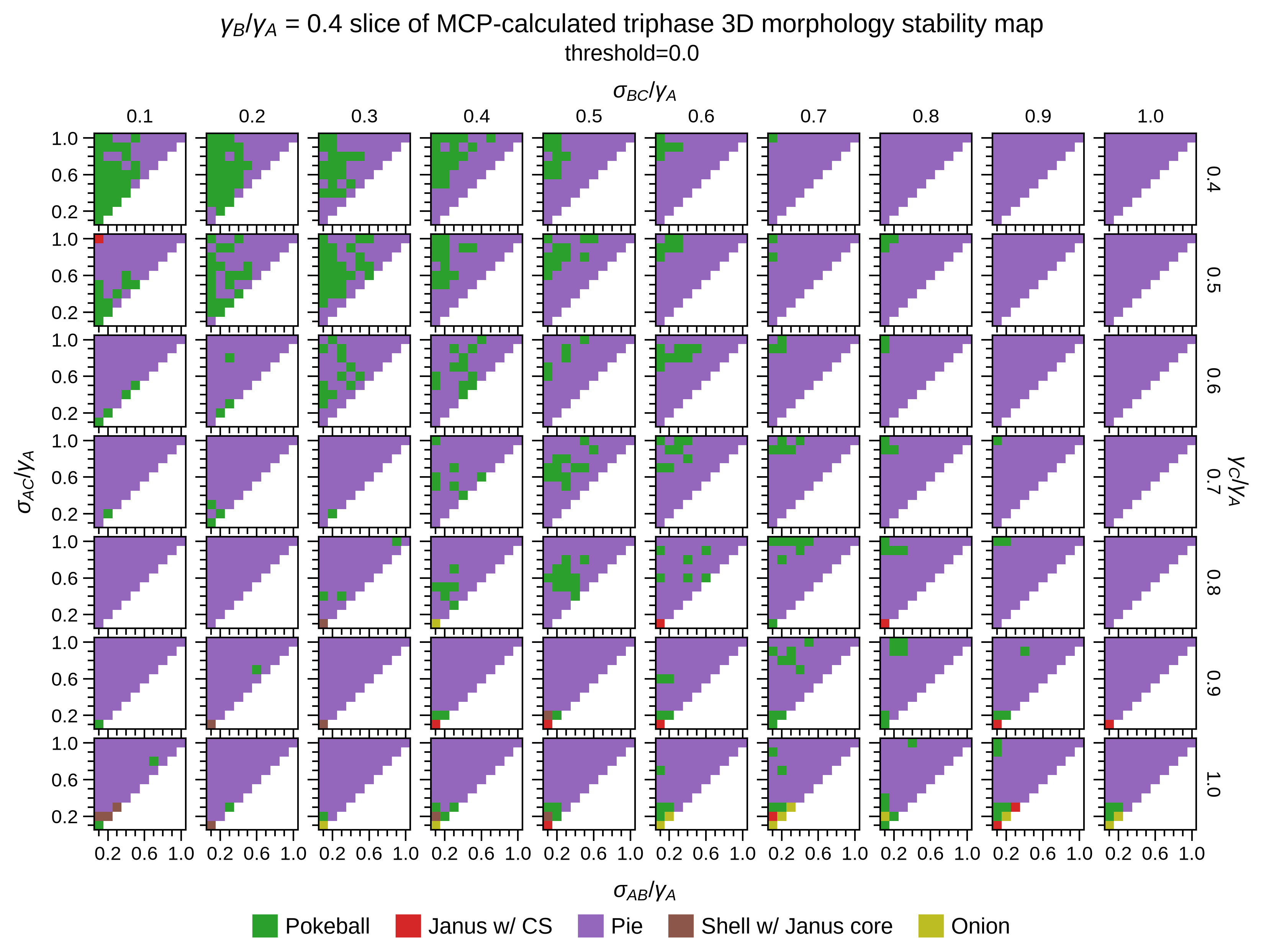}{fig:3p3d_04}\\[8pt]
\panel{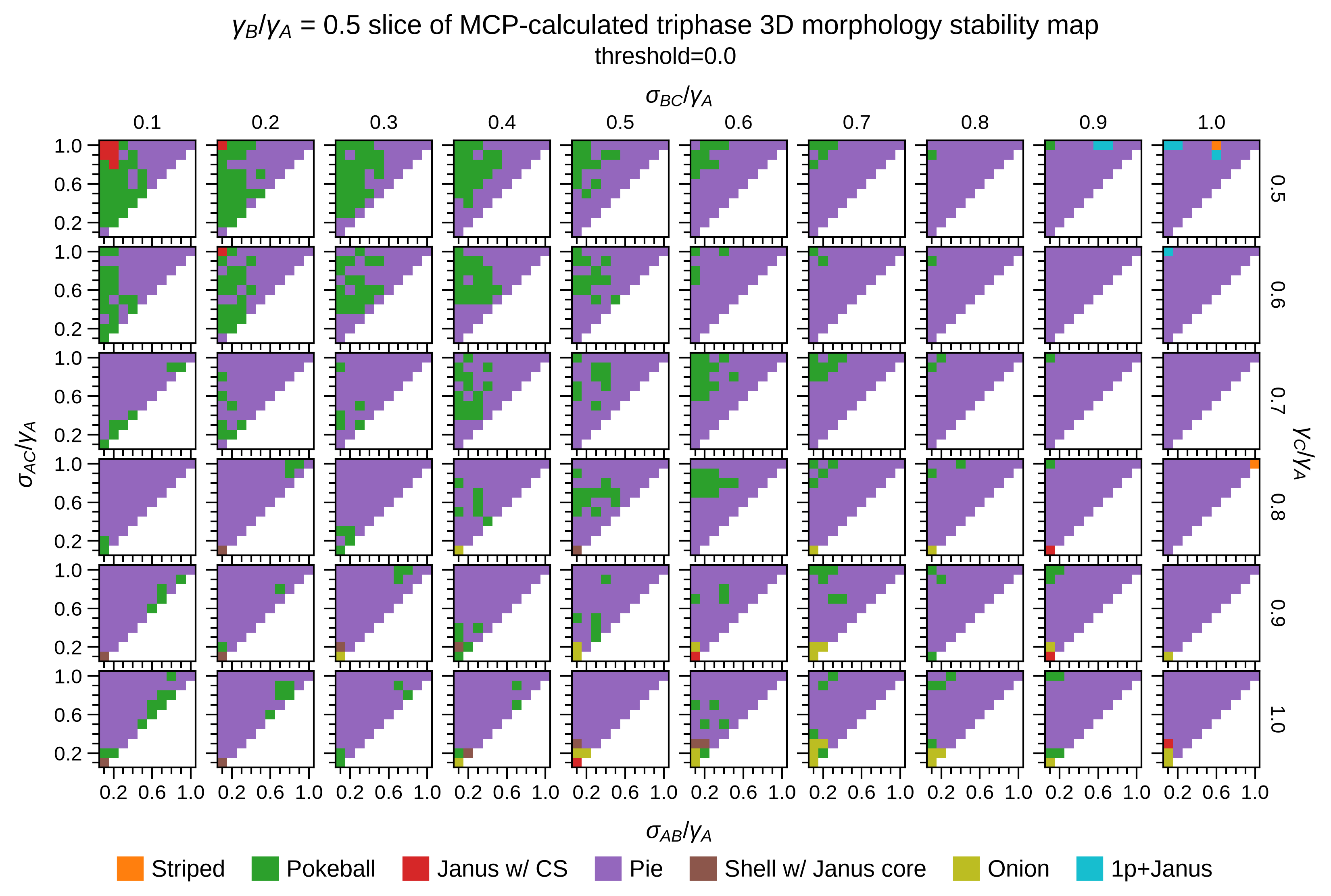}{fig:3p3d_05}\\[8pt]
\panel{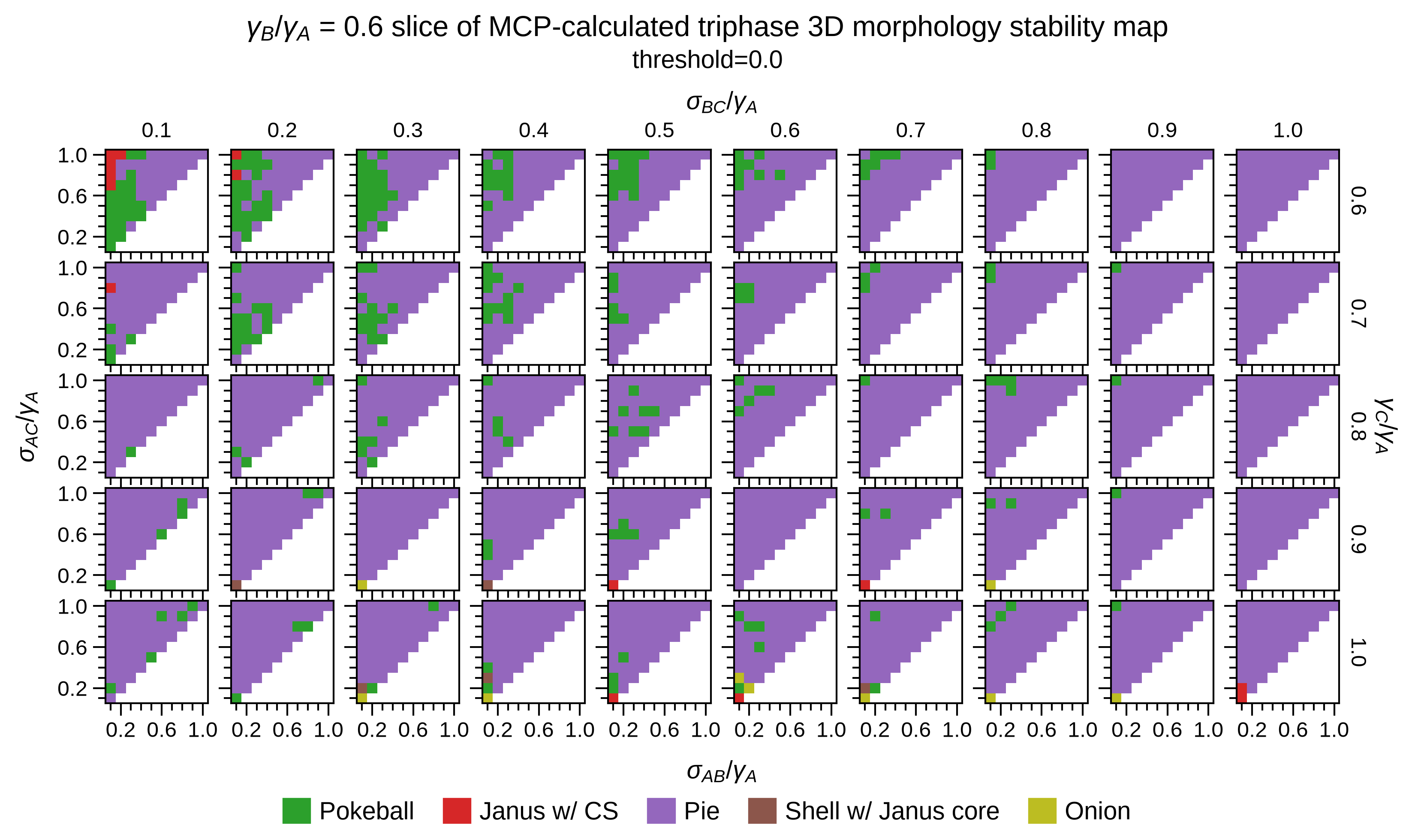}{fig:3p3d_06}\\[8pt]
\panel{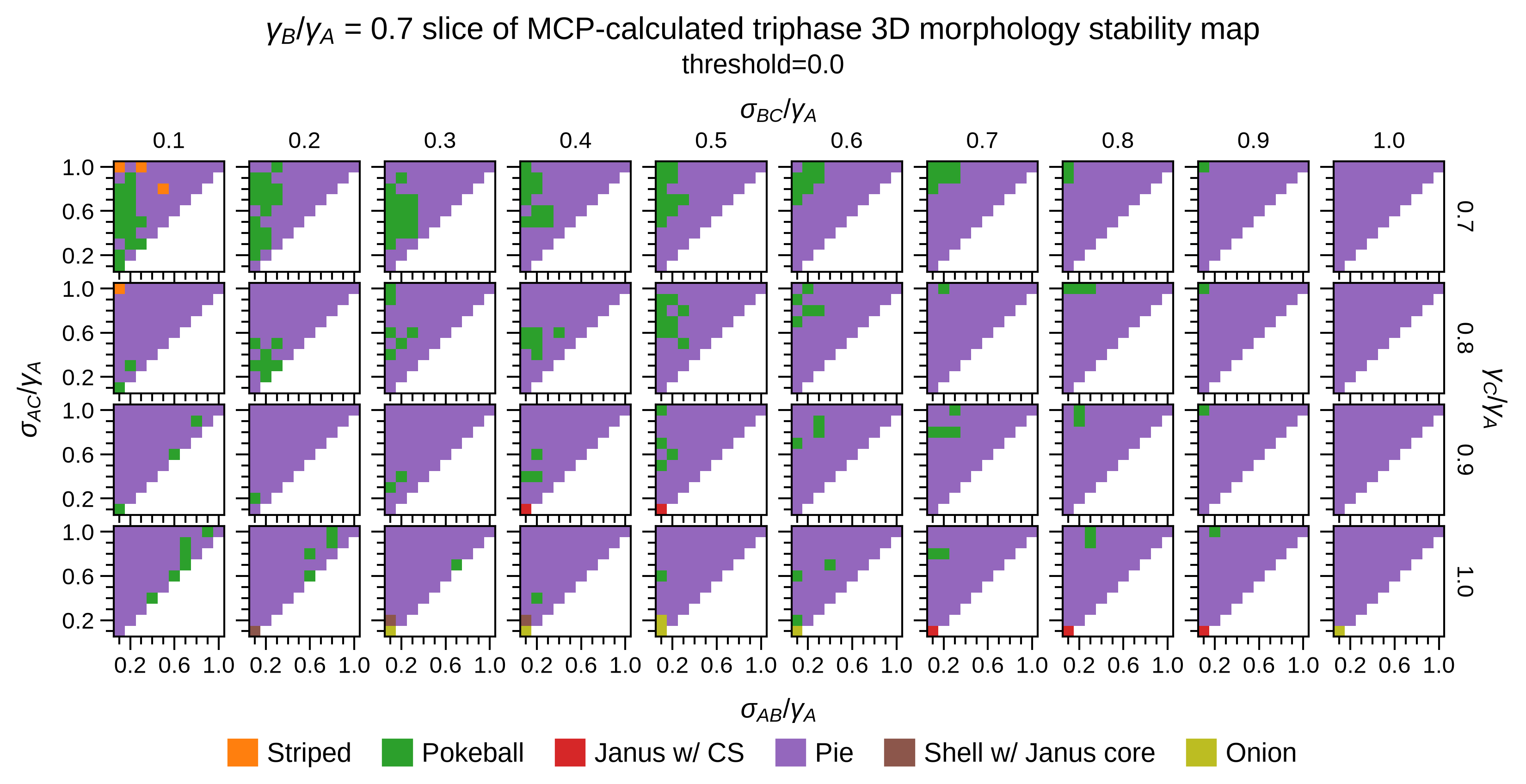}{fig:3p3d_07}\\[8pt]
\panel{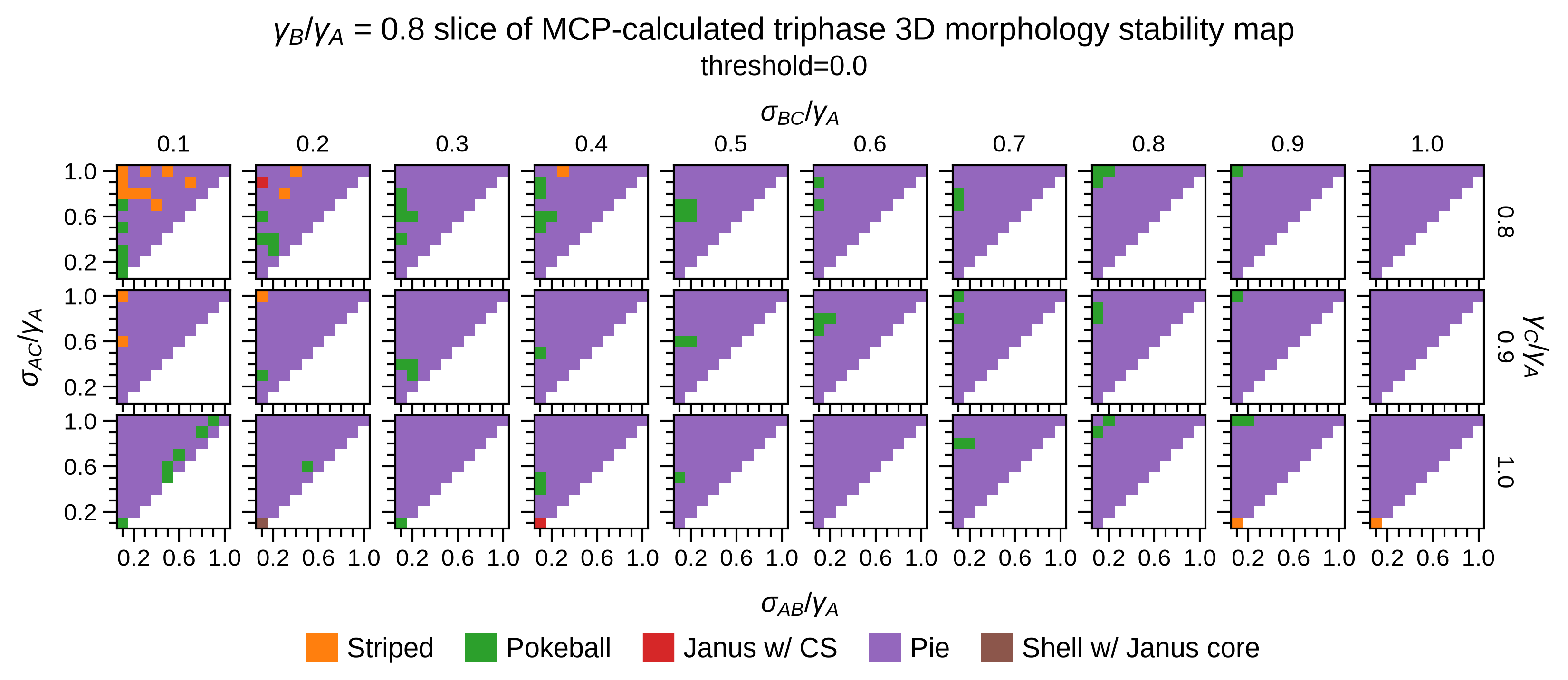}{fig:3p3d_08}\\[8pt]
\panel{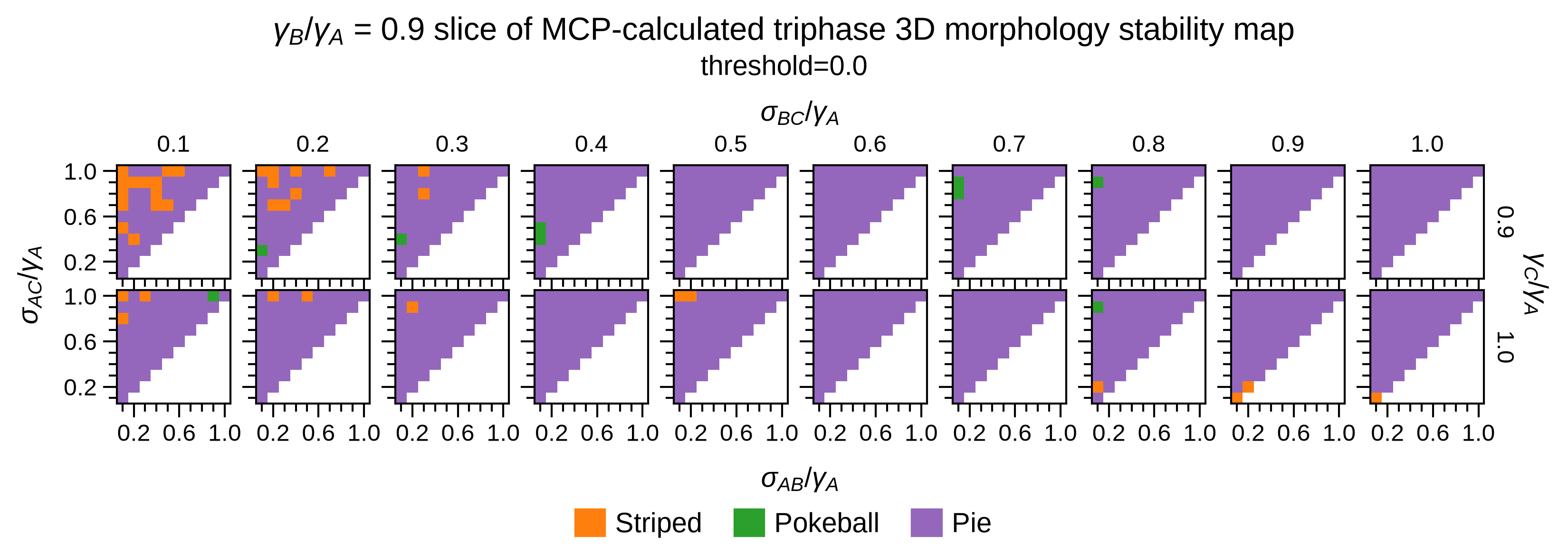}{fig:3p3d_09}\\[8pt]
\panel{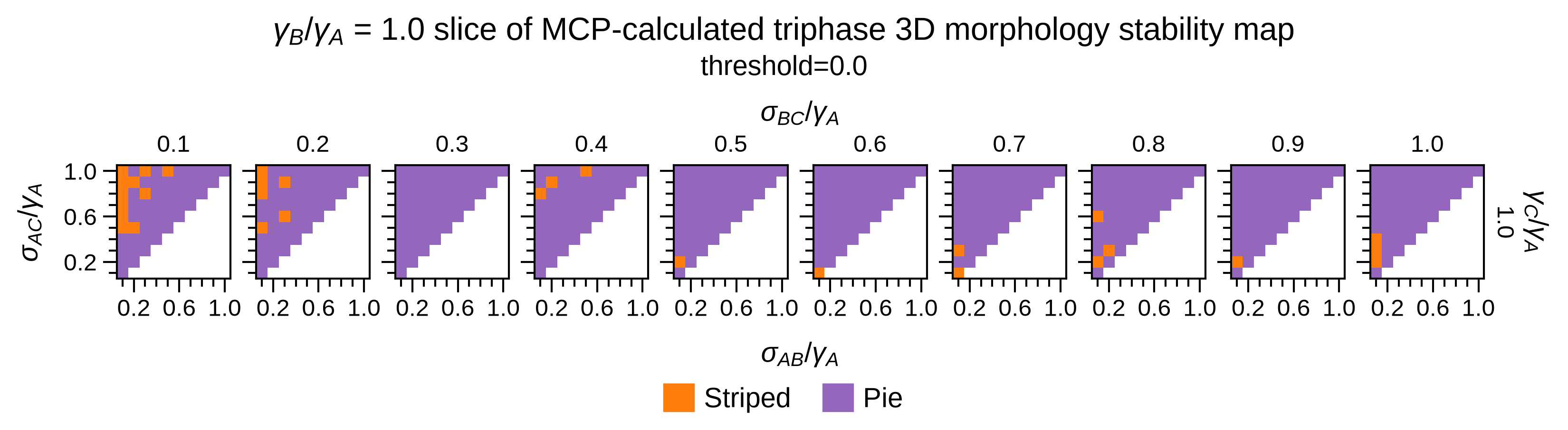}{fig:3p3d_10}
\end{center}

%% file: triphase_3d_loose_stability_map.tex
\newcommand{\figcompcaploose}{Stability map for triphase NP morphologies calculated from MCP on a 3D lattice using a domain-contact threshold of 2.0: two domains are considered not to share an interface if they have fewer than 3 NN and NNN interactions between them. A Venn morphology region appears at this threshold. Compare with Figure~\ref{fig:3p3d_stability_map}, which uses a threshold of 0.0.}

\setcounter{panelctr}{0}

\begin{center}
\captionof{figure}{\figcompcaploose}\label{fig:3p3d_loose_stability_map}

\panel{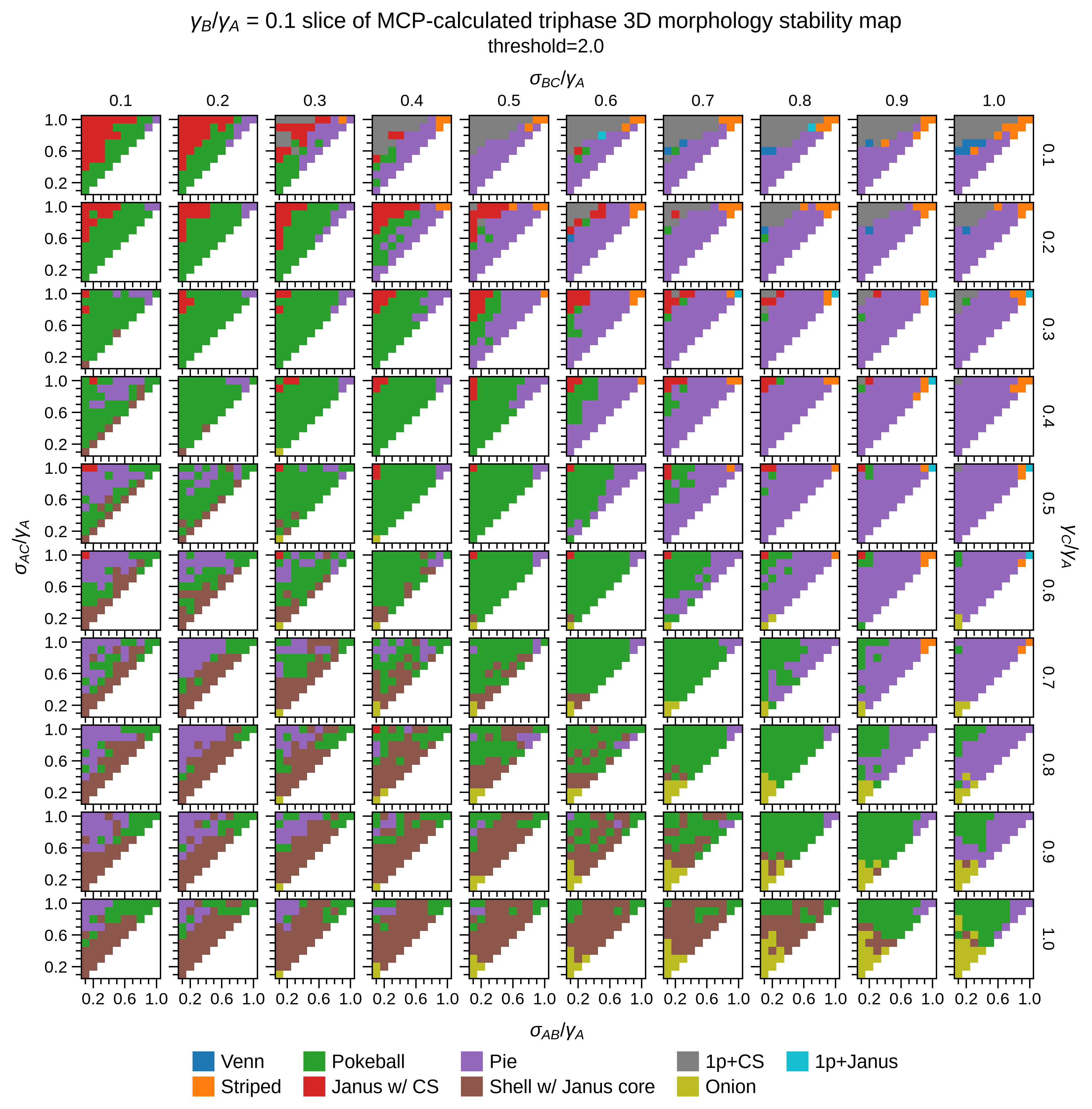}{fig:3p3d_loose_01}\\[8pt]
\panel{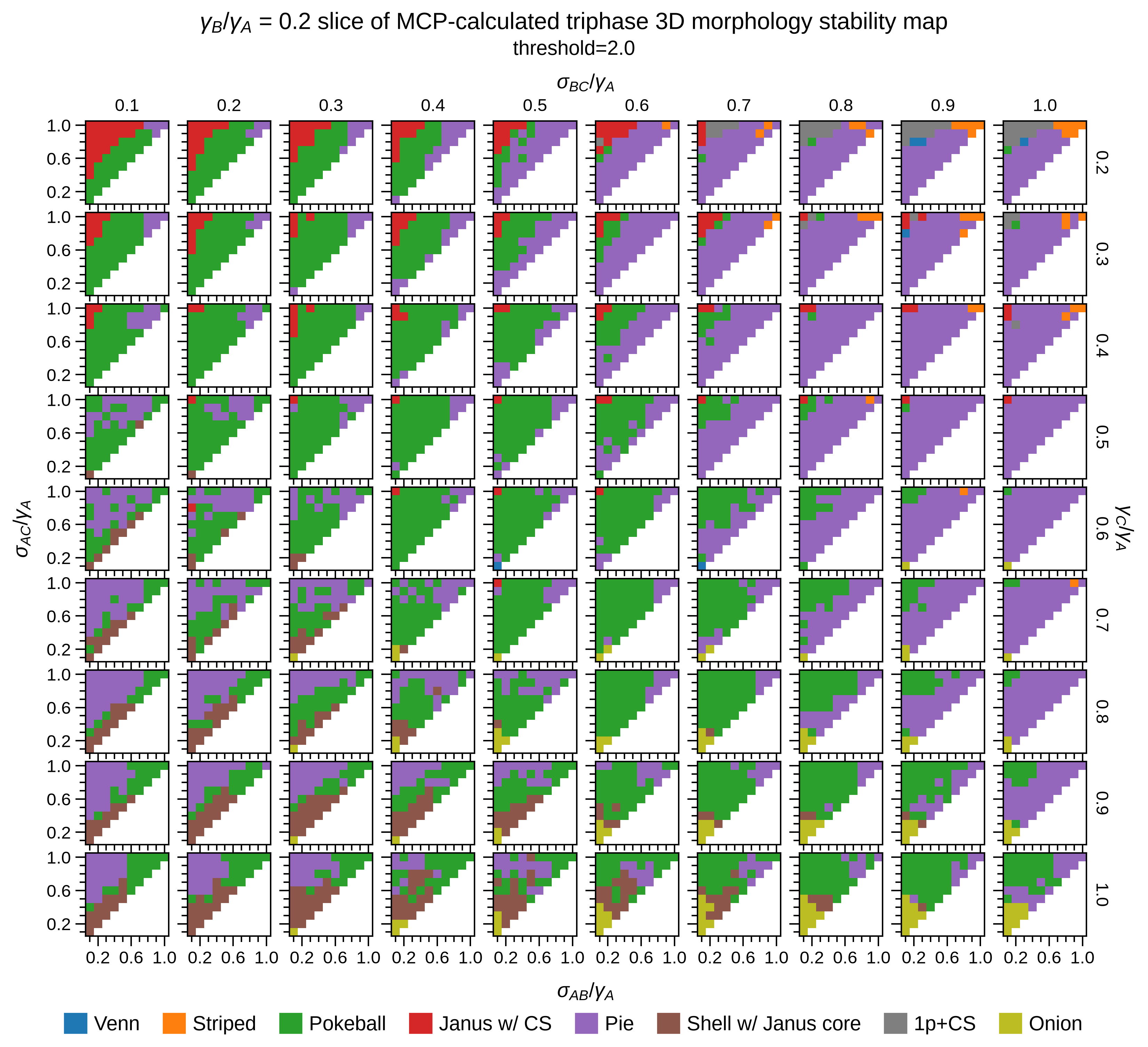}{fig:3p3d_loose_02}\\[8pt]
\panel{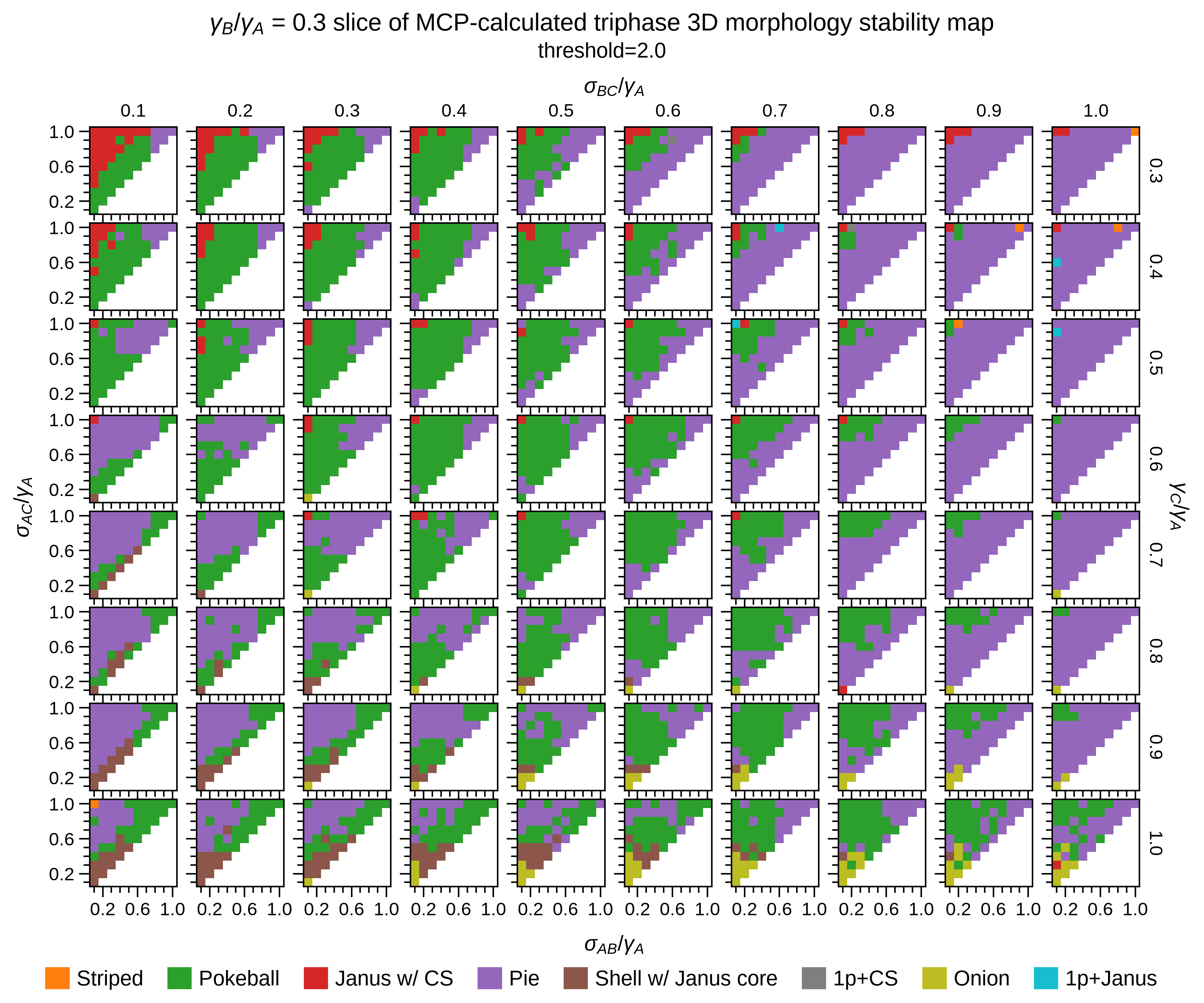}{fig:3p3d_loose_03}\\[8pt]
\panel{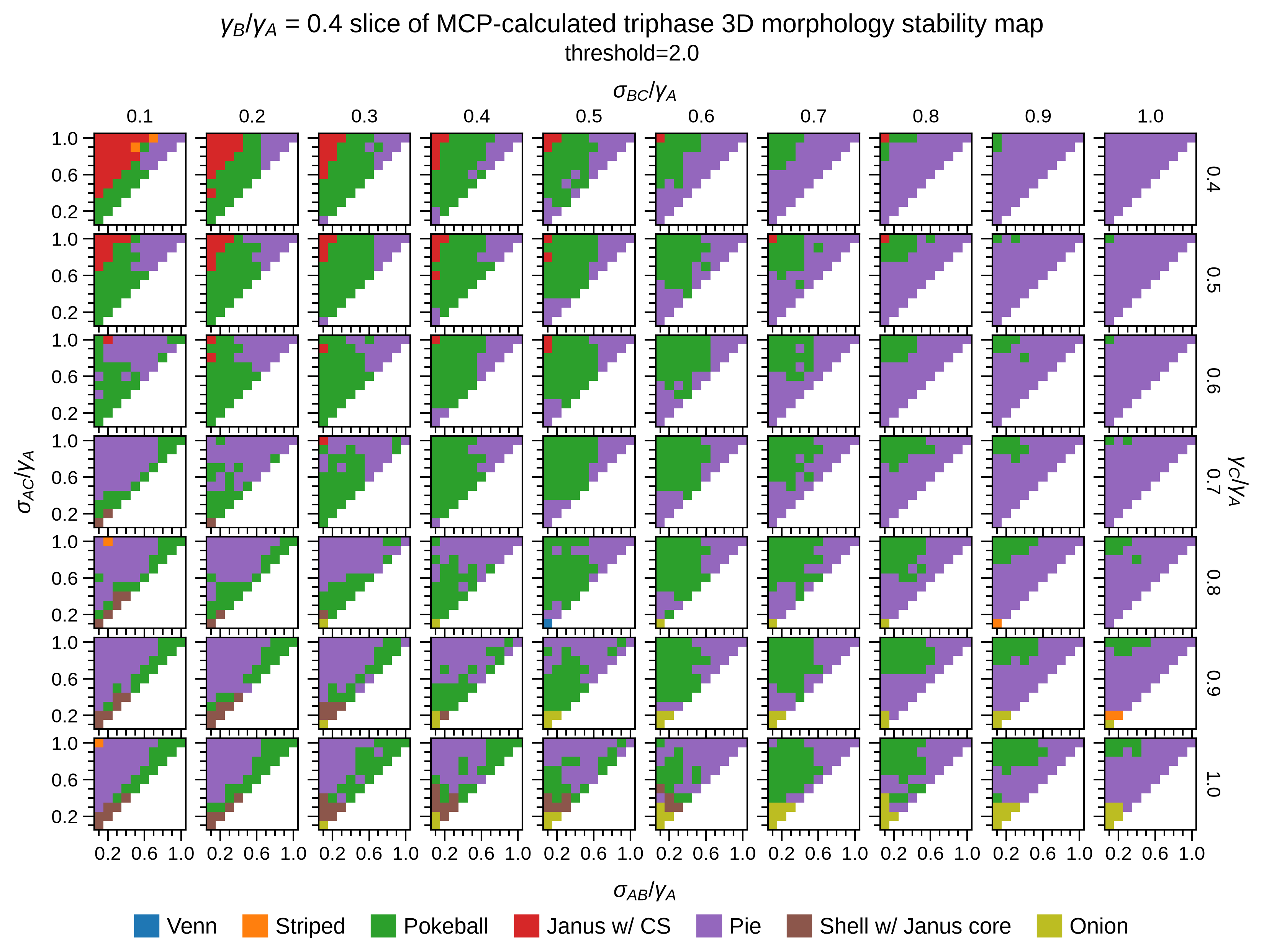}{fig:3p3d_loose_04}\\[8pt]
\panel{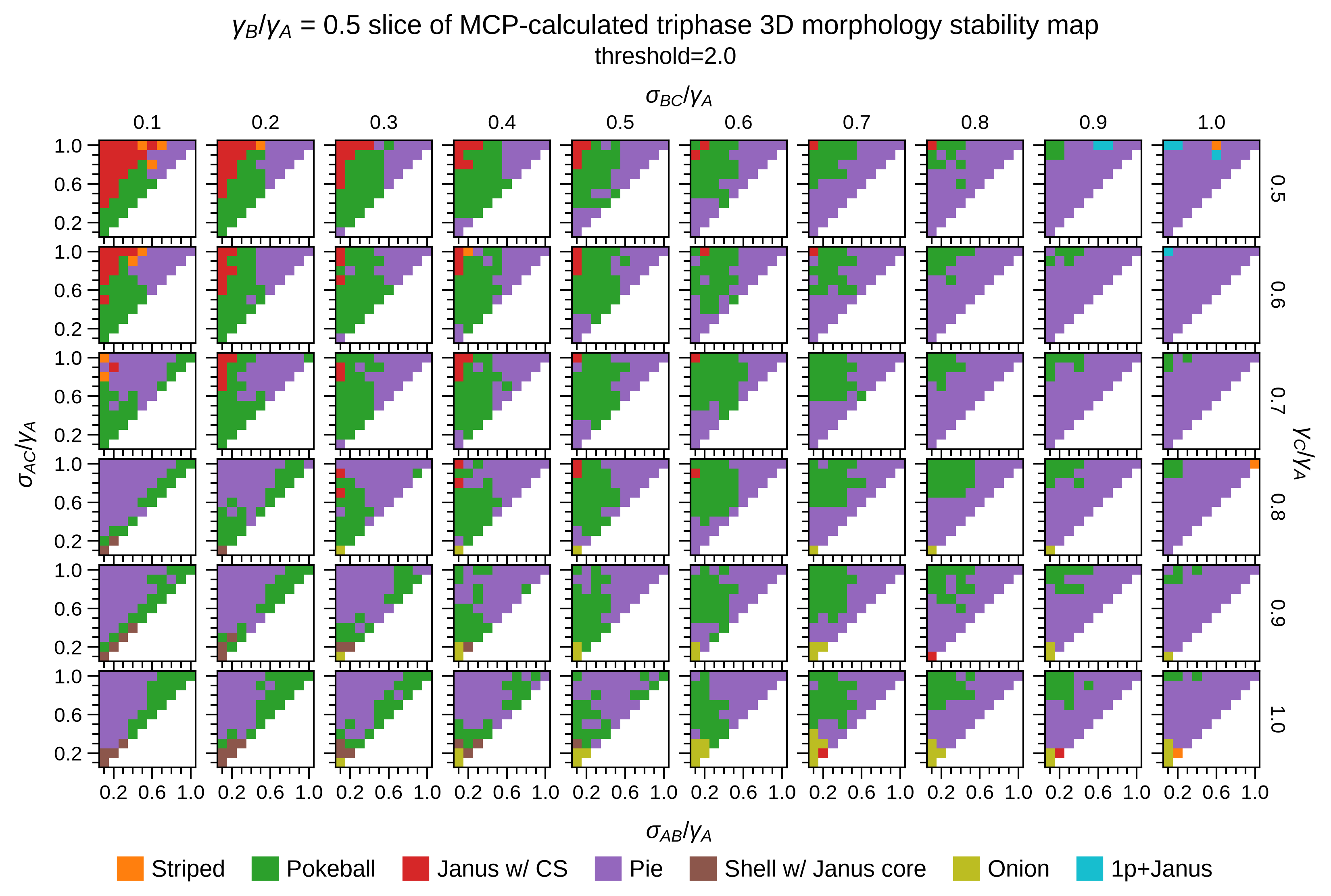}{fig:3p3d_loose_05}\\[8pt]
\panel{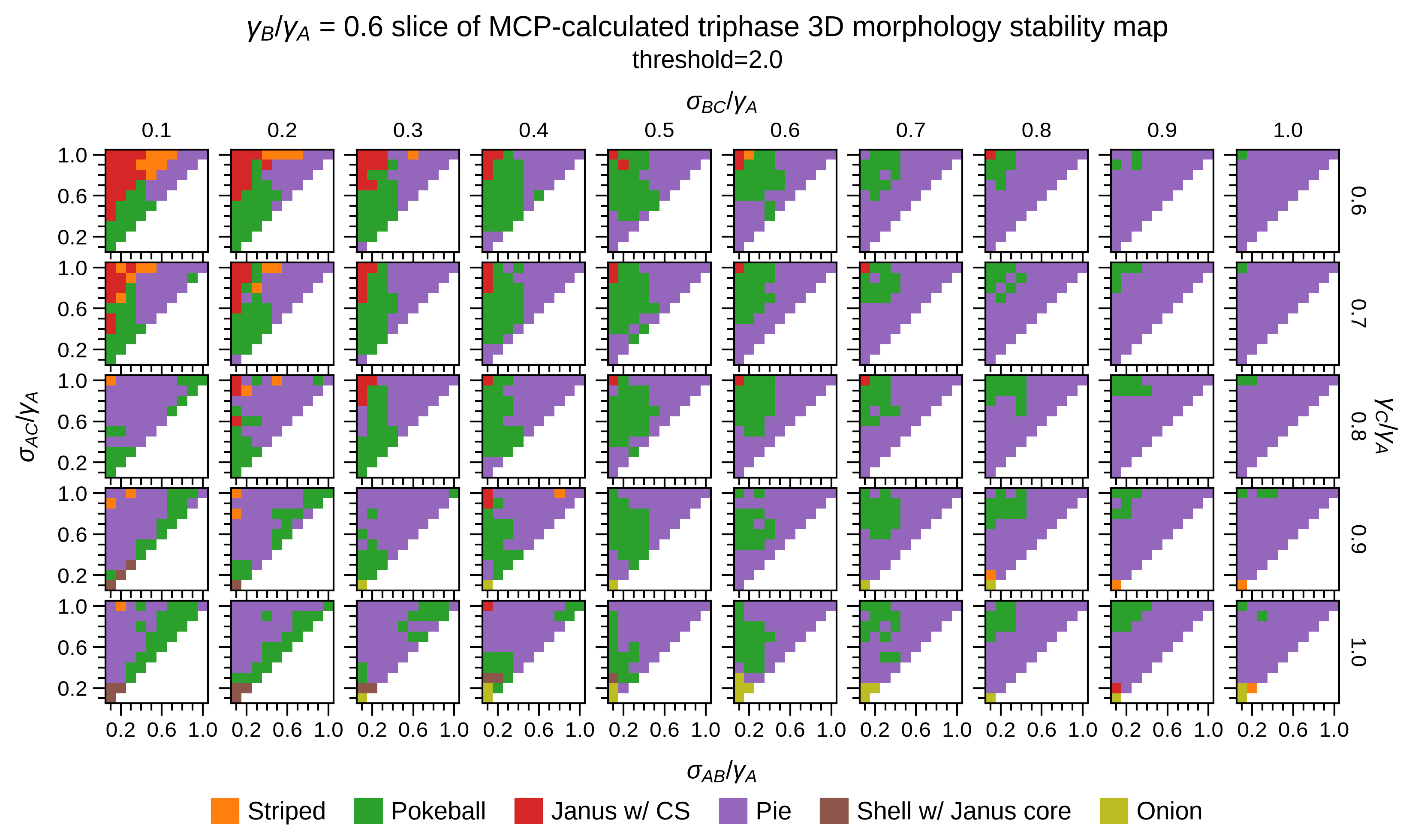}{fig:3p3d_loose_06}\\[8pt]
\panel{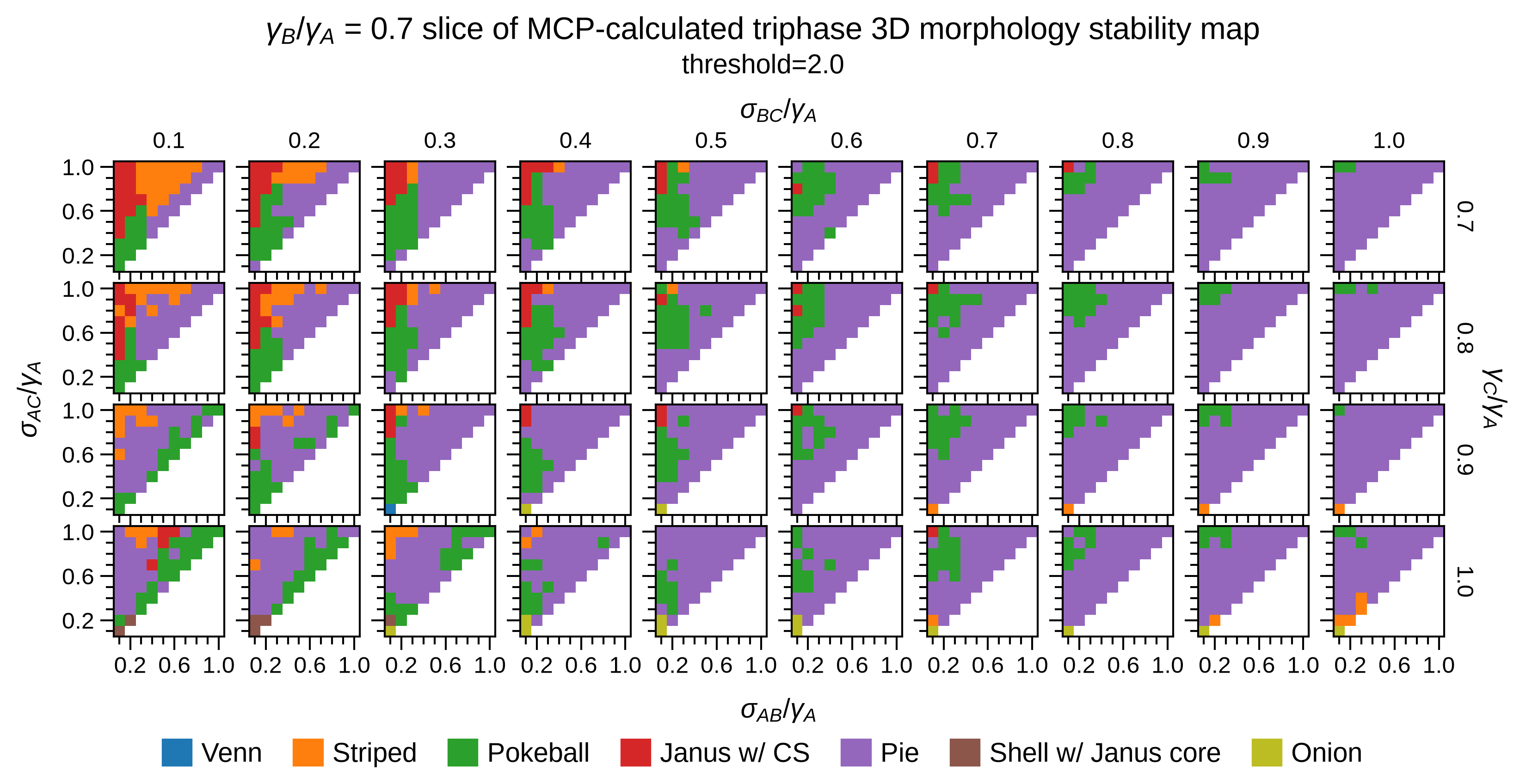}{fig:3p3d_loose_07}\\[8pt]
\panel{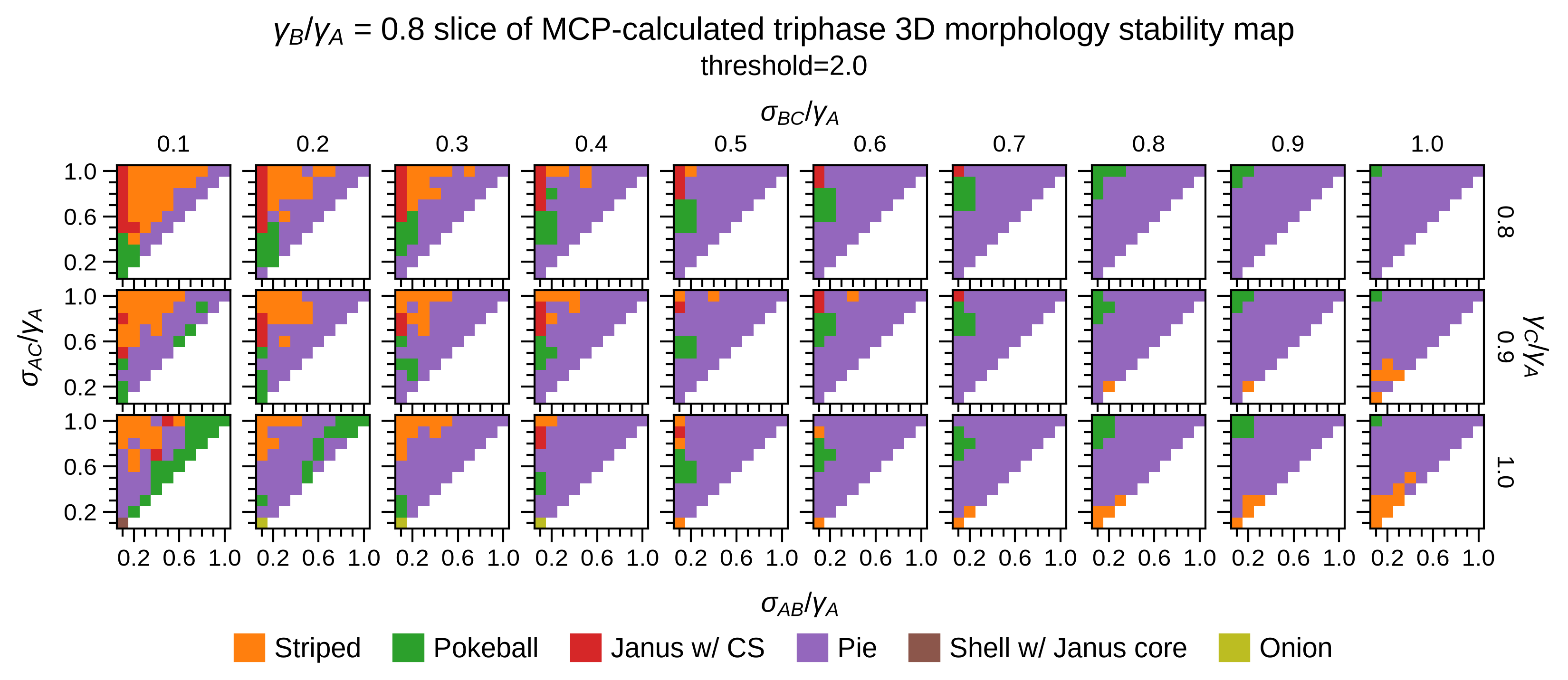}{fig:3p3d_loose_08}\\[8pt]
\panel{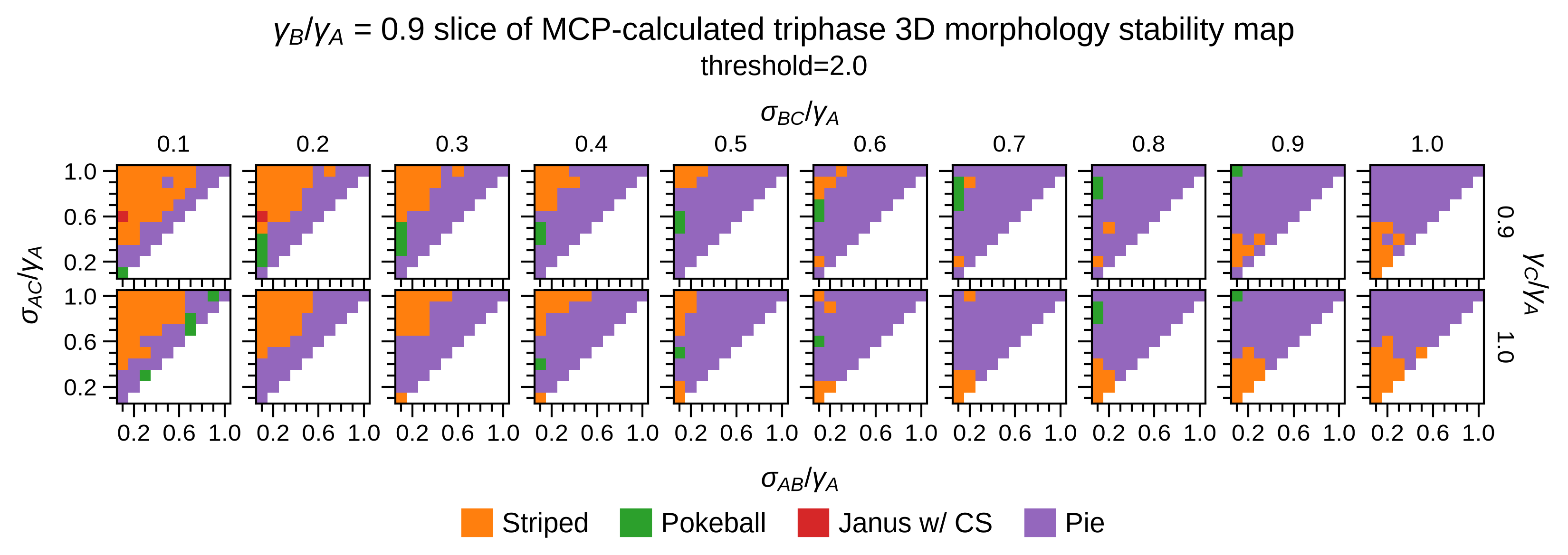}{fig:3p3d_loose_09}\\[8pt]
\panel{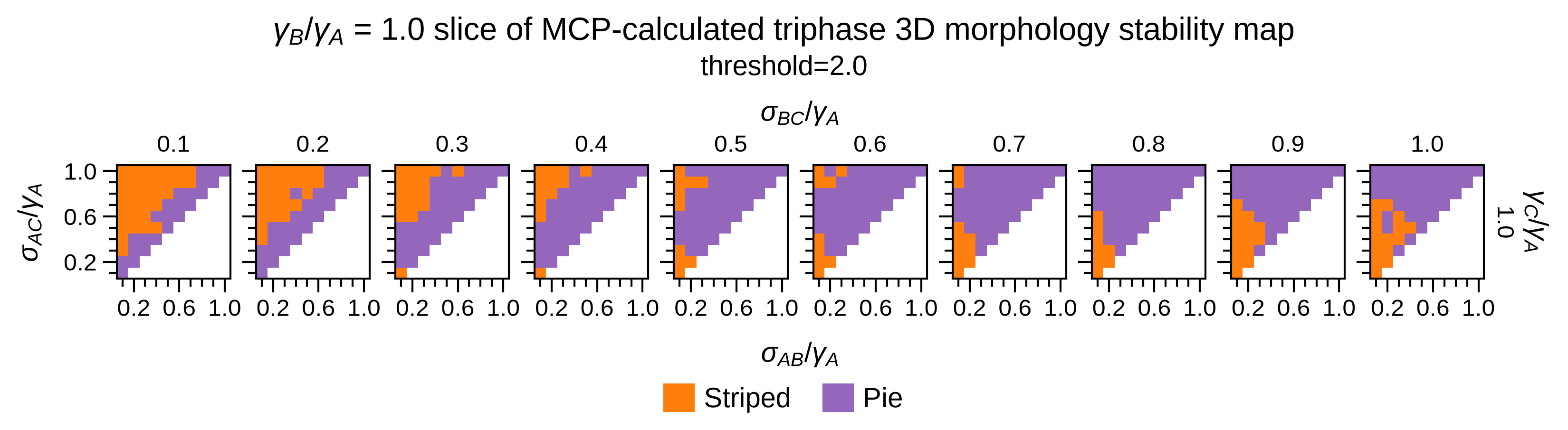}{fig:3p3d_loose_10}
\end{center}